\documentclass[]{JHEP3}

\usepackage{epsfig}

\def\al{\alpha} \def\be{\beta} \def\ga{\gamma} \def\de{\delta}
\def\ep{\epsilon}   
\def\th{\theta}   \def\ka{\kappa}
\def\la{\lambda}   
\def\si{\sigma}   
   
\def\om{\omega}   \def\Th{\Theta}
\def\La{\Lambda}   
  \def\mn{{\mu\nu}}

 \def\frac#1#2{{\textstyle{{#1}\over
{#2}}}} 
\def\lsim{\mathrel{\rlap{\lower4pt\hbox{\hskip1pt$\sim$}}
\raise1pt\hbox{$<$}}}
\def\gsim{\mathrel{\rlap{\lower4pt\hbox{\hskip1pt$\sim$}}
\raise1pt\hbox{$>$}}} \def\sqr#1#2{{\vcenter{\vbox{\hrule height.#2pt
\hbox{\vrule width.#2pt height#1pt \kern#1pt \vrule width.#2pt} \hrule
height.#2pt}}}}
\def\square{\mathchoice\sqr66\sqr66\sqr{2.1}3\sqr{1.5}3}
\def\beq{\begin{equation}} \def\eeq{\end{equation}}
\def\beqa{\begin{eqnarray}} \def\eeqa{\end{eqnarray}}
\def\eq#1{Eq. (\ref{#1})}

\title{Mimicking dark matter through a non-minimal gravitational coupling with matter}

\author{O. Bertolami \\ Departamento de F\'isica, Instituto Superior T\'ecnico, \\ Avenida Rovisco Pais 1, 1049-001 Lisboa, Portugal\\ Also at Instituto de Plasmas e Fus\~ao Nuclear, Instituto Superior T\'ecnico.\\ E-mail: \email{orfeu@cosmos.ist.utl.pt} }

\author{J. P\'aramos\\ Instituto de Plasmas e Fus\~ao Nuclear, Instituto Superior T\'ecnico, \\ Avenida Rovisco Pais 1, 1049-001 
Lisboa, Portugal \\E-mail:  \email{paramos@ist.edu}}

\date{\today}

\abstract{
In this study one resorts to the phenomenology of models endowed with a non-minimal coupling between matter and 
geometry, in order to develop a mechanism through which dynamics similar to that due to the presence of dark matter is 
generated. As a first attempt, one tries to account for the flattening of the galaxy rotation curves as an effect of the non-(covariant) 
conservation of the energy-momentum tensor of visible matter. Afterwards, one assumes instead that this non-minimal coupling 
modifies the scalar curvature in a way that can be interpreted as a dark matter component (albeit with 
negative pressure). It is concluded that it is possible to mimic known dark matter density profiles through an appropriate 
power-law coupling $f_2 = (R/R_0)^n$, with a negative index $n$ --- a fact that reflects the dominance of dark matter at large 
distances. The properties of the model are extensively discussed, and possible cosmological implications are addressed.}


\begin{document}

%%%%%%%%%%%%%%%%%%%%%%%%%%%%%%%%%%%%%%%%%%%%%%%%%%%
%%%%%%%%%%%%%%%%%%%%%%%%%%%%%%%%%%%%%%%%%%%%%%%%%%%
\section{Introduction}\label{intro}

One of the main motivations of many modifications of gravity is the search for an alternative explanation for the problem of the 
flattening of the galaxy rotation curves \cite{dm1,dm2,dm3,dm4}. In the context of usual 
$f(R)$ models \cite{fR1,fR2,fR3}, this has led to studies considering power-law curvature terms $f_1(R) \propto R^n$ in the action, 
instead of the linear curvature depicted in the General Relativity (GR) action. For $n>1$, it is found that an addition to the Newtonian 
potential of the form $\Delta \Phi(r) = -GM(r/r_c)^\be / 2r$, with $\be$ a function of $n$ and $r_c$ being a parameter characteristic of 
each galaxy, may be used to match several rotation curves: the best fit indicates that $\be = 0.817$, which arises from the choice $ n 
= 3.5$ \cite{CapoLSB}.

In another study, it is found that a metric describing a flat asymptotic orbital velocity $v $ may be 
obtained by assuming that a logarithmic factor affects the curvature term, $f_1(R) = f_0 R(1 + v \log R) $; this 
may be approached by a power-law with an exponent close to unity (given the non-relativistic condition $v 
\ll 1$), $f_1(R) \approx f_0 R^{1 + v^2}$ \cite{LoboLog}. Notice that this approach exhibits no 
galaxy-dependent parameters, and yields a universal asymptotic velocity $v$, in contrast with the 
Tully-Fisher and Faber-Jackson relations, which roughly-speaking posit a dependence of the visible 
mass $M$ of a galaxy with a power $v^m$, for spiral ($m=4$) and elliptical galaxies ($m=6$), respectively.

Usual $f(R)$ models are mostly considered as low-energy phenomenological models that attempt to grasp some of the possible implications of a more complete theory: indeed, it is known that one-loop renormalization of GR requires the introduction of higher order terms in the curvature in the Einstein-Hilbert action functional; furthermore, other available invariants --- such as contractions of the Ricci or of the Riemann tensor --- may also arise when quantum corrections arising from string theory are considered (see Ref. \cite{Sotiriou} for a thorough discussion).

This said, it is interesting to also consider another fundamental aspect common to some extensions of GR: the existence of a non-minimal coupling between matter and geometry. Indeed, while GR presents us the minimal coupling $\sqrt{-g} \mathcal{L}_m$ (a tensorial density of weight $1$) in the action, more evolved forms may arise in the context of scalar-tensor theories (see discussion of Appendix B) if matter is represented by a scalar field, or due to the effect of one-loop vacuum-polarization in the formulation of Quantum Electrodynamics in a curved spacetime \cite{QED}.

The structure of this study is as follows: in section \ref{model}, the non-minimal gravitational 
coupling 
model \cite{f2model} is introduced and its main features are discussed; session 
\ref{geo} 
presents a proposal to account for the flattening of the galaxy rotation curves, resorting to non-geodesical motion present in the 
model. In section \ref{metric} a more evolved scenario is 
studied, 
where the non-minimal coupling of geometry with a null dust matter distribution originates a 
dynamical 
effect that can be interpreted as dark matter. The dynamical features of this approach are then explored and, in section \ref{sectionmimicking}, applied to two well-known dark matter distributions: the Navarro-Frenk-White (NFW) cusped density profile and the isothermal sphere model.

\section{The model}
\label{model}

Following the introductory discussion, one postulates the following action for the 
theory \cite{f2model},

\beq S = \int \left[ {1 \over 2}f_1(R) + [1 + \la f_2(R) ] \mathcal{L}_m \right] \sqrt{-g} d^4 x ,
\label{action} \eeq

\noindent where $f_i(R)$ (with $i=1,2$) are arbitrary functions of the scalar curvature $R$, $
\mathcal{L}_m$ 
is the Lagrangian density of matter and $g$ is the metric determinant. The contribution of the 
non-minimal 
coupling of $f_2$ is gauged through the coupling constant $\la$, which has dimensions $[\la] = 
[f_2]^{-1}$. 
The standard Einstein-Hilbert action is recovered by taking $f_2=0$ and $f_1= 2 \ka (R - 2 \La)$, 
where $
\ka = c^4 /16 \pi G$ and $\La$ is the cosmological constant (from now on, one works in a unit 
system 
where $c= 1$).

Variation with respect to the metric $g_\mn$ yields the modified Einstein equations of motion, 
here 
arranged as

\beq \label{EE0} \left( F_1 + 2 \la F_2 \mathcal{L}_m \right) R_\mn - {1 \over 2} f_1 g_\mn =  
\left( \square_\mn - g_\mn \square \right) \left(F_1 + 2 \la F_2 \mathcal{L}_m \right) + \left( 1 + \la f_2 \right) T_\mn, \eeq

\noindent where one defines $\square_\mn \equiv \nabla_\mu \nabla_\nu$ for convenience, as 
well as 
$F_i(R) \equiv f_i'(R)$, and omitted the argument. The matter energy-momentum tensor is, as 
usual, defined as

\beq T_\mn = -{2 \over \sqrt{-g}} {\de \left(\sqrt{-g} \mathcal{L}_m \right) \over \de g^\mn } ,
\eeq

\noindent so that the trace of \eq{EE0} reads

\beq \label{trace0} \left( F_1 + 2 \la F_2 \mathcal{L}_m \right) R - 2 f_1 = \\  -3 
\square \left(F_1 + 2 \la F_2 \mathcal{L}_m \right) + \left( 1 + \la f_2 \right) T  .\eeq

The Bianchi identities, $\nabla^\mu G_\mn = 0$ imply the non-(covariant) conservation law

\beq \nabla^\mu T_\mn = {\la F_2 \over 1+ \la f_2} \left( g_\mn \mathcal{L}_m - T_\mn \right) 
\nabla^\mu R 
, \label{non-cons} \eeq

\noindent which is interpreted as due to an energy-momentum exchange between matter and geometry due to the 
non-trivial 
$f_1(R)$ and $f_2(R)$ terms \cite{scalar}.

Since a full study of the joint effect of a non-trivial $f_1(R) $ and $f_2(R) $ is too involved, we focus 
our attention on the latter, thus setting $f_1(R) = 2 \ka R $ (discarding the cosmological constant $
\Lambda$); this reduces \eq{EE0} to

\beq \label{EE}  \left( 1+ {\la \over \ka} F_2 \mathcal{L}_m \right) R_\mn - {1 \over 2} R g_\mn 
=   {\la \over \ka} \left( \square_\mn - g_\mn \square \right) \left( F_2 \mathcal{L}_m 
\right) + {1 
\over 2 \ka}  \left( 1 + \la f_2 \right)  T_\mn, \eeq

\noindent and, from the trace, the equivalent of \eq{trace0},

\beq \label{trace}  \left( 1 - {\la \over \ka} F_2 \mathcal{L}_m \right) R =  3{\la \over \ka} 
\square \left( F_2 \mathcal{L}_m \right) - { 1 \over 2 \ka}  \left( 1 + \la f_2 \right) T  .\eeq

\section{Flattening of the galaxy rotation curves due to non-geodesical motion}
\label{geo}

As discussed in section \ref{intro}, the flattening of the rotation curves of galaxies has been studied 
in the context of a non-trivial curvature term $f_1(R) \propto R^n$ \cite{CapoLSB,LoboLog}. These 
approaches share with several other similar studies the assumption that motion of test particles is geodesic. Clearly, although most 
extensions of GR ``inherit'' its metricity, which states that spacetime is endowed 
with a metric $g_\mn$ describing geodesics, motion is geodesical only if the energy-momentum tensor $T
\mn$ is conserved --- and, as \eq{non-cons} shows, this may not be always the 
case. Hence, one may first attempt to solve the problem of the flattening of the galaxy rotation curves 
through the effect of this non-geodesical nature of motion.

One begins by assuming that matter is a perfect fluid, with an energy-momentum tensor described by

\beq T_\mn =\left( \rho +p\right) U_{\mu }U_{\nu }+pg_{\mu \nu } , \eeq

\noindent where $\rho$ is the energy density, $p(\rho)$ is the pressure (given by a suitable 
equation of 
state) and $U_\mu$ is the four-velocity, satisfying the normalization condition $U_\mu U^\mu = 
-1$; \eq{non-cons} then translates into an additional force,

\beq \label{force0}
f^{\mu}={1 \over \rho +p} \Bigg[{\la F_2 \over 1+\la f_2}\left({\cal L}_m+p\right)\nabla_\nu
R+\nabla_\nu p \Bigg] h^\mn,
\eeq

\noindent with the projection operator $h_\mn = g_{\mn}-U_{\mu}U_{\nu}$ (so that 
$h_{\mn}U^{\mu }=0$ and 
the extra force is orthogonal to the four-velocity).

Classically, the Lagrangian density ${\cal L}_m$ of a perfect fluid has several equivalent, on-shell 
expressions: for instance, ${\cal L}_{m0} = -\rho$, ${\cal L}_{m1} = p$ and ${\cal L}_{m2} = -na$, where $n$ is 
the particle number density, and $a$ is the physical free energy, defined as $a=\rho/n-Ts$, with $T$ the fluid 
temperature and $s$ the entropy per particle. However, it was shown that, in the presence of a non-minimal 
gravitational coupling with matter, this degeneracy is lifted and one has to consider the original, bare 
Lagrangian density ${\cal L}_m = -\rho$ \cite{fluid}.

Furthermore, one may consider the non-relativistic case, whereas matter is described by a pressureless 
dust distribution for which $p \ll \rho c^2$; from a more rigorous viewpoint, the 
thermodynamical 
formulation of an action functional for a perfect fluid and subsequent relation $p = n (\partial \rho / 
\partial 
n) - \rho$ implies that a null dust distribution is characterized by an equation of state of the form $
\rho 
\propto n $. The energy-momentum tensor simplifies to $T_\mn = \rho U_\mu U_\nu $, with trace 
$T = -
\rho$.

For simplicity, one takes a static, spherically or cylindrically symmetric case. This {\it Ansatz} and 
the 
above choices yield

\beq \label{force}
f^{\mu}= - {\la F_2 \over 1+\la f_2} R'(r) h^{\mu r },\eeq

\noindent where the prime denotes differentiation with respect to the radial coordinate $r$.

Finally, one assumes that the velocity only has only a tangential component, $ U_\th = 
v(r)$, with 
the time component $U_0$ given by the normalization condition $U_\mu U^\mu = -1 $ --- which 
is valid for 
the spherical or cylindrical scenarios; this yields $ h^{\mu r} = g^{\mu r} $.

Before proceeding, one points out that the goal of this section suggests that the effect 
of the non-minimal gravitational coupling to matter is non-perturbative, $\la f_2(R) \gg 1$; this would 
yield a radically different set of equations of motion (\eq{EE}) than those of GR, and would prevent the 
classical identification $R = - 8 \pi G T $. However, this stems from a liberal use of what ``perturbative'' 
means. Indeed, one may reasonably aim for a non-perturbative modification of the geodesical motion --- which 
depend on the factor $\la F_2 / (1 + \la f_2)$, while still maintaining a perturbative regime for the metric, that 
is, equations of motion which closely resemble those of GR. This is achieved if

\beq |\la f_2(R)| \ll 1 ~~,\qquad  \la \left[F_2(R) \rho \right]'' \ll \rho, \label{pert} \eeq

\noindent having used $T = {\cal L}_m =-\rho$.

These two relations guarantee that \eq{trace} reads $R \approx -8\pi G T = 8 \pi G \rho$; assuming 
this, one may consider that the geodesical motion of a test particle is described by the usual Newtonian 
potential --- given that the non-relativistic regime is valid for a typical galaxy, $a_N \equiv GM/r^2$, $M$ being the 
visible mass of the galaxy, assuming one is sufficiently far away from its center --- while the non-geodesical 
force per unit mass$f^r$ leads to the flattening of the rotation curve; sufficiently far away from the center of the 
matter distribution, the latter should dominate, $|f^r| \gg a_N $. Assuming an asymptotic value $v(r) 
\rightarrow v_\infty$, one gets

\beq |f^r| \gg a_N \rightarrow {v_\infty^2 \over r} \gg {GM \over r^2} \rightarrow  r \gg {GM \over v_\infty^2} ,
\eeq

\noindent which, using typical values $M \sim 10^{10}~M_\odot$ and $v_\infty \sim 200~km/s$, 
yields 
$r \gtrsim 1~kpc$.

In the Newtonian limit, valid for the region under consideration $r \gtrsim 1~kpc$, one assumes a 
diagonal metric, and \eq{force} for an extra force that is purely radial. As previously discussed, one assumes that 
the field equations \eq{EE} are only weakly perturbed, so that one may substitute $R = 8 \pi 
G \rho$; this, together with condition $\la f_2 \ll 1$ and $g^{rr} \simeq 1$ yields

\beq f^r \simeq - \la F_2 R'(r) = -\la f_2'(\rho(r)) =  -{v_\infty^2 \over r} , \eeq

\noindent after using the chain rule for differentiation; given that in the subsequent sections the possibility of non-horizontal flattening 
(that is, non-vanishing slope for $v^2(r)$) will also be addressed, one considers instead a more general equation,

\beq  \la f_2'(\rho(r)) =  {v_\infty^2 \over r} \left( {r_* \over r}\right)^\be, \eeq

\noindent where $r_*$ is an unknown normalization constant and $\be$ is the outer slope of the rotation curve.

The solution then reads

\beq \la f_2(\rho(r)) =
\cases{-{v_\infty^2 \over \be} \left({r_* \over r} \right)^\be & , $\be \neq 0$
\cr  v_\infty^2 \log \left( {r \over r_*} \right) &,  $\be = 0$
},
 \label{sol}
\eeq

\noindent where $r_*$ is used in the $\be =0$ case as an integration constant ( i.e. its value is not shared between the two 
branches). Without loss of generality, one may set $\la = 
v_\infty^2 \approx 
10^{-12} $.

In order to assess the dependence of $f_2$ on the curvature, one simply needs the 
expression for $
\rho(r)$ with the above equation; assuming that, sufficiently far away from the galactic core, the 
visible 
matter density profile takes the form $\rho(r) \approx \rho_0 (a/ r)^m $ (with $m\neq 0$), one gets $r = a \sqrt[m]{\rho_0/\rho}$. 
Substituting into \eq{sol}, one has

\beq f_2(\rho) =
\cases{-{1 \over \be} \left({\rho \over \rho_*} \right)^{\be/m} & $\be \neq 0$
\cr -{1 \over m} \log \left( {\rho \over \rho_*} \right)  & $\be = 0$
},
\eeq

\noindent after defining $\rho_* =\rho_0(a/r_*)^m$. Recalling that $R \approx 8 \pi G \rho$, one 
gets

\beq f_2(R) =
\cases{-{1 \over \be} \left({R \over R_*} \right)^{\be/m} &, $\be \neq 0$
\cr -{1 \over m} \log \left( {R \over R_*} \right)  &, $\be = 0$
},
\label{logR}
\eeq

\noindent which exhibits the characteristic curvature $R_* = 8 \pi G \rho_* $. This yields the condition

\beq F_2(R) = -{1 \over m R}\left(R \over R_* \right)^{\be/m} , \eeq

\noindent for all $\be$.

One must now show that the non-minimal coupling given by \eq{logR} satisfies the perturbative conditions 
Eqs. (\ref{pert}):

\beq {v_\infty^2 \over \be} \left({r_* \over r} \right)^{\be} \ll 1 ~~,~~ \be \neq 0 \qquad , \qquad
v_\infty^2 \left| \log \left({r_* \over r} \right) \right| \ll 1 ~~ ,~~ \be =0, \label{pert2a}\eeq

\noindent and

\beq v_\infty^2 \be(1+\be) \left({r \over a} \right)^m \left({r_* \over r} \right)^\be \left({r_0 \over r} \right)^2 \ll 1 , \label{pert2b} \eeq

\noindent with the additional lengthscale $r_0 = 1/\sqrt{8\pi G \rho_0} \propto r_s$ (where the latter is the Schwarzschild radius of the 
galaxy). For a given visible matter density profile $\rho(r)$ characterized by values of $r_0$ and $a$, the former conditions (\ref
{pert2a}) bound the yet unspecified integration constant $r_* $ and, consequently, $R_*$.

Since $v_\infty^2 \simeq 10^{-12}$, and assuming that $\be,~m \sim O(1)$, one concludes that condition \eq{pert2a} is valid within 
the domain $ 1 ~kpc \lesssim r \lesssim 50 ~kpc$ (where the flattening of the rotation curves is measured) for a wide range of values 
of $r_*$: one could set, e.g., $r_*= 1~kpc$.

Regarding the validity of \eq{pert2b}, the value $r_* = 1~kpc$ is also sufficient to satisfy it within the domain $1~kpc \leq r \leq 
50~kpc$, since one expects $r_0 \sim r_s \ll 1~kpc$ and the $(r/a)^m$ factor (with $a$ typically of the order of $1 -10~kpc$) is 
suitably suppressed by the $v_\infty^2 = 10^{-12}$ term.

Hence, the 
assumption of a perturbative regime for the equations of motion \eq{EE} is self-consistent --- 
although the 
deviation from geodesical motion is purportedly relevant, dominating the Newtonian acceleration 
and leading to the flattening of the galaxy rotation curve.

Two remarks are in order: firstly, one should note that the obtained 
expression 
for the non-minimal coupling $f_2(R)$ in the flattened rotation curve case $\be = 0$ is strikingly similar to that of Ref. \cite{LoboLog}, 
where the 
non-trivial curvature term reads $f_1(R) = R[1+v^2 \log(R/R_*)] \approx R (R/R_*)^{v_\infty^2}$. Following 
those authors, one may suggestively rewrite the geometric coupling with matter depicted in \eq{logR}, 
so that action \eq{action} reads

\beq S = \int \left[ 8\pi G R + \left( {R_* \over R } \right)^\alpha \mathcal{L}_m \right] \sqrt{-
g} d^4 x ,
\label{action2} \eeq

\noindent where one defines $\al = v_\infty^2 / m$ or, conversely, one obtains the asymptotic velocity of the flattened region of the 
rotation curve through $v_\infty^2 = \sqrt\al m$.

Secondly, it is clear that this approach also displays the caveat of Ref. \cite{LoboLog}: assuming the same outer slope $m$ 
for the visible matter density profile $\rho$ of galaxies, the 
obtained 
flattening for the galaxy rotation curve is universal, that is, $v_\infty$ and $\be$ should be fixed parameters of the model, 
unadjustable 
for different galaxies. This issue might be alleviated in a more realistic approach where one considers different outer slopes $m
$ for distinct galaxies, but it would be troublesome and somewhat artificial to reconcile models for the visible matter 
density with the corresponding asymptotic velocities $v_\infty$ and different outer slope behaviour of available rotation curves (that 
is,the variety of observed values for $\be$).

One could conceivably invoke more complicated, yet unknown hidden dynamics within the model so to account for the variation from 
galaxy to galaxy --- including different density profiles, spatial symmetry and morphology. Nevertheless, this remains a noticeable 
disadvantage of this approach --- and serves as an encouragement for considering alternatives.

\subsection{Logarithmically decaying galaxy rotation curves}

In the previous paragraphs, the issue of explaining the flattening of galaxy rotation curves by resorting to the non-geodetic motion 
arising from a non-minimal coupling between matter and geometry was addressed. For generality, this was performed assuming that 
the reported flattening does not imply on an actual asymptotic behaviour of $v(r)$, and that inverse power law decays $v^2(r) \propto 
r^{-\be} $ are allowed. In this subsection, one further considers the possibility of a logarithmic decay in the outer region of the galaxy 
rotation curves, of the form of 

\beq \label{vrlog} v^2(r) = v_\infty^2  {r_* \over r} \log \left( {r \over r_*} \right). \eeq

\noindent As can be easily seen, this leads to an enclosed ``dark matter'' distribution of the form $ M_{dm}(r) \propto v^2(r) \times r 
\propto \log( r /r_*)$, so that the corresponding density is given by 

\beq 4\pi r^2 \rho_{dm}(r) = M'_{dm}(r) \rightarrow \rho_{dm}(r) \propto r^{-3}. \eeq

\noindent This scenario is significant, as the obtained dark matter density coincides with the outer profile corresponding to the NFW 
model, which shall be object of further scrutiny in the following sections (see also Appendix A).

Following the steps outlined in the previous paragraphs, one writes

\beq  \la f_2'(\rho(r)) =  {v_\infty^2 \over r} {r_* \over r } \log \left({r \over r_*}\right), \eeq

\noindent with solution

\beq \la f_2(\rho(r)) = -v_\infty^2 {r_*\over r} \left[1 + \log \left({r \over r_*}\right) \right]. \label{sollog} \eeq

\noindent As before, one sets $\la = v_\infty^2 \approx 10^{-12} $. Substituting the expression $r = a \sqrt[m]{\rho_0/\rho}$ and the 
classical approximation $R \approx 8 \pi G \rho$  into \eq{sollog} yields

\beq f_2(R) = - \left({R \over R_*}\right)^{1/m} \left[1 + {1 \over m} \log \left({R_* \over R}\right) \right]. \eeq

The perturbative conditions Eqs. (\ref{pert}) thus read

\beqa  v_\infty^2 \left({r_* \over r}\right) \left|1 + \log \left({r  \over r_*}\right) \right|& \ll & 1 ,\\ \nonumber { v_\infty^2 \over m } \left| 3 + 
\log \left({r_*\over r}\right) \right| \left({r \over a}\right)^m  \left({r_0 \over r}\right)^2  \left({r_* \over r}\right)  & \ll & 1 .\eeqa

\noindent Following the discussion of \eq{pert2a} and (\ref{pert2b}), it is straightforward to check that these conditions are satisfied 
within the domain $1~kpc < r < 10~kpc$ for a wide value of $r_*$,  i.e. $r_* = 1~kpc$.

\section{Pressureless dust with non-minimal gravitational coupling}
\label{metric}

In the following, one assumes that the motion is fairly close to geodesical, posit an appropriate non-minimal 
coupling $f_2(R)$ and ascertain under which conditions a dark matter component can be mimicked. 
As in the 
previous session, one isolates the effect of a non-minimal geometric coupling with matter by 
considering 
the usual scenario $f_1(R) = 2\ka R$. Furthermore, one assumes that the effect of the non-(covariant) conservation law \eq{non-cons} is negligible, so that test particles follow geodesics described by the metric $g_\mn$. This possibility will 
be scrutinized later on.

One ultimately aims to establish a relationship between the behaviour due to dark matter and the one arising from the model arising from \eq{action}. 
Moreover, one aims to do so by mimicking accepted density profiles for the dark 
component 
of a galaxy. This reflects the overwhelming evidence towards some yet unexplained phenomena 
that 
behaves as non- or weakly interacting matter (asides from gravitational), and the willingness of the 
authors to 
account for those results, instead of simply conjuring another hypothetical candidate among the so-called 
``dark matter zoo'' of possibilities. This marks a departure from previous efforts, which also imply the flattening of the rotation curves, 
but cannot be interpreted in terms of an actual dark matter density profile.

For generality, one considers a power law gravitational coupling of the form

\beq f_2(R) = \left( {R \over R_0} \right)^n, \label{f2} \eeq

\noindent with $n$ a yet unspecified exponent and $R_0$ a characteristic scale of the non-minimal 
coupling, given this choice for $f_2$, one freely sets $\la = 1$, so that there are only two model 
parameters, $n$ and $R_0$.

Notice that \eq{f2} does not exclude a more elaborate form for the non-minimal 
coupling: if one assumes that it is written as a series

\beq f_2(R) = \sum_p \left({R\over R_{0p}}\right)^p ,\eeq

\noindent summed over the rational exponents $p$ of the ``full'' model (integer or not), then the choice 
of \eq{f2} 
may be regarded as an approximation valid where the $p=n$ term is dominant. Likewise, this 
study is not in opposition to the findings of Ref. \cite{matter}, which approached the possibility of including a 
linear 
coupling $f_2 = R/R_1$ --- and, through the calculation of its impact on solar observables, it was found 
that $R_1 
\gg 8 \pi G \rho_\odot$, with $\rho_\odot$ the central density of the Sun.

One considers that matter within a galaxy is modeled by a pressureless perfect fluid,  i.e., 
dust; this is 
characterized by a density $\rho$, so that the energy-momentum tensor becomes

\beq T_{\mu \nu } = \rho U_{\mu }U_{\nu } \rightarrow T = -\rho. \eeq

\noindent Using the Lagrangian density ${\cal L}_m =  - \rho$ \cite{fluid}, \eq{EE} reads

\beq \label{EE1} \left[ 1 - {n \over \ka} \left({R \over R_0} \right)^n {\rho \over R} \right] R_\mn 
- {1 \over 
2} R g_\mn =  {n \over \ka} \left(  g_\mn \square -\square_\mn \right) \left[ \left({R 
\over 
R_0} \right)^n {\rho \over R} \right] + {1 \over 2 \ka}  \left[ 1 + \left({R \over R_0} 
\right)^n 
\right] \rho U_\mu U_\nu, \eeq

\noindent and \eq{trace} becomes

\beq \label{trace1} R = {1 \over 2 \ka} \left[ 1 + (1 - 2n) \left({R \over R_0} \right)^n \right] \rho 
-  {3n \over \ka} \square \left[ \left({R \over R_0} \right)^n {\rho \over R} \right]  , 
\eeq

The added terms induced by the non-minimal gravitational coupling $f_2(R)$ may be interpreted 
as 
contributing to the dark matter halo density profile $\rho_{DM}$. To see how this mimicking 
behaviour may 
arise, one first assumes that the non-minimal terms dominate very far from the galactic core 
(where the rotation curve flattens). Hence, in this exterior region, one may rewrite \eq{trace1} as

\beq \label{trace2} R \simeq  {1 -2n\over 2 \ka} \left({R \over R_0} \right)^n \rho - 3 {n\over \ka} \square 
\left[ \left({R \over R_0} \right)^n {\rho \over R} \right]  . \eeq

\subsection{Static solution}

Since the visible matter density $\rho$ is assumed to be known, solving the above equation amounts to 
obtaining 
the relation $R = R(\rho)$. It is easy to verify that the classical identification $R = 2 \ka \rho$ fails, 
since 
one would obtain a differential equation for $\rho$, which might not be satisfied by the considered 
visible matter density 
profile. By the same token, one concludes that a solution defined implicitly by the expression

\beq R = {1 -2n\over 2 \ka} \left({R \over R_0} \right)^n \rho ,\eeq

\noindent is self-consistent, since the gradient term vanishes everywhere; for this reason, one dubs 
this as 
a ``static solution''. This yields

\beq \label{solution} R = R_0 {\left[ {1 - 2n \over 2 \ka} {\rho \over R_0}\right]}^{1/(1-n)}= R_0 
{\left[ (1 - 2n ) 
{\rho \over \rho_0}\right]}^{1/(1-n)} ,\eeq

\noindent introducing the characteristic density $\rho_0 \equiv R_0/2\ka$.

Recall that the objective of this study is to obtain an effect that mimics a dark 
matter profile $\rho_{dm}$. One is tempted to extract this directly through $ 
\rho_{dm} = 2 
\ka R$; this implies that the dominance of the non-minimal gravitational coupling $f_2(R)$ effects 
is 
equivalent to the condition $\rho_{dm}>\rho$, valid when the galactic rotation curve flattens. 
However, this 
identification assumes that the effect of the non-minimal coupling $f_2(R)$ may be interpreted as 
a 
pressureless dust distribution; to qualify this, one should first analyze the full field Eqs. 
(\ref{EE1}), 
assuming solution \eq{solution} and a perfect fluid distribution for the dark matter component one 
wishes to 
identify, namely

\beq T_{(dm)\mn} =\left( \rho_{dm} +p_{dm} \right) V_{\mu }V_{\nu }+p_{dm} g_{\mu \nu } , \eeq

\noindent where $\rho_{dm}$ is the density, $p_{dm}$ the pressure and $V_\mu$ the four-velocity of the 
mimicked dark matter component.

\beqa \label{EE2} R_\mn &=& \left( {1 -2n \over 2}g_\mn + U_\mu U_\nu \right) {R \over 1-4n} 
\rightarrow \\ \nonumber 
\nonumber R_\mn - {1 \over 2} R g_\mn &=& \left( n g_\mn + U_\mu U_\nu \right) {R \over 1-4n} 
\equiv {1 \over 2 \ka} \left[ \left( \rho_{dm} +p_{dm} \right) U_\mu U_\nu +p_{dm} g_{\mu 
\nu } \right] 
. \eeqa

\noindent By assuming that the four-velocity of the mimicked dark matter component is equal to 
that of 
ordinary matter, $V_\mu = U_\mu$, one obtains the pressure and density of the former:

\beq \label{rhoandp} \rho_{dm} = {1-n \over 1-4n}2 \ka R  \qquad,\qquad p_{dm} = {n \over 1 - 4n} 2 \ka R, \eeq

\noindent and, substituting \eq{solution},

\beq \label{rhodm} \rho_{dm} = {1-n \over 1-4n} \rho_0 {\left[ (1 - 2n ) {\rho \over 
\rho_0}\right]}^{1/(1-n)} . 
\eeq

Thus, one concludes that, sufficiently far away from the galactic core, where the rotation curve 
has 
flattened, the effect of the non-minimal gravitational coupling with ordinary matter mimics a dark 
matter 
component which is dragged by ordinary matter and amounts for a non-vanishing pressure and an equation 
of state of 
the form

\beq \label{EOS} p_{dm} = {n \over 1-n} \rho_{dm}. \eeq

\noindent This dark matter-like component may have a negative density $\rho_{dm} < 0 $, if $1/4 < n 
< 1$; 
although strange, this is not a pathology of the model, as the curvature in the considered outer 
region will 
remain positive, $R > 0$.

The above treatment was pursued with no regard concerning the problematic values that the 
exponent $n$ 
might take on. Indeed, one sees that $n = 1$ produces a matter distribution with no density and 
negative 
pressure, while $n = 2$ leads to a zero curvature solution $R=0$ (see \eq{solution}) and $n=1/4$ 
yields a 
singular \eq{EE2}. A correct treatment of this cases is not developed here, as one shows 
in the following section that adequate dark matter density profiles may be mimicked by the non-minimal 
coupling with an exponent $n$ different from the discussed values.

\subsection{Mimicking dark matter density profiles}
\label{profiles}

In the context of dark matter models, it is assumed that, at the region where the galaxy rotation curve has flattened, both visible and 
dark matter might be modelled by a power law density profile, characterized by the outer slopes $m$ 
and $m'$, respectively (see Appendix A for a thorough discussion):

\beq \label{approx} \rho \approx \rho_v \left({a \over r}\right)^m ~~,\qquad \rho_{dm} \approx 
\rho_d \left({a' \over r } \right)^{m'}, \eeq

\noindent so that, in this outer region, they are related by a power law of the form $\rho_{dm} 
\propto \rho^p $, defining the slope ratio $p \equiv m' / m$. The constant $\rho_v$ is specific to each visible matter distribution $\rho$, 
and $\rho_d$ will be related to this quantity and to the previously defined $\rho_0$.

This relation is crucial to our study, since a similar power law was obtained in the previous 
session: one now resorts to \eq{rhodm} to provide a translation between the parameters $R_0$, 
$n$ of the non-minimal coupling model and the mimicked dark matter density profile quantities 
$a'$, $m'$ and $\rho_d$, assuming a given visible matter density profile characterized by $\rho_v$, $m$ and $a$.

Firstly, one obtains the following relationship between the exponent $n$ and the outer slopes $m$ 
and $m'$,

\beq \label{exprel} {1 \over 1-n} = {m' \over m} \equiv p \rightarrow n = 1 - {m \over m'} = 1- 
{1\over p},
\eeq

\noindent  which yields the power-law discussed above,

\beq \rho_{dm} = {1 \over 4 + 3p } \left( {2 \over p}-1 \right)^p \rho_0^{1-p} \rho_v^p \left({a\over r}
\right)^{m'}, \label{result1} \eeq

\noindent implying that

\beq \rho_d = {1 \over 4+3p} \left({2 \over p}-1 \right)^p \rho_0^{1-p} \rho_v^p. \eeq

\noindent Since, in this study, 
one assumes that dark matter is a dynamically generated effect arising from the non-minimal 
gravitational 
coupling $f_2(R)$, the length scale $a'$ is inherited from the visible matter 
density, so 
that $a'=a$. However, one should note that the inner behaviour of our model no longer has 
\eq{solution} as a solution, since one can no longer rely on the condition $\rho_{dm} > \rho$. This 
shall be 
discussed in the subsequent session.

\subsection{Discussion}
\label{discussion}

In order to verify the consistency of the mechanism proposed in the previous section, one now ascertains 
the 
behaviour of the trace Eq. \ref{trace1}, without the assumption of solution \eq{solution} nor 
the outer 
region approximation $\rho_{dm} \gg \rho$. To do so, one first rescales all relevant 
quantities; one 
considers the length scale $r_0 = 1/\sqrt{R_0}$ and the already used characteristic density $
\rho_0 = 
2 \ka R_0$ (so that, if $n < 0$, GR corresponds to $R_0 = 0$, that is, $ r_0 \rightarrow \infty
$). 
These define the dimensionless quantities

\beq \label{defdim} \bar{a} \equiv {a \over r_0}\qquad ,\qquad  \bar{r} \equiv {r \over r_0} \qquad, \qquad\th 
\equiv 
\left({\rho \over \rho_0} \right)^{1/(1-n)}  \qquad ,\qquad  \varrho \equiv {2\ka R \over \rho_0 \th} ={1 \over \th }{R \over R_0} , 
\eeq

\noindent and one also defines the dimensionless visible matter density $\th_* = \rho / \rho_0 = 
\th^{1-n}$, 
for convenience. Introducing the above into \eq{trace1}, one obtains

\beq \label{traceeq} \varrho = \th^{-n} + (1 - 2n) \varrho^n - {6n \over \th} \bar{\square} 
\varrho^{n-1}  , 
\eeq

\noindent where $\bar{\square} = r_0^2 \square = \square / R_0$ is the rescaled D'Alembertian 
operator. In 
this form, the classical identification $R = - 2\ka \rho $ reads $ \varrho = \th^{-n}$, while the 
solution 
\eq{solution} becomes $ \varrho = (1-2n)^{1 /( 1-n )}$.

For simplicity, one assumes that the galaxy is Newtonian, that is, that one may disregard 
relativistic 
contributions to the gradient term in the above equation. This said, one writes the D'Alembertian in 
spherical 
coordinates, assuming the spherical symmetry imposed by the visible matter density $\rho(r)$ (it 
is a 
simple matter to extend this to cylindrical coordinates and axial symmetry):

\beq \label{square} \square = {d^2 \over d \bar{r}^2} + {2 \over \bar{r}} {d \over d \bar{r}}. \eeq

Introducing the inverse coordinate $y = \bar{r}^{-1} = r_0/r$, one obtains the simplified form

\beq \bar{\square} = y^4 {d^2 \over d y^2}, \eeq

\noindent so that \eq{traceeq} becomes

\beq \label{tracefinal} \varrho = \th^{-n} + (1 - 2n) \varrho^n - {6n \over \th} y^4 {d^2 \varrho^{n-1} 
\over d 
y^2 } . \eeq

\noindent One cannot straightforwardly extract the GR result $\varrho = \th^{-n}$ from the above 
equation, 
as there is no model parameter allowing one to take the appropriate limit. Notice, however, that 
in GR $R_0 = 0 $ and $r_0$ diverges, so that the equation vanishes trivially.

The programme consists in introducing the visible matter density $\th_*(y) = \th^{1-n}(y)$ into the 
above, 
and extract the rescaled curvature $\varrho$, from which one might infer the effect of the non-
minimal 
coupling $f_2(R)$, as well as the validity of solution \eq{solution}. To do so, one resorts to the 
Hernquist 
density profile (with inner slope $\ga = 1$ and outer slope $m = 4$, see Appendix A), written as

\beq \label{Hernquist} \rho(r) = {M \over 2\pi }{a \over r} {1 \over \left(r+a \right)^3} , \eeq

\noindent so that $\rho_v = M/(2\pi a^3)$ in \eq{approx}, where $M$ is the total visible mass of 
the galaxy. 
This yields

\beqa \label{visible} \th_*(y) &=& 2\de \bar{a} {y^4 \over (1 + \bar{a} y )^3} , \\ \nonumber \th(y) 
&=& (2\de \bar{a})^{1/
(1-n)} {y^{4/(1-n)} \over (1 + \bar{a} y )^{3/(1-n)}} = (2\de \bar{a})^{m'/4} {y^{m'} 
\over (1 + 
\bar{a} y )^{3m'/4}} , \eeqa

\noindent using relation $n = 1 - 1/p = 1 - 4/m'$ and defining the dimensionless parameter

\beq \de = {M \over 4 \pi r_0^3 \rho_0} = {r_s \over r_0},\eeq

\noindent with $r_s = 2GM/c^2$ being the Schwarzschild radius of the galaxy. Clearly, in the large $r$ limit $
\th 
\propto y^{m'}$ is proportional to the desired dark matter density profile.

In order for the assumption of negligible general relativistic corrections to be self-consistent (expressed in the flat spacetime form of 
the D'Alembertian operator, \eq{square}), one assumes that the visible matter density is sufficiently low, so that these corrections do 
not dominate the effect arising from the non-trivial gravitational coupling. This translates into the condition $\de < 1$, which shall be 
tested later.

\subsubsection{Characteristic density and background matching}
\label{subsubsectionmatching}
One can now infer the characteristic density $\rho_0$:

\beq \label{rhoinfty} \rho_0 = 2 \ka R_0 = {c^2 \over 8 \pi G r_0^2 } = \left({c \over r_0 H }\right)^2 {\rho_
\infty \over 3} = \left({2.5 ~Gpc \over r_0 }\right)^2 \rho_\infty  , \eeq

\noindent where $\rho_\infty= 3H_0^2/(8\pi G)$ is the critical density of the Universe, and $H_0 \simeq 70~(km/s)/Mpc$ is the 
Hubble constant. Notice that, in a cosmological setting, the non-trivial coupling $(R/R_0)^n $ is given by $( \rho_\infty / 
\rho_0)^n $, so that the obtained relation \eq{rhoinfty} might support a previous claim: that the proposed form $f_2(R) = (R_0/R)^n$ 
could hint at a unified description of dark matter (as presented here) and dark energy, responsible for the acceleration of the expansion rate of the Universe, which is relevant in a cosmological context.

This avenue of research is not followed here, but one should nevertheless point out that this might 
account for a feature of the mechanism described in this work: not only it gives rise to a ``dark 
matter' component with adequate density profile, but it also endows it with negative pressure \eq{EOS} 
(for $n < 0$) --- an attribute usually associated with dark energy. 

This feature is not dependent on the particular geometry here adopted, and is not exclusive to the proposed power-law coupling $f_2(R)$ (with negative exponent). Indeed, it can be shown that the ``curvature pressure'' associated with the effect of either $f_1(R)$ or $f_2(R)$ involves terms in the corresponding derivatives $F_1(R)$ and $F_2(R)$ (see e.g. Ref. \cite{curvature} for a proposal of usual $f(R)$ theory in a cosmological context). As a result, the non-minimal coupling scenario yields a negative ``pressure'' if $F_2(R) < 0 $: in particular, if it takes a power-law form $f_2(R) \propto R^n$ with a negative exponent $n$.

Hence, as the visible matter density $\rho$ asymptotically falls to its background cosmological value, so does the dark matter component $\rho_{dm}$, thus 
presenting a possible smooth transition from the local galactic scenario to the averaged, large 
scale cosmological setting. Other models of unification of dark energy and dark matter include the Chaplygin gas model \cite
{Chaplygin1, Chaplygin2, Chaplygin3, Chaplygin4} and some particular field theory constructions \cite{Rosenfeld}.

This may be used to signal the distance $r_\infty$ where the total galactic density profile (including dark matter) blends with 
the averaged cosmological value, defined by $ 2 \ka R(r_\infty) \equiv \rho_\infty $. Assuming the ``static'' solution $\varrho = 
(1-2n)^{1 /( 1-n )}$, this yields

\beq \th(y_\infty) = {\rho_\infty\over \rho_0} (1 - 2n )^{-1/(1-n)}  , \eeq

\noindent which, in the large $r$ limit, reads

\beqa \label{matchingr} && (2\de \bar{a})^{1/(1-n)} y_\infty^{4/(1-n)} = {\rho_\infty\over \rho_0} (1 - 2n )^{-1/(1-n)} 
\rightarrow \\ 
\nonumber && y_\infty = \left[{3^{1-n} \over 2(1-2n) \de \bar{a} } \right]^{1/4} \left( {r_0 
H \over c} 
\right)^{(1-n)/2} \rightarrow \\ \nonumber && r_\infty = \left[{2\over 3} \times 3^n (1-2n) 
\right]^{1/4} \left( {c \over r_0 H } \right)^{(1-n)/2} \left( r_s r_0^2  a\right)^{1/4} .  \eeqa

One can interpret the distance $r_\infty$ as an indicator of the radius of the mimicked ``dark'' matter halo; for consistency, it should 
vanish when the non-minimal coupling \eq{f2} is negligible,  i.e. $R_0 \rightarrow 0$ if the exponent $n$ is negative or $R_0 
\rightarrow \infty$ for $n>0$; this is trivially satisfied, since $r_\infty \propto r_0^{n/2} = R_0^{-n} \rightarrow 0$ (excluding the 
constant coupling $n=0$).

\subsection{Crossover between visible and dark matter dominance}
\label{subsectioncrossover}

The key feature of the present proposal is that the behaviour of the non-minimal 
gravitational 
coupling, which enables the scaling law \eq{rhodm}, (cf. below \eq{defdim}) yields an appropriate dark matter-like 
component 
from a given visible matter density profile. The model parameter $R_0$ introduces both the 
length scale $r_0 $ as well as a density scale $\rho_0$; given that the discussion of dark matter is related to the 
flattening of 
the galaxy rotation curve in the outer region, where it is assumed that $r_0$ has a clearer 
physical 
meaning --- it should signal the crossover distance $r_c$ between the classical solution $R = 2\ka 
\rho 
\rightarrow \varrho = \th^{-n}$ and the outer solution $ R = R_0 \left[ (1 - 2n ) (\rho / \rho_0) 
\right]^{1/(1-n)} 
\rightarrow \varrho = (1 - 2n )^{1/(1-n)}$. This should occur when both solutions are similar,

\beq \th^{-n} = (1 - 2n)^{1/(1-n)} \rightarrow \th_*(y) = (1-2n)^{-1/n}. \eeq

\noindent Assuming that this crossover occurs about the outer region of the galaxy, so that $r_c > 
a $, one 
might still approximate the visible matter density profile by its outer slope profile, obtaining

\beqa  \label{rc} \th_*(y) & = & 2 \de r_0^3 {a \over r_c^4} = (1-2n)^{-1/n} \rightarrow \\ \nonumber 
r_c^4 & = 
& 2 (1-2n)^{1/n} \de r_0^3 a = 2 (1-2n)^{1/n} r_s r_0^2 a. \label{r_c} \eeqa

Hence, the crossover radius $r_c$ is not simply proportional to $r_0$, as one might naively 
expect, but 
involves a sort of ``geometric mean'' of all the length scales present. This translates into the 
dimensionless 
crossover coordinate $y_c$, given by

\beq y_c^{-4} = 2 (1-2n)^{1/n} \de \bar{a} . \label{yc} \eeq

\subsection{Behaviour of the gradient term}
\label{subsectiongradient}

\subsubsection{Inner region ($r < a < r_c$) behaviour}

The discussion above assumes that the gradient term on the {\it r.h.s} of \eq{tracefinal} is 
negligible --- which is valid for the dark matter dominated region, $\varrho \sim  (1 - 2n )^{1/(1-n)}
$. However, one should analyze the relevance of the gradient term present in \eq{tracefinal} as one 
approaches the 
inner regions of the galaxy, $y \rightarrow \infty$, since in this region the power-law \eq{approx} is 
no longer 
an accurate description of the visible matter density profile, and large gradients may arise. From \eq{tracefinal}, one sees that 
the gradient term is negligible if

\beq \label{gradient} \left| 6n{y^4 \over \th}  {d^2 \varrho^{n-1} \over d y^2 } \right| \ll \left| \th^{-n} \right| 
\rightarrow \left| 6n {y^4 \over \th_* } {d^2 \th_*^n \over d y^2 } \right|  \ll 1. \eeq

At distances $r<a$, the Hernquist density profile reads

\beq \rho(r) = {M \over 2 \pi a^2 r} \rightarrow \th_*(y) = 2 \de {y \over \bar{a}^2}. \eeq

In the interior region where the visible matter density is ruled by its inner slope behaviour, $r_c < 
a $
( $\bar{a} y_c > 1$), the consistency condition \eq{gradient} reads

\beq \label{gradient2} \left| 6n {y^4 \over \th_*} {d^2 \th_*^n \over d y^2 } \right| = \left| 6n^2 ( n-1 ) 
y^{n+1} 
\left( {2 \de \over \bar{a}^2 } \right)^{n-1} \right| \ll 1. \eeq

\noindent One first studies the $n > -1$ case, thus defining the quantities $y_{k(in)}$ and $r_{k(in)}$

\beqa \label{kin} y &\ll & y_{k(in)} \equiv \left[ {1 \over 6n^2 (1-n)} \left( {2 \de \over \bar{a}^2 } 
\right)^{1-n} \right]^{1/(n+1)} \rightarrow \\ \nonumber r &\gg& r_{k(in)} \equiv \left[ 6n^2 (1-n) 
\left( {a^2 \over 2 r_s} \right)^{1-n} r_0^{2n} \right]^{1/(n+1)}.   \eeqa

\noindent As one approaches the center of the galaxy, $y \rightarrow \infty$, the relevance of the 
gradient term of \eq{tracefinal} increases, so that eventually one gets $ y > y_{k(in)} $(at shorter distances one would have to include 
relativistic corrections of the form $r_s/r$, which are not considered here).

Although not exhibited here, numerical results show that this behaviour is not a direct consequence 
of the visible matter inner density profile: instead of the inner slope of the Hernquist distribution, one 
might choose a density profile that flattens towards the galactic core (that is, so that $( d\rho/dr )|
_{r=0} = 0$ --- see e.g. Ref. \cite{Burkert}); however, since $|d\rho / dr|$ increases outwards, the gradient term in 
the {\it r.h.s} of the equation of motion \eq{tracefinal} would eventually come to dominate, albeit at 
a different region. The bottom line is that, for any given visible matter density profile $\rho$, the 
gradient term will eventually dominate below a threshold $r < r_{k(in)}$.

However, the argument implicitly assumes that the Newtonian regime extends down to these core 
regions; although one does not pursue a full treatment of the relativistic regime (when $r \lesssim 
r_s$), one expects the behaviour of the gradient term to be significantly different in these region, 
since the D'Alembertian operator is no longer of the form $y^4 d^2 / dy^2$ (in particular, 
corrections of the form $r_s/r = \de y$ arise, which are deferred to a future study). Hence, one 
may  circumvent this issue by ascertaining that $r_{k(in)}<r_s \rightarrow \de y_{k(in)} > 1 $. This 
yields the condition

\beq \label{gradient_nneq-1}  { \de \over \bar{a}^{1-n} } =  { r_s r_0^{-n}\over a^{1-n} } > -n
\sqrt{3\times 2^n(1-n)} ,\eeq

In the case $n=-1$, condition \eq{gradient2} reads 

\beq \label{gradient_n-1}  {\de \over \bar{a}^2 } = {r_s r_0 \over a^2} > \sqrt{3} , \eeq

\noindent which coincides with \eq{gradient_nneq-1}.

\subsubsection{Tracking behaviour of gradient solution}
\label{subsubsectiontracking}

As discussed above, a suitable choice of the model parameter $r_0$ ensures that the gradient term on the {\it r.h.s.} of 
\eq{tracefinal} dominates only within the relativistic region $r < r_s$, where the Newtonian approximation breaks down. 
However, a numerical analysis (which will be displayed later, for two relevant cases) indicates that it is impossible to further 
tweak the model so that this gradient term will not become dominant after the cross-slope region $r\sim a$, while still mimicking the 
dark matter behaviour. Hence, one concludes that the ``static'' solution \eq{solution}, given by $ \varrho = 
(1-2n)^{1 /( 1-n )}$ is not valid: in particular, the discussion of the dominance of the gradient term in the outer region $a < r < 
r_c$ is flawed, as the crossover distance $r_c$ assumed for its validity. 

Hence, instead of the crossover $r_c$, one must consider the distance $r_k > a$ above which the gradient term dominates (or, 
equivalently, a dimensionless coordinate $y_k) < 1/\bar{a}$); one may write it as $r_k= \xi a \rightarrow y_k = 1/(\xi \bar{a})$ 
(analogously to the relation $r_c = \ep a$ used to study the ``static'' solution).

Using the outer slope Hernquist profile, $\th_*(y) = 2 \de  \bar{a} y^4$, \eq{gradient} reads (for negative $n$)

\beq   \left| 6n {y^4 \over \th_* } {d^2 \th_*^n \over d y^2 } \right| =  24 n^2 (1-4n) 
\left( 2 \de \bar{a} \right)^{n-1} y^{4n-2} \ll 1, \eeq

\noindent so that, when $y = y_k = 1/(\xi \bar{a})$, one has

\beq \label{xi} 12 \times 2^n n^2 (1-4n) \de^{n-1} \bar{a}^{1-3n} \xi^{2-4n} = 1, \eeq

\noindent which is used to set $r_0$, for a given $n$.

One must now look at the behaviour of \eq{tracefinal} when it is dominated by its gradient term, thus reading

\beq \label{tracekin}{d^2 \varrho^{n-1} \over d y^2 }  \approx - {\th \over 6n y^4  }  \varrho \approx = - {\left(2 \de \bar{a} y^{4n} 
\right)^{1/(1-n)} \over 6n}   \varrho ,\eeq

\noindent having substituted the outer slope Hernquist profile, $\th_*(y) = 2 \de  \bar{a} y^4$. This equation admits what one 
dubs as a gradient solution: clearly, this cannot be a constant solution $\varrho \sim {\rm const.}$ (like the ``static'' one $\varrho = 
(1-2n)^{1 /( 1-n )}$). 

Since one is interested in the behaviour of the above equation in the outer region $r > a$, it is tempting to resort to a series 
expansion around $y =0$. However, this would require the stronger assumption $r > r_0$, which may not be valid; in 
particular, if the characteristic density $\rho_0$ has an order of magnitude similar to the cosmological background $\rho_0$, 
\eq{rhoinfty} will yield $r_0 \sim 100~Gpc$, which is much larger than the length scale of a galaxy. However, one may instead define 
the coordinate $ z = \bar{a}y = a/r$ and the dimensionless function $\psi = \varrho^{n-1}$, rewriting \eq{tracekin} in the Emden-
Fowler form

\beq \label{EF} {d^2 \psi\over d z^2} = \Th z^{4n/(1-n)} \psi^{1/(n-1)}, \eeq

\noindent with

\beq \Th = -{\left( 2 \bar{a}^{3(1-2n)} \de \right)^{1/(1-n)} \over 6n} . \eeq

For a given $n$, the solution to this differential equation may be obtained implicitly, in a parameterized form $z=z(\tau,\Th,n),
\Psi=\varrho(\tau,\Th,n)$) \cite{difeqs}; instead of pursuing this possibility or analyzing some relevant cases (such as the $n=-1$ 
and $n=-1/3$ scenarios studied in a following section), one undertakes a more physically oriented, albeit less rigorous, 
exploration of the values of interest.

In order to do so, one may ascertain the order of magnitude of the {\it r.h.s.} of \eq{tracekin} as the gradient term begins to 
dominate, at $y = y_k = \xi a$. One considers the classical GR solution $ \varrho = \th^{-n}$, valid up to this point, 
that is

\beq \label{tracekin2}  \left( {d^2 \varrho^{n-1} \over d y^2 }\right)_{y=y_k} = - {\th^{1-n}(y_k )\over 6n y_k^4  } = - {\th_*(y_k) 
\over 6n y_k^4  } = -{\de \bar{a} \over 3n }.\eeq

For typical values of $a=1-10~kpc$ and $r_s =  10^{-3}-10^{-1}~pc$ (corresponding to galaxies of approximately 
$10^{10}-10^{12}$ solar masses), condition \eq{gradient_nneq-1} yields $r_0 \gtrsim 1~Gpc$, much larger than $r_s $ and 
$a$; hence, one concludes that the {\it r.h.s.} of the above equation is much smaller than unity, independently of $\xi$. After 
the gradient term of \eq{tracefinal} begins to dominate, the dimensionless quantity $\varrho$ is essentially ruled by the 
differential equation 

\beq {d^2 \varrho^{n-1} \over d y^2 } \approx 0, \eeq

\noindent and, since $\varrho$ decreases, one concludes that  the gradient solution to \eq{tracefinal} is approximately 
constant when $\bar{a}y \rightarrow 0$ --- and is similar to the ``static'' solution $ \varrho = (1-2n)^{1 /( 1-n )}$.

\subsection{Tully-Fisher law}

The Tully-Fisher is an empirical relation between the intrinsic luminosity $L$ and a power of the rotation velocity 
$v_\infty$ of a spiral galaxy (with the related Faber-Jackson relation for elliptic ones). Since the luminosity is proportional to 
the visible mass, $L \propto M$,  it may be written as $M \propto v_\infty^\sigma$, with $3 < \sigma \lesssim 4$ for spiral galaxies. 

The velocity width $v_\infty$, measured in the outer region of a galaxy, depends upon the total mass distribution within the 
delimited zone. This may be obtained by integrating the profile $2 \ka R = \rho_0 \th \varrho$, after numerically solving 
\eq{tracefinal}; recall that, from \eq{rhoandp}, this also includes the effect of the negative pressure arising from the non-minimal 
coupling $f_2(R)$. Clearly, \eq{rhodm} indicates that the mass profile of the visible and dark matter components will 
differ substantially. Actually, the total mass corresponding to the NFW dark matter density profile is infinite.

This said, one can still get a crude estimate of the form of the power-law relating $M$ and $v_\infty$, deferring a more 
elaborate numerical estimate to the subsequent session. One assumes that $M_{dm}(r) \propto r$ at large 
distances (leading to a flat rotation curve), and $M_{dm}\gg M$. This yields the familiar $v_\infty^2 \propto M_{dm}$ relation 
that, resorting to the scaling \eq{rhodm}, yields

\beq \label{TF} M \propto v^{2(1-n)} . \eeq

For the scenarios considered in the following session, one obtains $M \propto v^{8/3}$ (for $n=-1/3$) and $M \propto v^4$ 
(for $n=-1$).

\subsection{Conservation of energy-momentum tensor}

In section \ref{geo}, one has first attempted to justify the flattening of the galactic rotation curves as due to the non-conservation of 
the energy-momentum tensor of matter,  i.e., through the non-geodesical motion. However, one is now exploring a mimicking 
behaviour of the model under scrutiny, so that a dark matter-like behaviour emerges as a consequence of the dynamics of \eq
{tracefinal}. Hence, one is assuming that matter follows geodesics, as explicitly stated in the beginning of section \ref{metric}. After 
the 
previous characterization of the model, one is able to ascertain the validity of this assumption.

One begins by rewriting \eq{force0}, with $p=0$ for pressureless matter, the Lagrangian density is taken to be ${\cal L}=-\rho$ and 
the adopted non-minimal coupling $\la f_2(R) = (R/R_0)^n$:

\beq \label{force01}
f^{\mu}=  -n { \left( {R \over R_0}\right)^n \over 1+\left( {R \over R_0}\right)^n} \left[ \log\left({R \over R_0} \right)\right]_{,\nu} h^
\mn,
\eeq

\noindent with radial component 

\beq \label{force01r}
f^r =  -n { \left( {R \over R_0}\right)^n \over 1+\left( {R \over R_0}\right)^n} \left[ \log\left({R(r) \over R_0} \right)\right]',
\eeq

\noindent assuming spherical symmetry and the Newtonian approximation $h_{rr} = g_{rr} - (v_r/c)^2 \sim 1$.

\subsubsection{Non-physical result arising from the ``static'' solution}

From the discussion presented in subsection \ref{subsectiongradient}, one recalls that the gradient term in \eq{tracefinal} will 
become dominant at distances $r>r_k$, so that the crossover between visible and dark matter dominance does not occur at 
the distance $r_c$ discussed in subsection \ref{subsectioncrossover}. This is not only a numerical mishap, but a crucial 
component of the model presented here: if the ``static'' solution  $\varrho = (1-2n)^{1 /( 1-n )}$ is valid for the outer region 
$r>a$, the dominance of dark matter would be equivalent to the condition $(1-2n) \varrho^n \gg \th^{-n}$ or, from definitions 
\eq{defdim}, $(R/R_0)^n \gg 1/(1-2n)$: if this was the case, \eq{force01} would read

\beq \label{force01rnot}
f^r=  -n \left[ \log\left({R \over R_0} \right)\right]' = -n {R'\over R} .
\eeq

\noindent Recall that, in the outer region $r > a$ where the gradient solution to \eq{tracefinal} tracks the constant ``static'' solution  $
\varrho = (1-2n)^{1 /( 1-n )}$, so that $R \propto r^{-m'}$ follows the desired dark matter density profile. Hence, the above equation 
would lead to a force $f_r = n m' / r$ (pointing towards the center of the galaxy, since $n<0$). Clearly, for small $|n|$ and $m'$, this 
would lead to a flattening of the rotation curve on the onset of the outer region $r \gtrsim a$, with a asymptotic velocity $v_\infty /c = 
\sqrt{-nm'} \sim 1$, in blatant violation of the observed value $v_\infty \sim 100~km/s \sim 10^{-3}c$; asides from this non-physical 
result, it would undermine the primordial objective of the proposed programme: to obtain a dynamical effect from the non-minimal 
coupling that mimics dark matter, so that geodesical motion produces the flattening of the galaxy rotation curve.

\subsubsection{Geodesical motion from the ``dynamical'' solution}

As discussed above, the crossover between visible and dark matter occurs not due to the onset of condition $(R/R_0)^n 
> 1 $, but because the gradient term in \eq{tracefinal} becomes dominant. Hence, one can instead take $(R/R_0)^n < 1$ and 
consider for \eq{force01r} the approximation

\beq \label{force01rdyn}
f^r =  -n \left({R \over R_0}\right)^{n-1} \left({R(r) \over R_0} \right)'.
\eeq

\noindent Clearly, the pathological result stemming from \eq{force01rnot} is suppressed in this case. Aiming at 
comparison with the already defined dimensionless quantities, one rewrites it as 

\beq
f^r =-n \left( \varrho \th \right)^{n-1} \left( \varrho \th \right)'  \approx {4n \over 1-n} \left( 2\de\bar{a} 
\right)^{1/(1-n)} {c^2 \over r} \left( { r_0 \over r } \right)^{4n/(1-n)}, \eeq

\noindent after using $\th= \th_*^{1/(1-n)}$, the outer Hernquist density profile $\th_* \approx 2 \de \bar{a}(r_0/r)^4$ and the 
asymptotic result $\varrho \approx {\rm const.}$. In order for this force to dominate the Newtonian gravitational force $f_N = - 
GM/r^2 = - (r_s/2) (c/r)^2$ (which is smaller than the overall gravitational force, including the dynamically generated dark 
matter), one must have

\beqa && {8n \over 1-n} \left( 2\de\bar{a} \right)^{1/(1-n)} \left( { r_0 \over r } \right)^{4n/(1-n)} > -{ r_s \over r} \rightarrow  \\ \nonumber 
&& r 
>  r_0 \left[ {1\over 2} \left( -{1-n \over 8n}\right)^{1-n} \left({r_s \over r_0}\right)^{-n} {r_0\over a} \right]^{1/(1-5n)}  . \eeqa

\noindent For the already considered values of $a=1-10~kpc$ and $r_s =  10^{-6}-10^{-4}~kpc$, one concludes that it 
suffices that $r_0 > 200~kpc$ for the above lower bound to be larger than $1~Mpc$, thus ensuring that the extra force 
\eq{force01rdyn} arising from the non-conservation of the energy-momentum tensor is negligible at galactic scale.

\subsection{Energy conditions}
\label{energyconditions}

Following Ref. \cite{energy}, one discusses the relations required to satisfy the strong, null, weak, and dominant energy conditions 
(SEC, NEC, WEC and DEC, respectively); one considers $p=0$, $f_1 = 2 \ka R$ and $f_2 = (R/R_0)^n$; using the equality

\beq {f_1 \over f_2} - {f_1'+2{\cal L}_m f_2' \over f_2} R = \nonumber {2 n \rho \over 1 + \left( {R \over R_0} \right)^{-n} }, \eeq

\noindent the energy conditions read \cite{energy}

\beq \left(1 + { E_n \over 1 + \left( {R \over R_0} \right)^{-n} } \right)\rho \geq 0 ,\eeq

\noindent with 

\beq E_n=
\cases{-2n &, {\rm SEC}
\cr 0 & , {\rm NEC}
\cr 2n & , {\rm DEC}
\cr n & , {\rm WEC}
}.
 \label{cases}
\eeq

Clearly, both the NEC and SEC are always satisfied, for a negative power law $n< 0$. The WEC and DEC are also fulfilled for the 
ranges $-1 \leq n \leq 0$ and $-1/2 \leq n \leq 0$, respectively.

Moreover, as will be seen, the  condition $R > R_0$ is always valid at galactic scales (due to the dominance of the gradient term of 
\eq{trace2}) for the considered models $ n= -1$ (isothermal sphere profile) and $n=-1/3$ (NFW profile), so that $ \left( 1+ (R/R_0)^{-
n} \right)^{-1} \approx 0$; hence, the WEC and DEC are also satisfied for the isothermal sphere and NFW profiles.

Asides from these energy conditions, one should also verify whether the considered model does not give rise to the so-called 
Dolgov-Kawasaki instabilities, whereas small curvature perturbations could arise and expand uncontrollably, thus rendering the 
underlying theory unphysical \cite{DK}. From Refs. \cite{energy,viability}, one writes the mass scale $m_{DK}$ for these instabilities 
as

\beqa \label{mDK} m_{DK}^2 &=& {2\ka - 2 {\cal L}_{m} \left[ f'_2(R) + f_2''(R) R \right] + f_2'(R)T \over 2 {\cal L}_{m} f_2''(R) } = \\ 
\nonumber &=&- {2 \ka R \left( {R \over R_0} \right)^{-n} + n(2n-1) \rho \over 2n(n-1) \rho} R , \eeqa

\noindent using $T= {\cal L}_m = -\rho$. Since the non-minimal coupling mimics an additional, positive matter component, one has $ 
\rho \leq 2\ka R $; furthermore, the cosmological matching condition \eq{rhoinfty} hints that the characteristic density obeys $2 \ka 
R_0 \equiv \rho_0 \ll \rho$, so that one may write $R/R_0 \gg 1$. Since $n$ is a negative exponent, one may approximate the above 
equation by 

\beq m^2_{DK}  \approx  - \left( {R \over R_0} \right)^{-n} { \ka R^2 \over n(n-1) \rho } =  \left( {R \over R_0} \right)^{2-n} {\rho_0 \over 
\rho} { r_0^{-2} \over 2n(n-1) }  , \eeq

\noindent which, in terms of the dimensionless quantities defined in \eq{defdim}, reads

\beq \label{mDK2} m^2_{DK} \approx - \varrho^{2-n} \th {r_0^{-2}\over 2n(n-1) } . \eeq

\noindent Since $m^2_{DK}$ is negative, there is a potential concern about instabilities. These evolve at a length scale $r_{DK}$

\beq r_{DK} = m_{DK}^{-1} =  \sqrt{2n(n-1)} \left( {R_0 \over R} \right)^{1-n/2} \sqrt{\rho \over \rho_0}  r_0= \sqrt{2n(n-1)}\varrho^
{n-1/2} \th^{-1/2}  r_0 . \eeq

\noindent At large distances, when the dark matter-like component dominates and the scalar curvature is larger than the matter 
density, the dimensionless quantity $\varrho$ approaches a constant value (displaying the tracking behaviour of the ``gradient 
solution'', as discussed before). The remaining quantity $\th$ behaves proportionally to the mimicked dark matter distribution with an 
outer region profile  (see \eq{visible}), 

\beq \th(r) = (2\de \bar{a} )^{m'/4} (a/r)^{m'},\eeq

\noindent so that one may write

\beq r_{DK} =  \sqrt{2n(n-1)} {\varrho^{n-1/2} \over (2\de \bar{a} )^{m'/8}} \left({ r \over a}\right)^{m'/2} r_0 . \eeq

\noindent Thus, one finds that the typical length scale $r_{DK}$ grows linearly, for the $m'=2$ isothermal sphere scenario, or with $r^
{3/2}$, for the $m'=3$ NFW density profile.

The perturbative expansion leading to the mass scale \eq{mDK} assumes an initially small perturbation $\de R$ to a constant 
background curvature $R$; hence, one must check if the obtained length scale $r_{DK}$ is larger than the characteristic distance 
over which one cannot assume that the curvature is constant. The later is given by $L \equiv -R(r)/R'(r) = r/m' $.

Comparing the instability length scale $r_{DK}$ against the constant curvature domain $L$, one concludes that the validity of the 
perturbative expansion, embodied in the inequality $r_{DK} < L$, requires

\beqa  \sqrt{2n(n-1)} {\varrho^{n-1/2} \over (2\de \bar{a} )^{m'/8}} \left({ r \over a}\right)^{m'/2} r_0 < {r \over m'} \rightarrow \\ \nonumber 
\left({r\over a}\right)^{m'/2-1} < \varrho^{-n+1/2} (2\de \bar{a} )^{m'/8} {\bar{a} \over \sqrt{2n(n-1)}  m' } . \eeqa

\noindent Inserting the scaling relation $m' = m/(1-n)$ and the outer slope $m=4$ of the visible mass Hernquist density profile, one 
rewrites this as

\beq \label{insta} \left({r\over a}\right)^{1+n\over1-n} < \varrho^{-n+1/2} (2\de \bar{a} )^{1\over 2(1-n)} {\bar{a} \over 4 }\sqrt{n-1\over 
2n} . \eeq

\noindent This relationship shows that the perturbative expansion is invalid, in the case of the $n=-1$, isothermal sphere distribution, 
or limited to a distance $r< r_s$ (where the Newtonian regime breaks down), in the case of the $n=-1/3$, NFW profile. Hence, the instability detected via \eq{mDK2} is not 
physically relevant.

\section{Mimicking the dark matter density profile of spherical galaxies}
\label{sectionmimicking}

One now aims to further develop the mechanism described in the previous sections, by attempting to model the rotation curves of 
galaxies of type E0, which exhibit an approximate spherical symmetry adequate for the results obtained in this work. Since its 
manifest purpose is to avoid the intricacies related to the accurate modelling of galaxies, one opts for using seven 
readily available galaxy rotation curves, as provided in Ref. \cite{kronawitter, kronawitter2,Faber,Rix}, divided into visible and dark matter components of the following galaxies: NGC 2434, NGC 5846, NGC 6703, NGC 7145, NGC 7192, NGC 7507 and NGC 7626 galaxies.

In the discussion presented here, particular attention is paid to the scenarios whereby the proposed mechanism mimics either the 
NFW or the isothermal sphere dark matter density profiles (given by the $n=-1/3$ and $n=-1$ cases, respectively), due to its ubiquity 
in the literature. Moreover, the analytical approach taken assumed that the non-minimal coupling was given by a single power-law 
$f_2(R) = (R_0/R)^n$, thus yielding a natural explanation for either of the two profiles.

Conversely, it is known that some galaxies are much better fitted by the NFW profile, while others conform to the isothermal sphere 
model: hence, arguing for a single power-law coupling between matter and geometry amounts to defending a particular dark matter 
density profile --- a task clearly outside the scope of this work, and better left to the astronomy community.

Furthermore, the present framework offers a convenient way through which both dark matter density profiles may be implemented 
simultaneously, with no {\it a priori} need for any unnatural selection. One is thus led to conclude that the assumed single power-law 
coupling should be generalized to a sum over exponents $n$, hinting perhaps at the Laurent series expansion (since $n$ is not 
constrained to positive values) of a more evolved form for $f_2(R)$.

Given the motivation above, one undertakes the numerical task of modelling the rotation curves of the selected galaxies by 
combining both the NFW as well as the isothermal sphere dark matter density profiles,  i.e., choosing the non-minimal coupling

\beq \label{f2two} f_2(R) =\sqrt[3]{R_3 \over R} + {R_1 \over R} , \eeq

\noindent which, by taking the trace of the Einstein \eq{EE1}, leads to an equation analogous to \eq{trace2}:

\beq \label{tracetwo} R = {1 \over 2 \ka} \left[1 + 3{R_1 \over R} + {5 \over 3} \left({R_3 \over R}\right)^{1/3} \right] \rho + {1 \over \ka} 
\square \left( \left[3 {R_1 \over R} + \left({R_3 \over R}\right)^{1/3} \right] {\rho \over R} \right)  .\eeq

\noindent Notice that one can no longer resort to the identification of a ``static'' solution, as before: this is due to the non-linearity of 
the problem, so that adding the $n=-1$ and $n=-1/3$ non-minimal couplings does not translate into a sum of the solution obtained 
before. Likewise, one can no longer resort to the straightforward identification with a ``dark'' matter energy-momentum tensor 
endowed with an equation of state such as \eq{EOS}. Tentatively, one can posit that the effects of combining the two coupling 
amount to adding the corresponding mimicked ``dark'' matter density profiles, together with interaction effects arising from the 
mentioned non-linearity.

One now attempts to define the dimensionless quantities analog to \eq{defdim}, aiming at a generalization of the equation of motion 
\eq{traceeq}; since there are two lengthscales $r_1 = R_1^{-2}$ and $r_3 = R_3^{-2}$, one could adopt a normalization lengthscale 
$r_0= f(r_1,r_3)$, so that $r_0\rightarrow r_1$ when $r_3 \rightarrow +\infty$ ( i.e., the $R_3$ coupling is ``switched off''), and {\it vice-versa}. However, this would unnecessarily complicate the obtained equation of motion, specially since one may isolate one of the power-laws in the non-minimal coupling \eq{f2two},

\beq \label{f2two2} f_2(R) =\left({R_3 \over R}\right)^{1/3} \left[1 +{R_1 \over R_3}\left({R_3 \over R}\right)^{2/3}\right] = {R_1 \over R} 
\left[1 +\left({R_3 \over R_1}\right)^{1/3}\left({R \over R_1}\right)^{2/3}\right]  . \eeq

\noindent This said, one adopts the simpler normalization $r_0 = r_1$, thus defining the dimensionless quantities as in \eq{defdim}. 
Some algebra yields the equation of motion

\beq \label{eqmotiontwo} \varrho = \th + {3 \over \varrho} + {5 \over 3} \ep \left( {\th^2 \over \varrho} \right)^{1/3} + {2 \over \th} \bar
{\square} \left[ {3 \over \varrho^2} + \ep \left( { \th \over \varrho^2 } \right)^{2/3} \right], \eeq

\noindent with the parameter $\ep = (R_3/R_1)^{1/3} = (r_1/r_3)^{2/3}$ indicating the relative strength of the two couplings present in 
\eq{f2two}: the pure $n=-1/3$ NFW scenario is obtained when $\ep \rightarrow \infty$, while the $n=-1$ isothermal sphere profile is 
given by the limit $\ep \rightarrow 0$.

The programme for the numerical session is simple: for a given rotation curve, one uses the visible matter density profile $\rho$ as an 
input to \eq{eqmotiontwo}, adjusting the parameters $r_1$ and $r_3$ so that the obtained full (visible + dark) matter density profile 
yields the best fit available. Ideally, the model parameters $r_1$ and $r_3$ are universal, so that all rotation curves should  be fitted 
simultaneously. However, one instead proceeds by obtaining pairs for these quantities corresponding to each galaxy, and then 
discusses the obtained results: this allows for the identification of any abnormal case, possible trends, etc..

Following the previous sections, one attempts to model the visible matter density with a Hernquist distribution. As it turns out, the 
visible components of the selected rotation curves are not compatible with just one of these curves, but are very well fitted by two 
Hernquist profiles: each is characterized by a mass $M_i$ and lengthscale $a_i$ (i=1, 2), with $M_1< M_2$ and $a_1 < a_2$ --- 
thus modelling a core + diffuse matter distribution, with the latter dominating.

The results obtained from the numerical solution of \eq{eqmotiontwo} with the described approach are presented in Table \ref{table}, 
which displays the values for $M_1$, $M_2$, $a_1$ and $a_2$; the best fit values for the characteristic lengthscales $r_1$ and $r_3$ are depicted in Table \ref{tableboth}. These are illustrated by Figs.\ref{fig1} and \ref{fig2} (left): as can be seen, the composite 
non-minimal coupling \eq{f2two} provides close fits to all rotation curves in the outer region, with some galaxies exhibiting a 
discrepancy in the inner galactic region: this could be due to uncertainty in the original derivation of the rotation curves or enhanced 
effects due to the deviation from purely spherical symmetry (which should be more pronounced near the core). 

The density profiles are also plotted for visible matter and full visible + ``dark'' matter, together with the contribution arising from the 
gradient term present in \eq{eqmotiontwo} in Figs. \ref{fig1} and \ref{fig2} (right): the contribution from the latter is completely overlapped with the mimicked ``dark matter'' profile, indicating that it dominates the {\it r.h.s} of \eq{eqmotiontwo}. Although not presented, for brevity, it is straightforward to check that this dominance is supported by the criteria discussed in section \ref{subsectiongradient}; this verification is achieved by resorting to the naive identification $a=a_1$, $M=M_1$ ( i.e. considering only the long-range Hernquist profile that dominates the distribution of visible matter) --- and disregarding the non-linear effects arising from the interaction between the two power-laws present in the composite non-minimal coupling \eq{f2two}. This extrapolation attempts to reconcile the analytic results obtained for the simpler scenario addressed before the numerical fitting, and constitutes a qualitatively cross-validation of both efforts.

%%%%%%%%%%%%%%%%%%%%%%%%%%%%%%%%%%%%%%%%%%%%%%%%%%%%%%
%%%%%%%%%%%%%%%%%%%%%%%%%%%%%%%%%%%%%%%%%%%%%%%%%%%%%%

\TABULAR[h]{c|cc|cc|cc}{

Galaxy 		&$ M$	&	 $r_s $ &	 $a_1$ &$ M_1$ &  $a_2$& $M_2$  \\
(NGC) & ($10^{11}~M_\odot$) & ($10^{-2}~pc$) & ($kpc$) 	&  ($10^{11}~M_\odot$) & ($kpc$) & ($10^{11}
~M_\odot$) \\ \hline
2434 & 1.38 & 1.32 & 5.05 &1.14 &0.568 & 0.23    \\ 
5846 & 11.7 & 11.2 & 21.1 & 9.10& 2.36 &2.62 \\ 
6703 & 1.21 & 1.15 & 2.74 & 1.05 & 0.323 &0.15  \\ 
7145 & 0.83 & 0.796 & 4.02 & 0.706 & 0.508 & 0.125  \\ 
7192 & 1.43 & 1.37& 4.08 & 1.27 & 0.472 &0.161  \\ 
7507 & 2.52 & 2.41 & 4.70 &1.66 & 0.820 & 0.860 \\ 
7626 & 8.33 & 7.98 & 11.0 & 6.75 & 1.08 & 1.58 

}
{Relevant quantities for the selected set of galaxies: $M$ is the visible mass, $r_s$ the corresponding Schwarzschild radius, $a_1, 
~a_2$ and $M_1,~M_2$ the best fit values of the two-component Hernquist profile of the visible matter density.\label{table}}

%%%%%%%%%%%%%%%%%%%%%%%%%%%%%%%%%%%%%%%%%%%%%%%%%%%%%%
%%%%%%%%%%%%%%%%%%%%%%%%%%%%%%%%%%%%%%%%%%%%%%%%%%%%%%

%%%%%%%%%%%%%%%%%%%%%%%%%%%%%%%%%%%%%%%%%%%%%%%%%%%%%%
%%%%%%%%%%%%%%%%%%%%%%%%%%%%%%%%%%%%%%%%%%%%%%%%%%%%%%

\TABULAR[h]{c|cc|cc|cc}{

Galaxy & \multicolumn{4}{c|}{Composite} & \multicolumn{2}{c}{Single} \\
(NGC)& $r_1$ & $r_3$  &  $r_{\infty ~1 }$  & $r_{\infty ~3 }$   &  $r_1$ & $r_3$  \\ 
\hline 
2434 &  $\infty$ & 0.9&   0 & 33.1   &   4.1 & 0.9 \\
5846 & 37 & $\infty $ &   138 & 0    &  37 & 34.9 \\
6703 &  22 & $\infty$   &  61.2 & 0    &  22 & 26.2 \\
7145 & 22.3 & 47.3  &  60.9 & 14.2   &   19.9 & 23.8  \\
7192 &  14.8 & 24   &  86.0 & 18.3    &   14.5 & 7.3 \\
7507 &  4.9 & 2.9  &  178 & 31.1   &   4.3 & 1.1 \\
7626 &  28 & 9.6  &  124 & 42.5   &  16.0 & 7.1 

}      
{Best fit values for the characteristic lengthscales $r_1$ and $r_3$ for the composite and separate fits of the galaxy rotation curves with the $n=-1$ isothermal sphere and $n=-1/3$ NFW mimicked ``dark matter'' scenarios, together with background matching distances $r_{\infty~1}$ ($n=-1$) and $r_{\infty ~3}$ ($n=-1/3$) for the composite non-minimal coupling. $ r_i = \infty$ in the composite scenario indicates that the corresponding scale $R_i = 0$ . Units: $r_1$ ($Gpc$), $r_3$ ($10^5~Gpc$), $r_{\infty~1}$ and $r_{\infty~3}$ ($kpc$). \label{tableboth}}

%%%%%%%%%%%%%%%%%%%%%%%%%%%%%%%%%%%%%%%%%%%%%%%%%%%%%%
%%%%%%%%%%%%%%%%%%%%%%%%%%%%%%%%%%%%%%%%%%%%%%%%%%%%%%

%%%%%%%%%%%%%%%%%%%%%%%%%%%%%%%%%%%%%%%%%%%%%%%%%%%%%%
%%%%%%%%%%%%%%%%%%%%%%%%%%%%%%%%%%%%%%%%%%%%%%%%%%%%%%

\begin{figure}[h]
\begin{center}
$\begin{array}{cc}

	\epsfxsize=7.5cm
	\epsffile{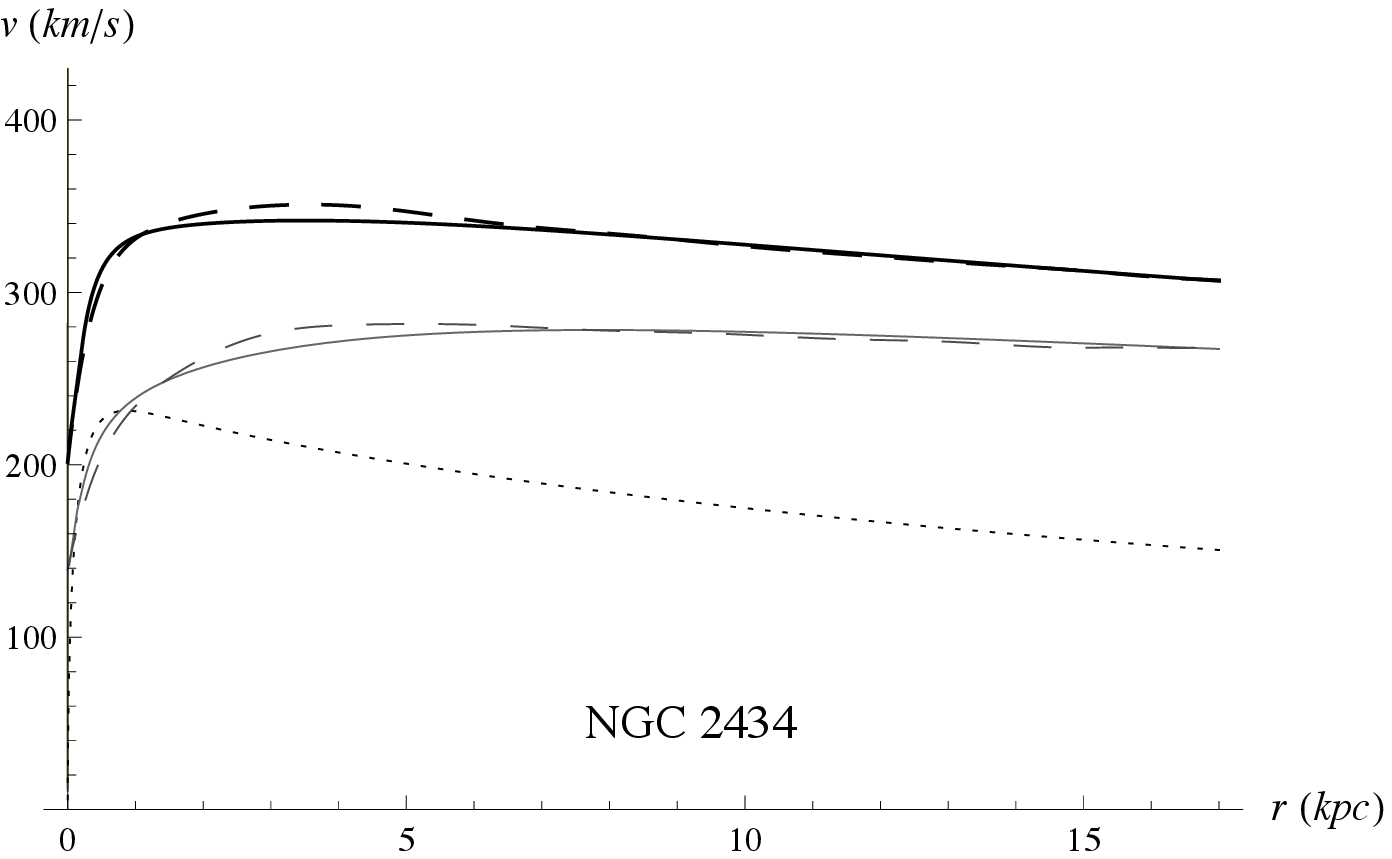} &
	\epsfxsize=7.5cm
	\epsffile{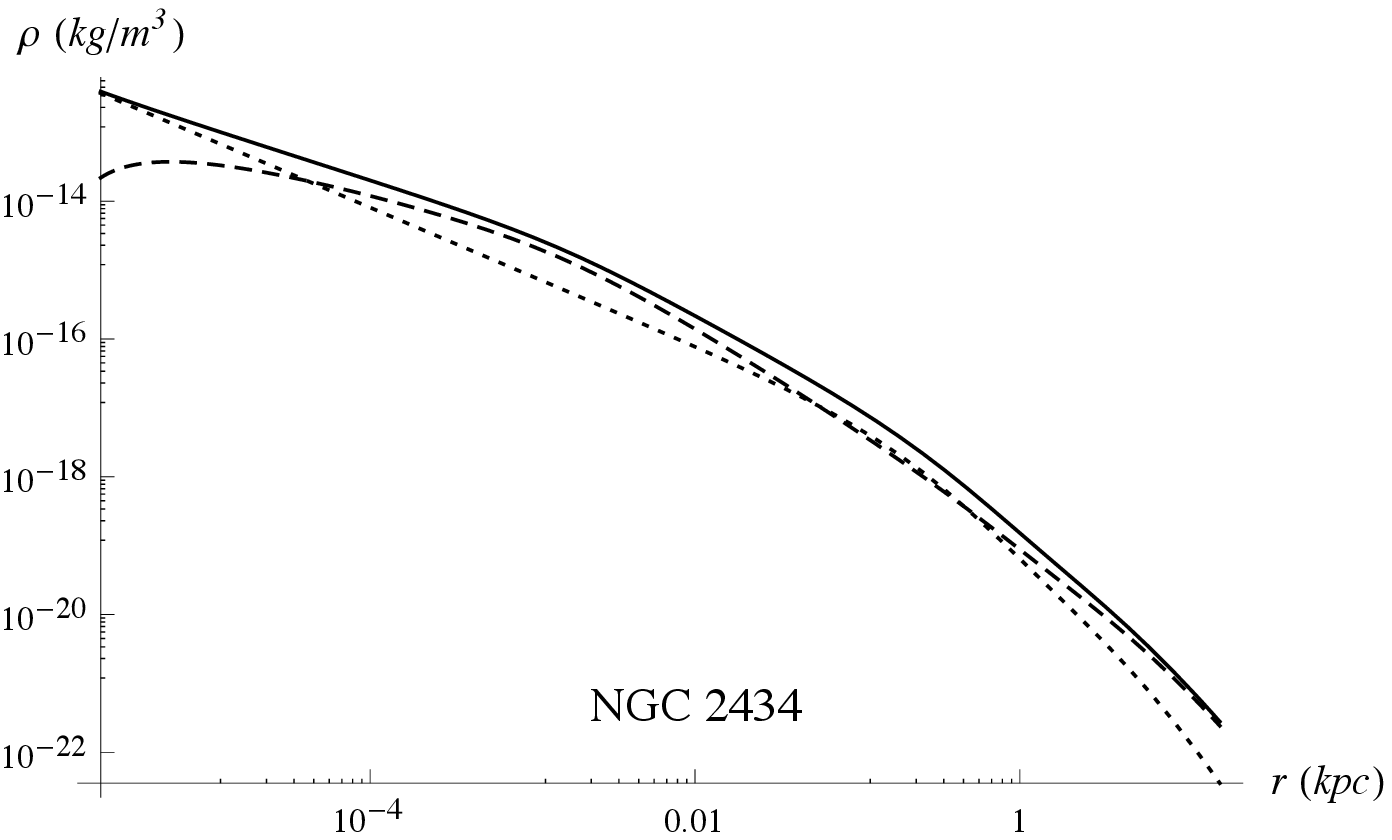} \\
	
	\epsfxsize=7.5cm
	\epsffile{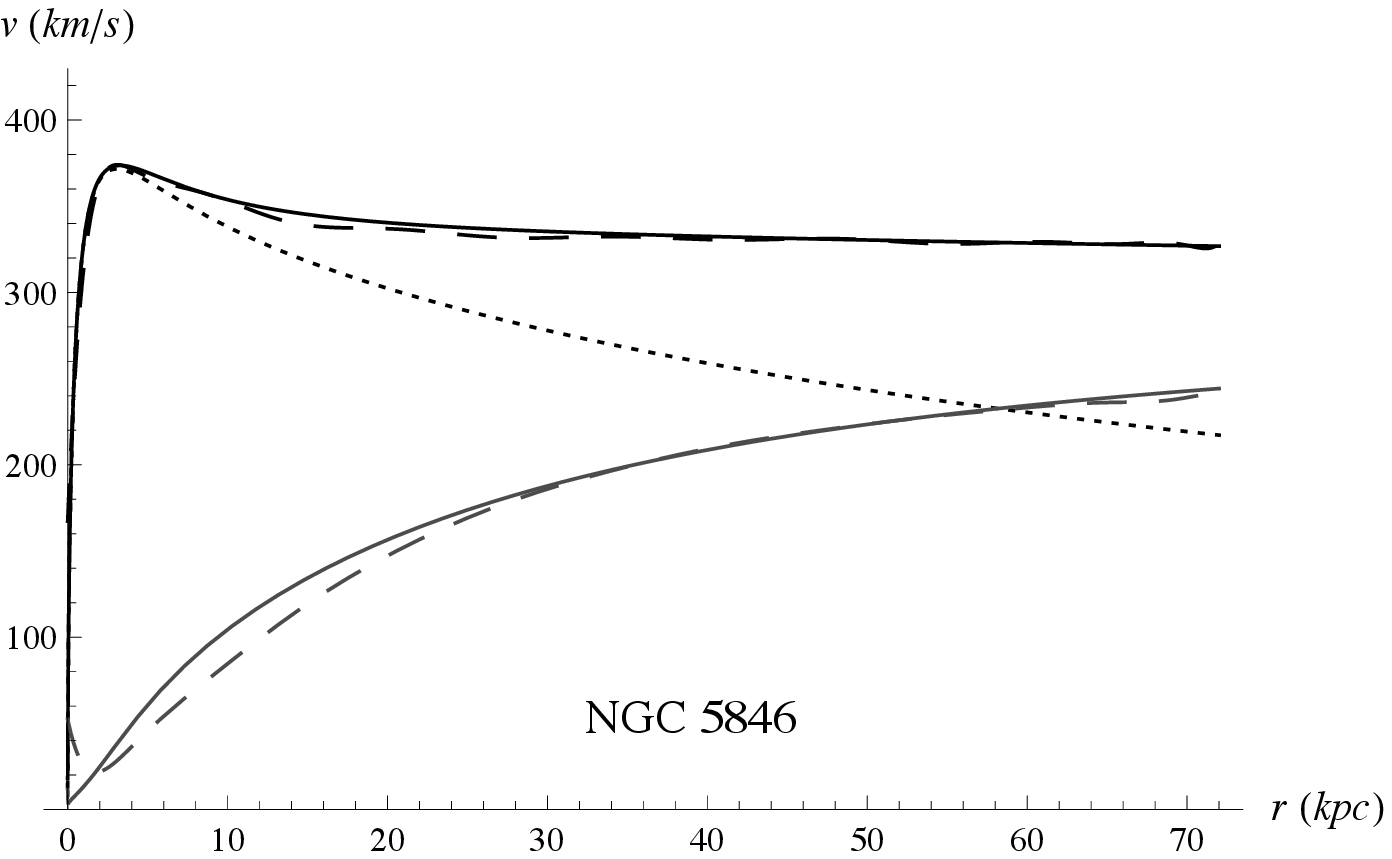} &
	\epsfxsize=7.5cm
	\epsffile{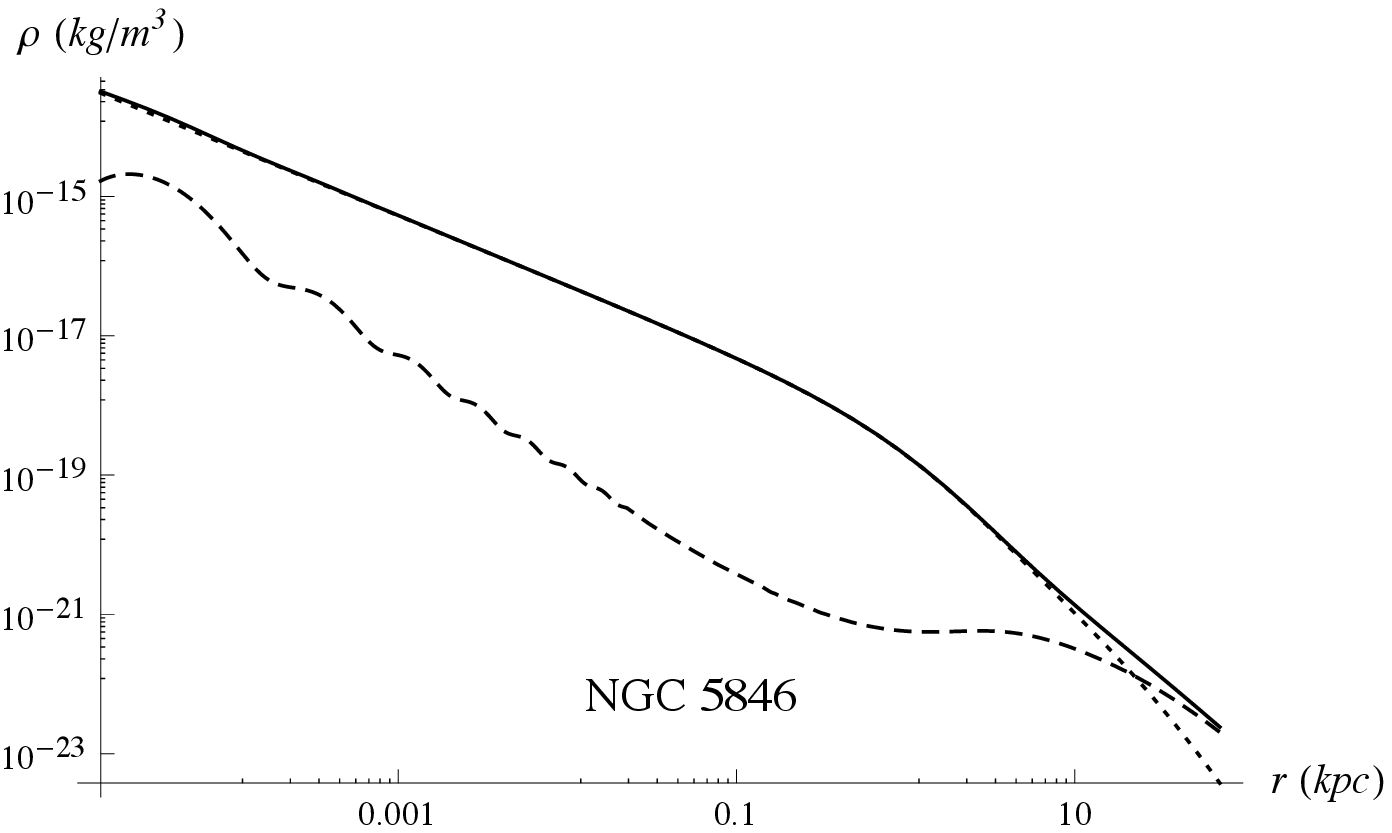} \\
	
	\epsfxsize=7.5cm
	\epsffile{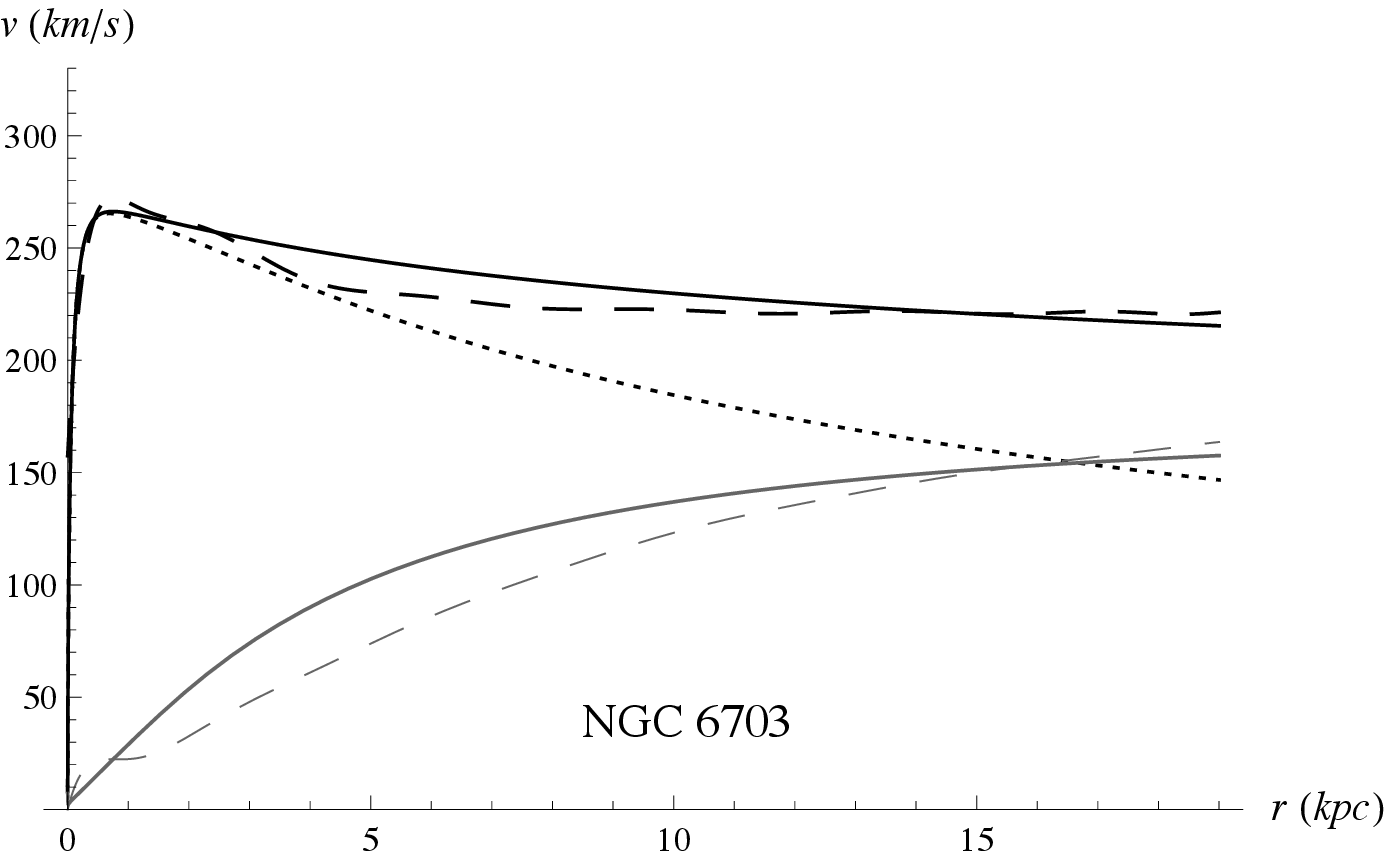} &
	\epsfxsize=7.5cm
	\epsffile{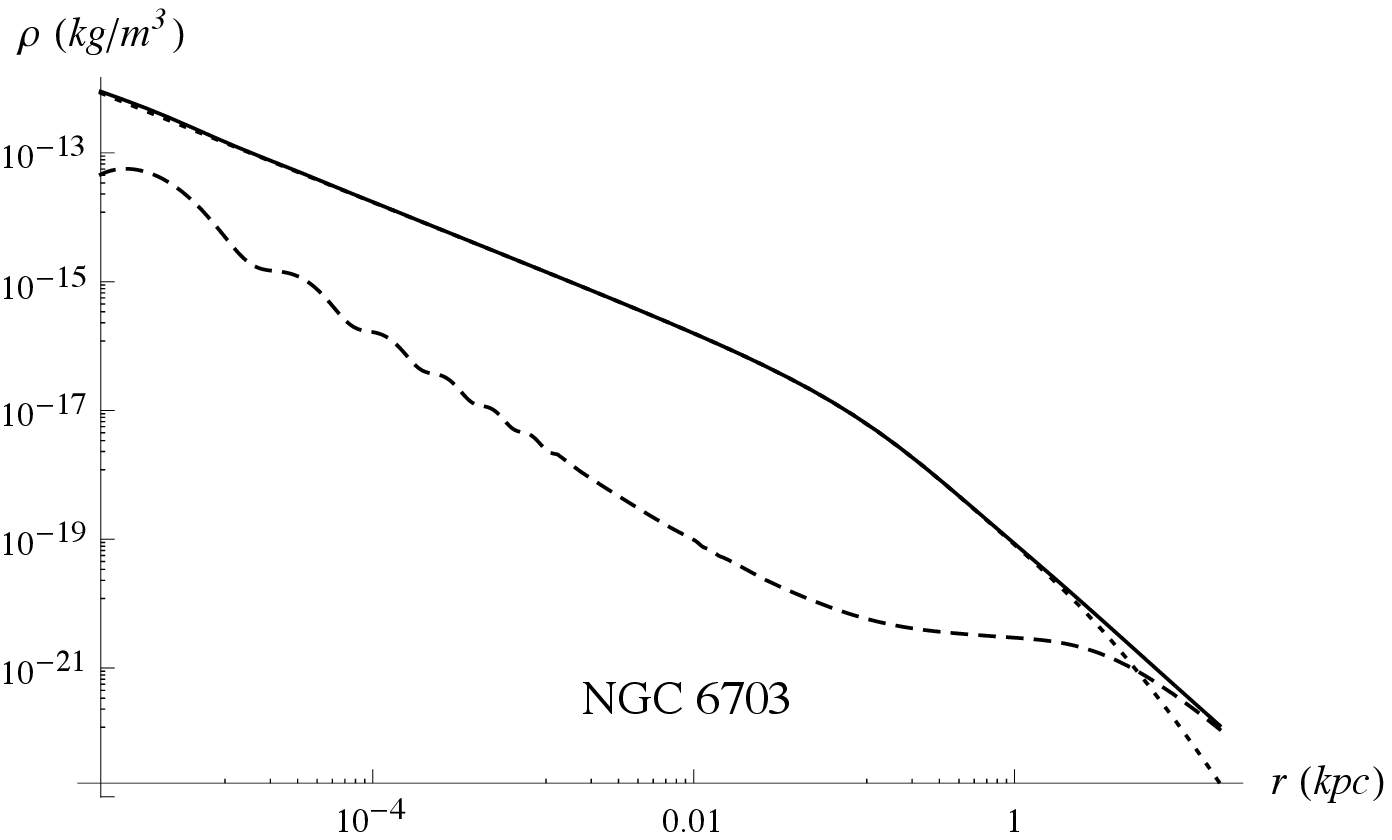} \\
		
	\epsfxsize=7.5cm
	\epsffile{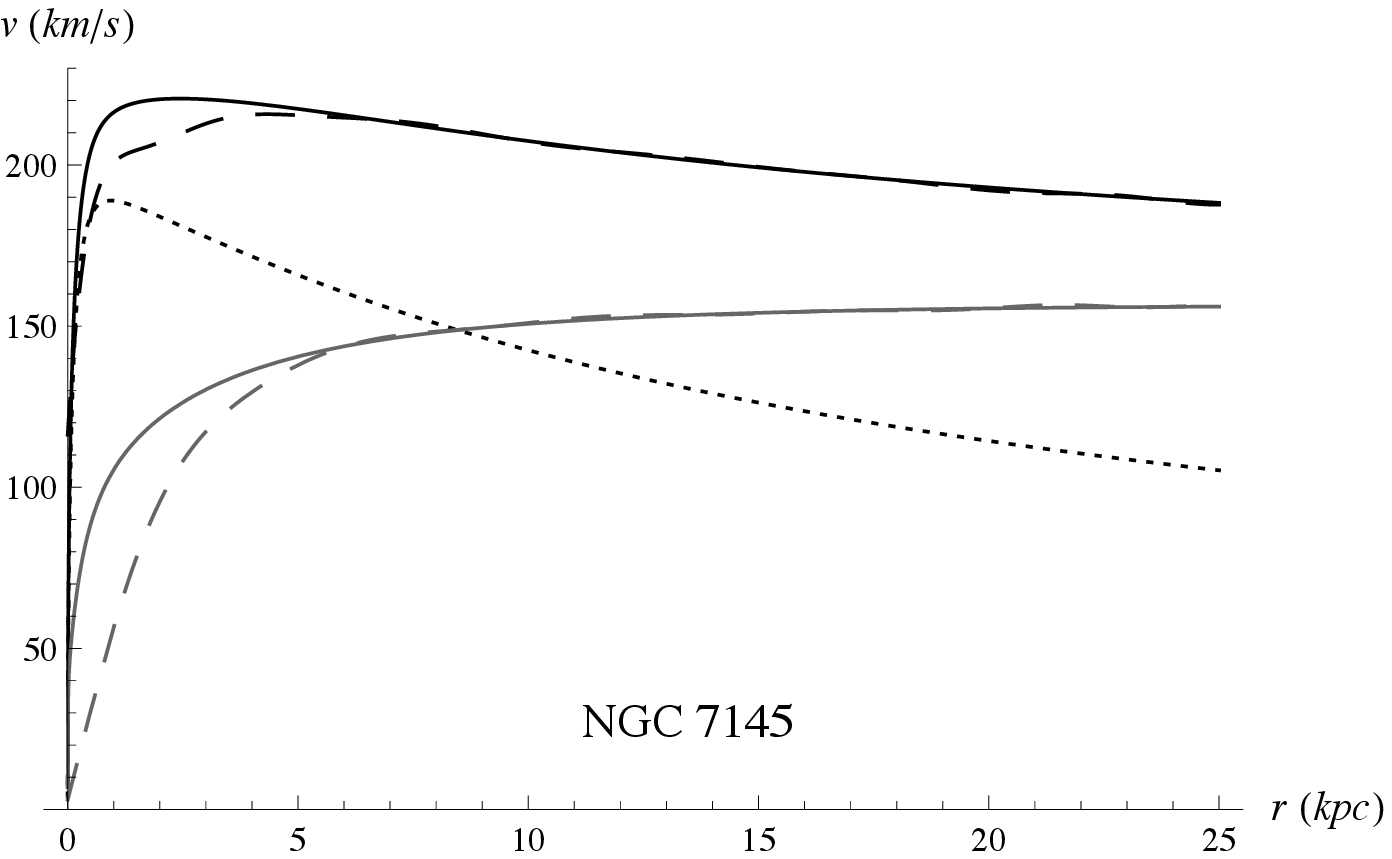} &
	\epsfxsize=7.5cm
	\epsffile{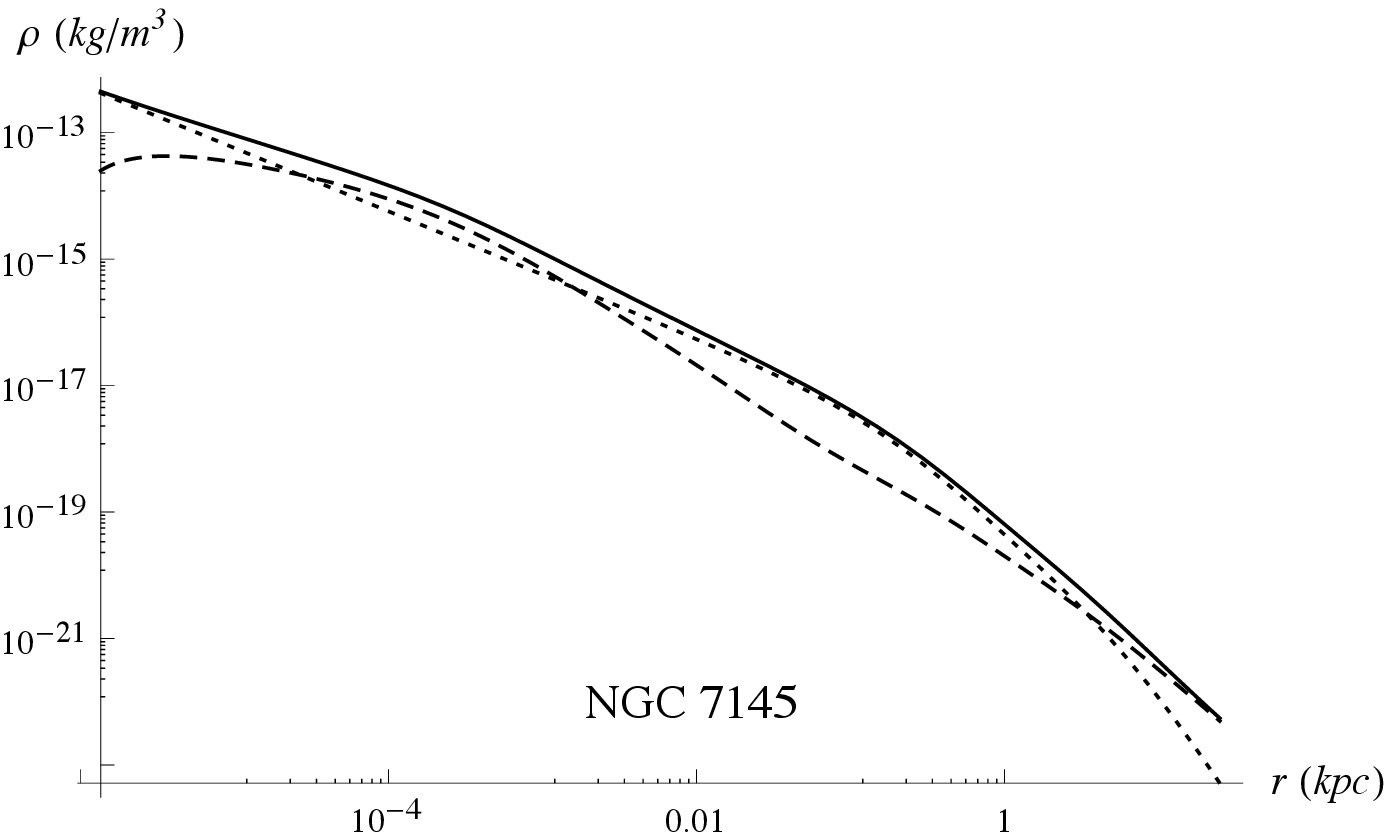} \\

\end{array}$
\end{center}
\caption{{\bf Left}: Observed rotation curve (dashed full), decomposed into visible (dotted) and dark matter (dashed grey) contributions  \cite{kronawitter}, superimposed with the mimicked dark matter profile (full grey) arising from the composite non-minimal coupling and resulting full rotation curve (full).
{\bf Right}: Log-Log profile of the visible matter density (dotted), mimicked ``dark matter'' contribution (dashed) and the sum of both components (full).}

\label{fig1}
\end{figure}

%%%%%%%%%%%%%%%%%%%%%%%%%%%%%%%%%%%%%%%%%%%%%%%%%%%%%%

\begin{figure}[h]
\begin{center}
$\begin{array}{cc}

	\epsfxsize=7.5cm
	\epsffile{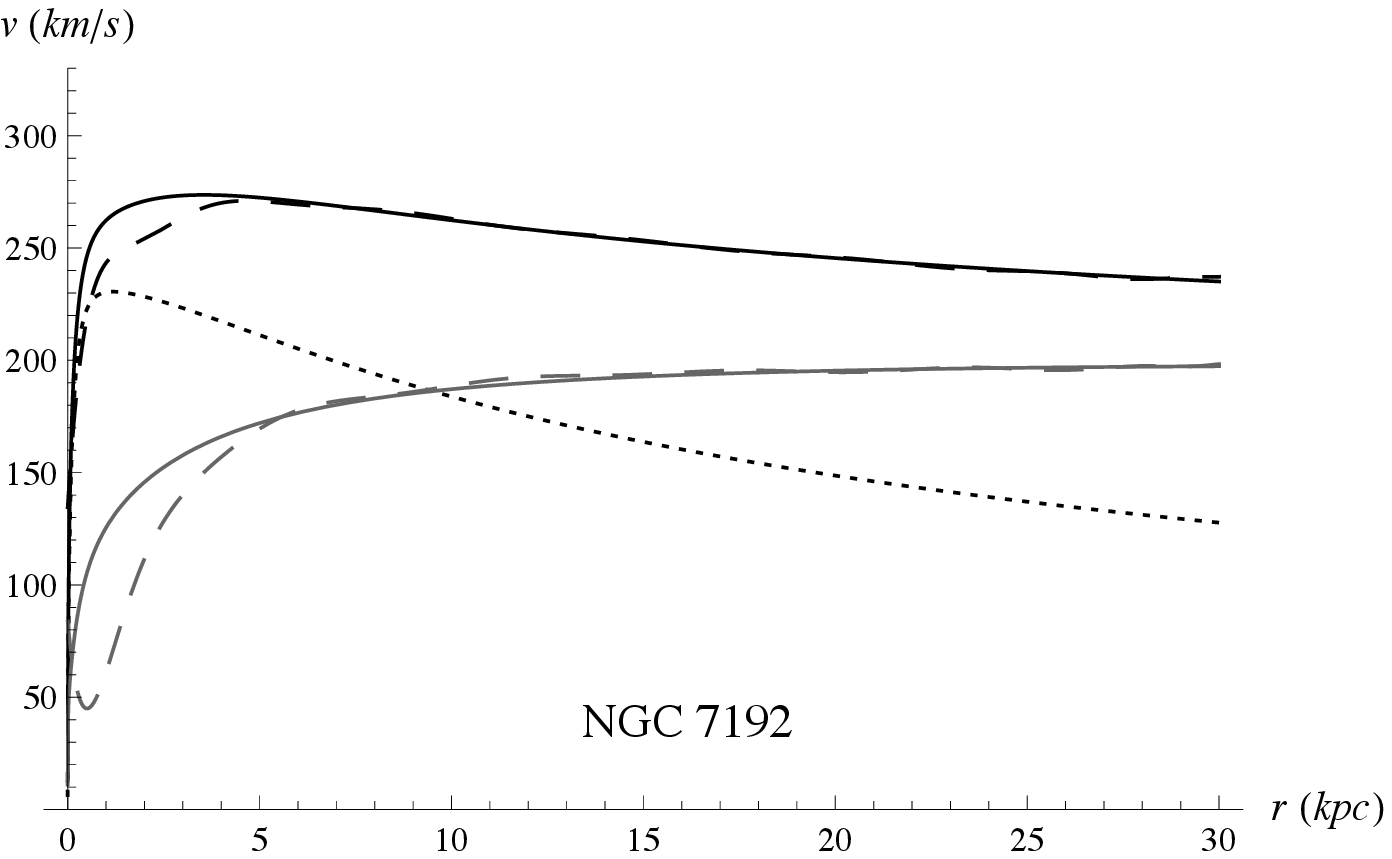} &
	\epsfxsize=7.5cm
	\epsffile{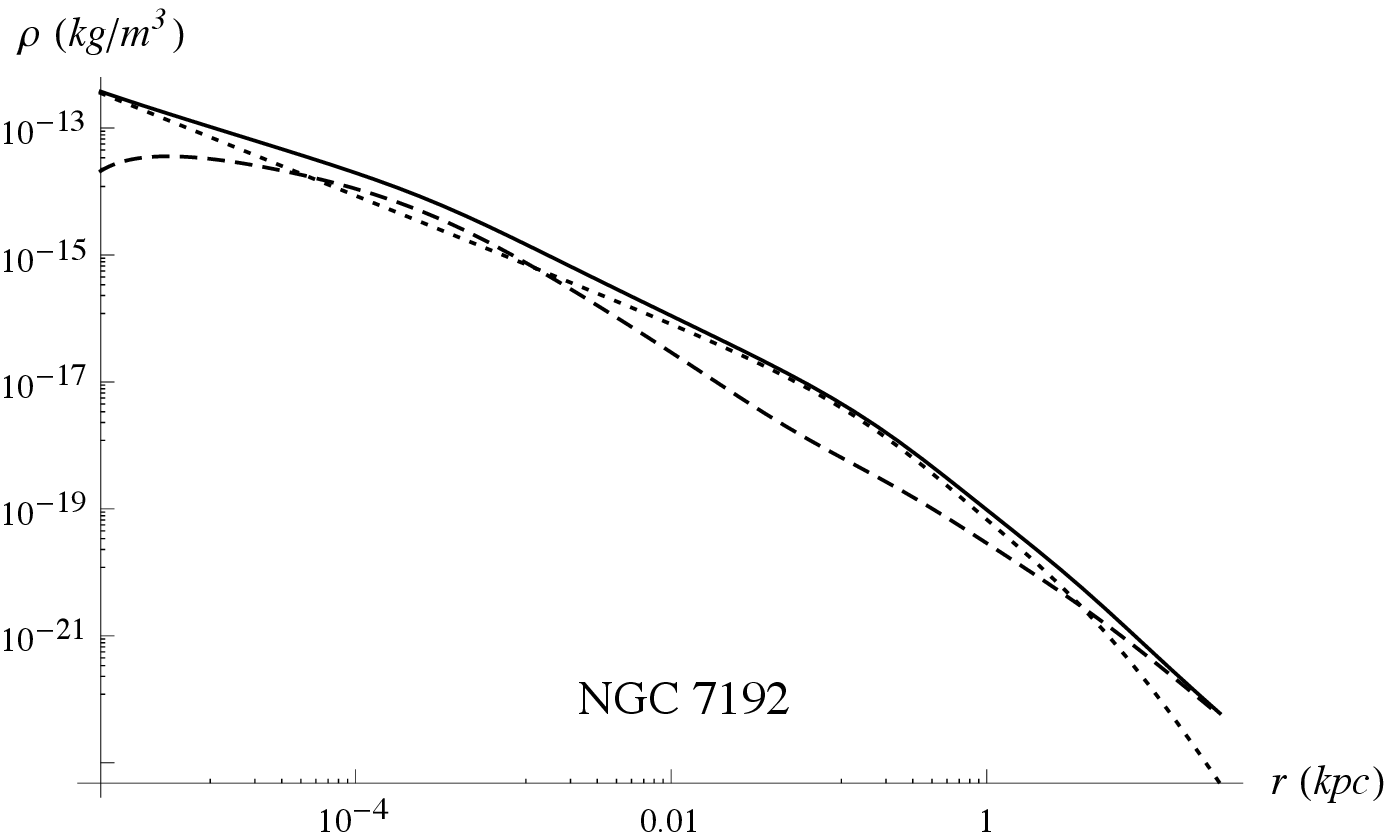} \\
		
	\epsfxsize=7.5cm
	\epsffile{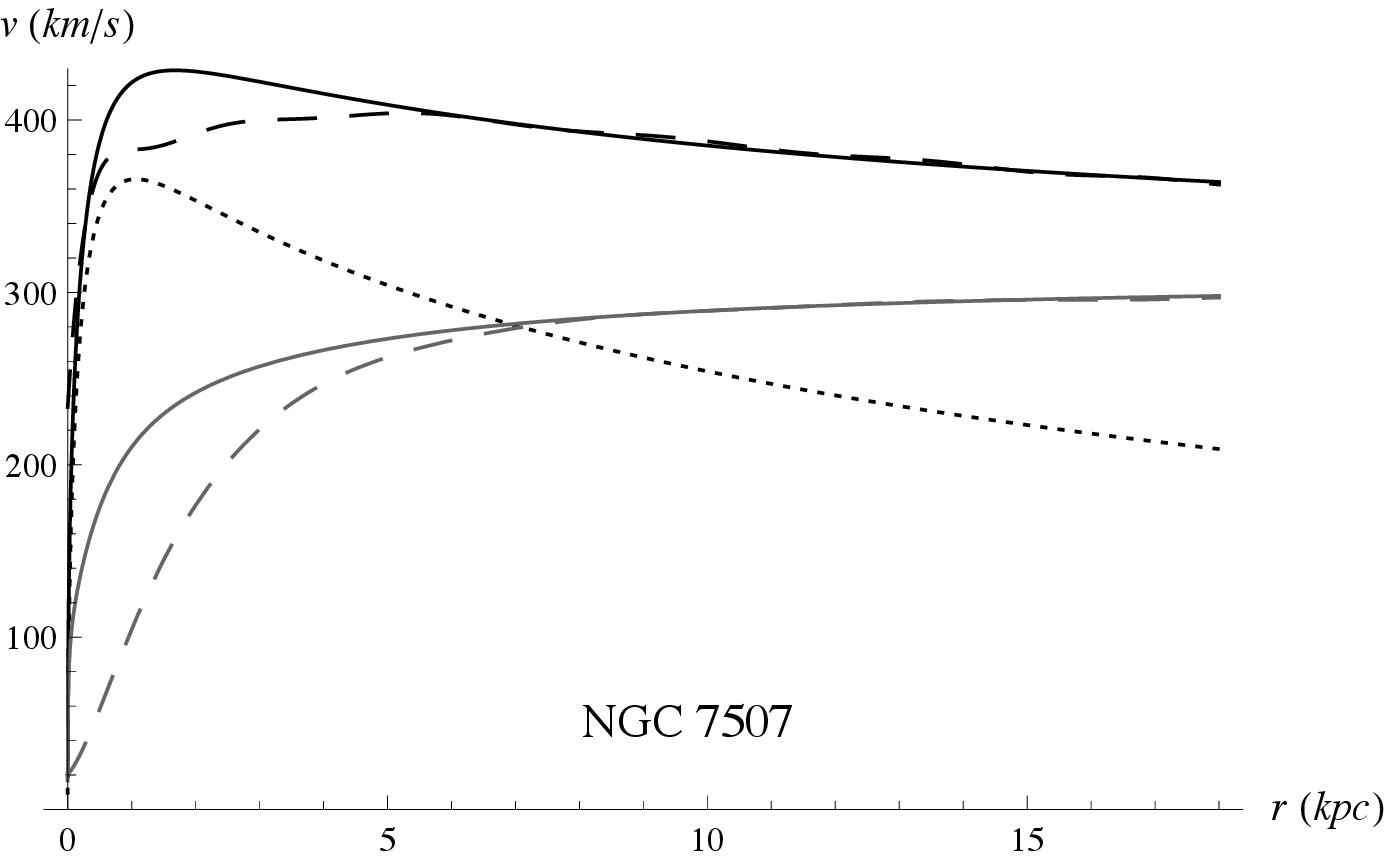} &
	\epsfxsize=7.5cm
	\epsffile{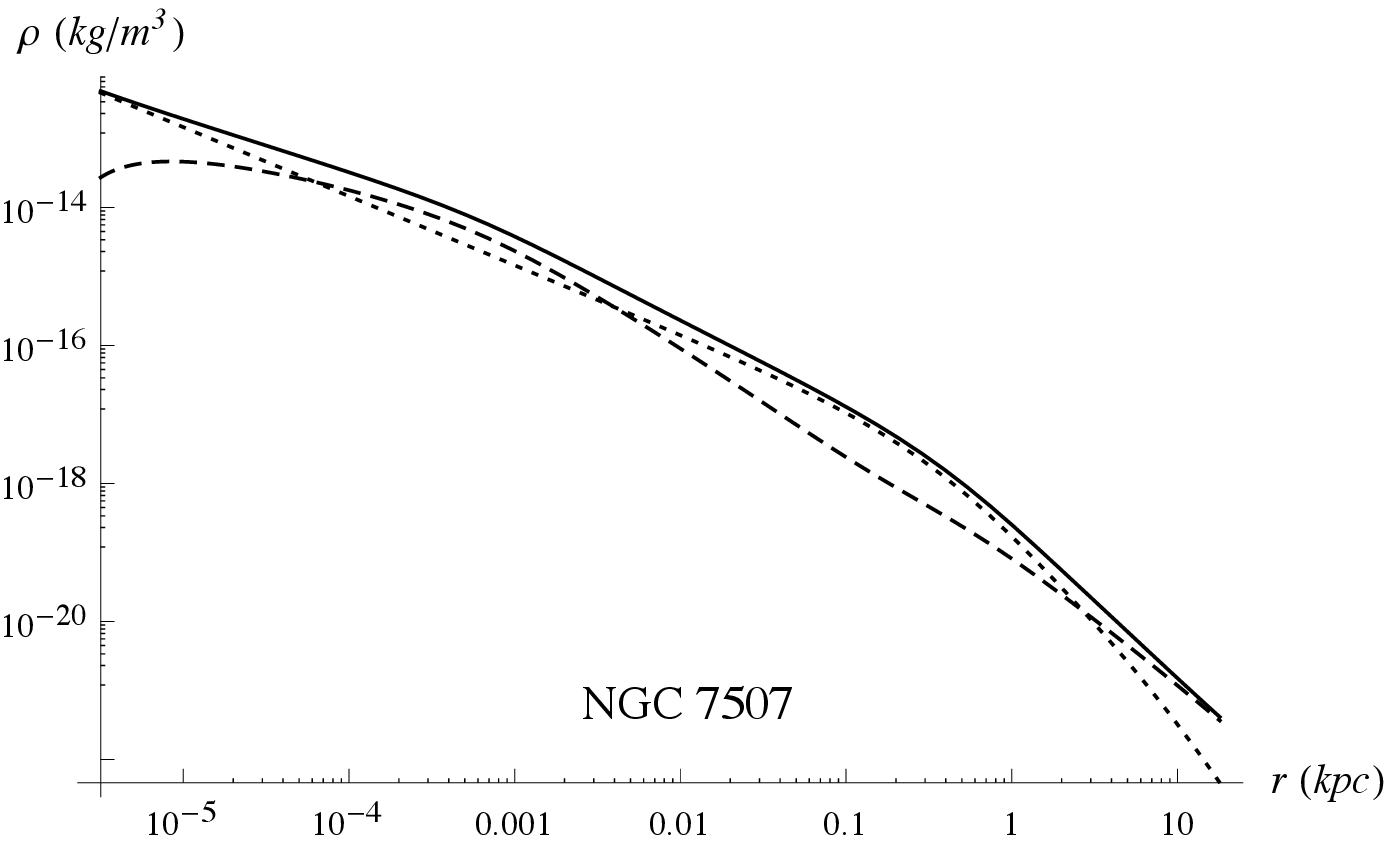} \\
		
	\epsfxsize=7.5cm
	\epsffile{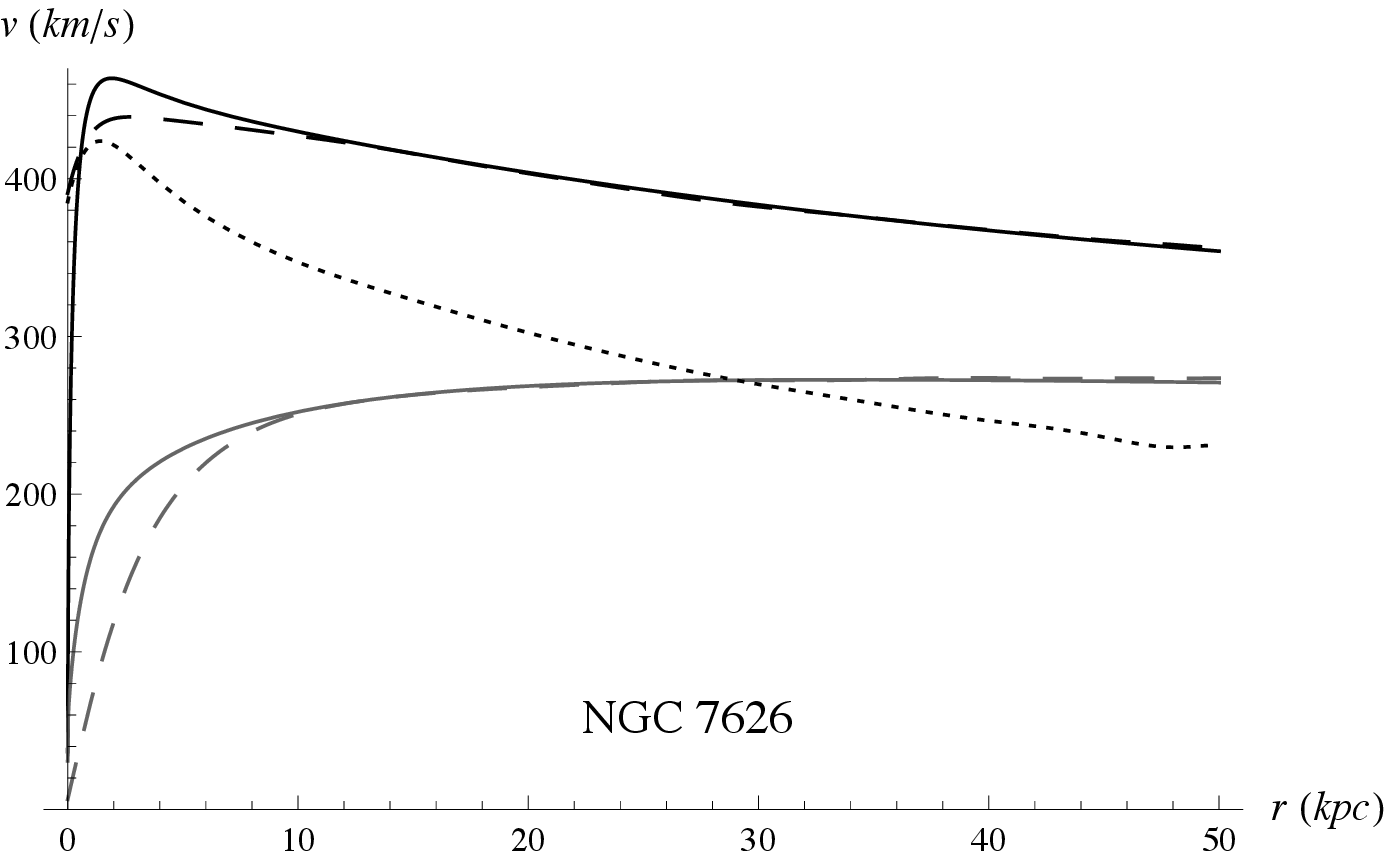} &
	\epsfxsize=7.5cm
	\epsffile{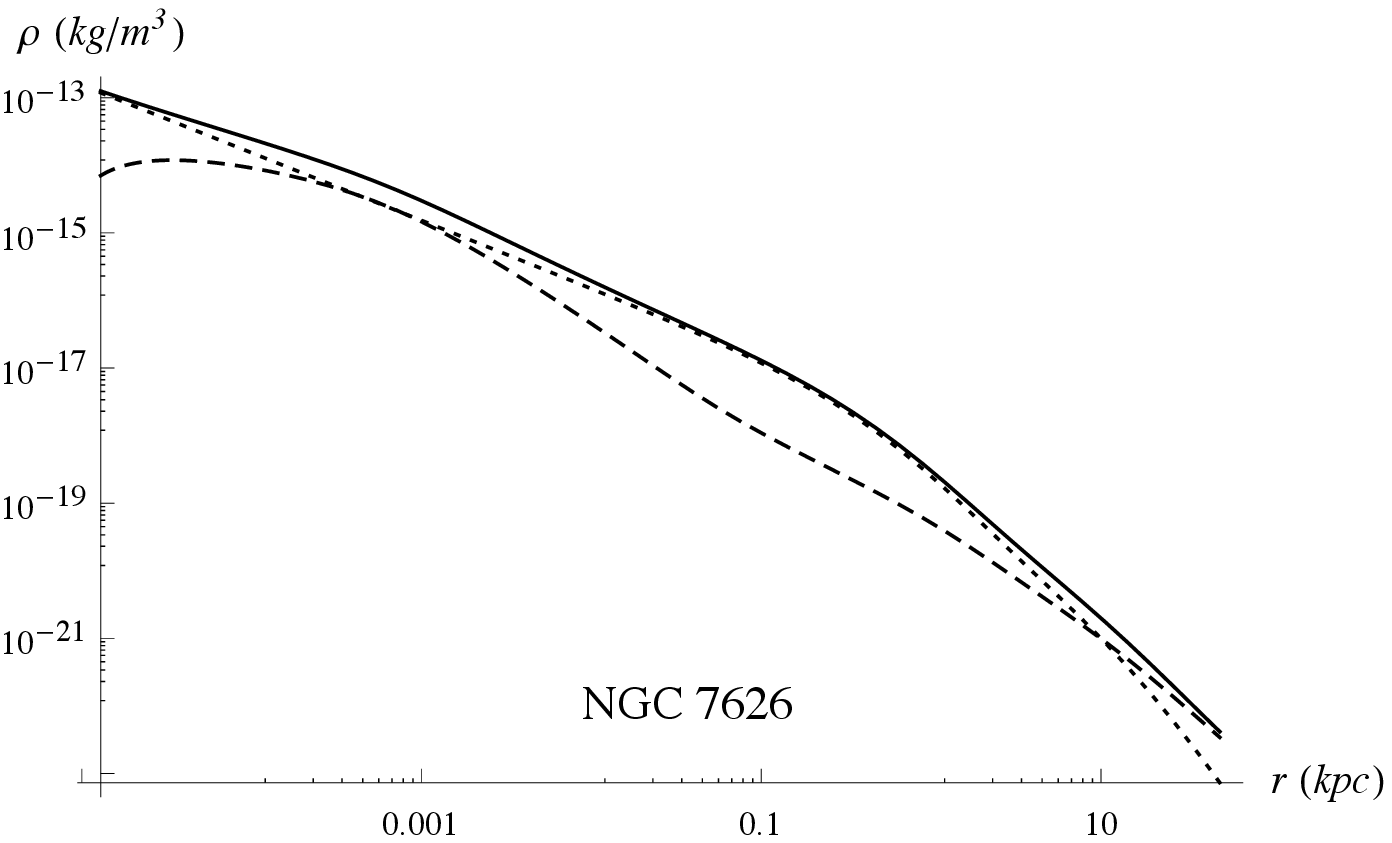} 

\end{array}$
\end{center}
\caption{Continuation of Fig. 1.}

\label{fig2}
\end{figure}

%%%%%%%%%%%%%%%%%%%%%%%%%%%%%%%%%%%%%%%%%%%%%%%%%%%%%%
%%%%%%%%%%%%%%%%%%%%%%%%%%%%%%%%%%%%%%%%%%%%%%%%%%%%%%

\subsection{Single power-law non-minimal coupling}

In this paragraph the results of fitting a single power-law non-minimal coupling $f_2(R)=(R/R_0)^n$ (with $n=-1/3$ or $n=-1$) to the 
rotation curves of the selected galaxies are presented. Table \ref{tableboth} indicates the values obtained from the best fit scenarios for 
the characteristic lengthscales $r_1$ and $r_3$; as in the composite coupling \eq{f2two}, these are of the order $r_1 \sim 10~Gpc$ 
and $r_3 \sim 10^5 ~Gpc$. The obtained fits, although of worse quality, exhibit a similar variation: an average $\bar{r}_1=16.8 ~Gpc
$ with a standard deviation $\si_1=10.5~Gpc$ and $\bar{r}_3 = 1.45 \times 10^6~Gpc$ and $\si_3 = 1.26 \times 10^6 ~Gpc$.

\begin{figure}[h]
\begin{center}
$\begin{array}{cc}

	\epsfxsize=7.5cm
	\epsffile{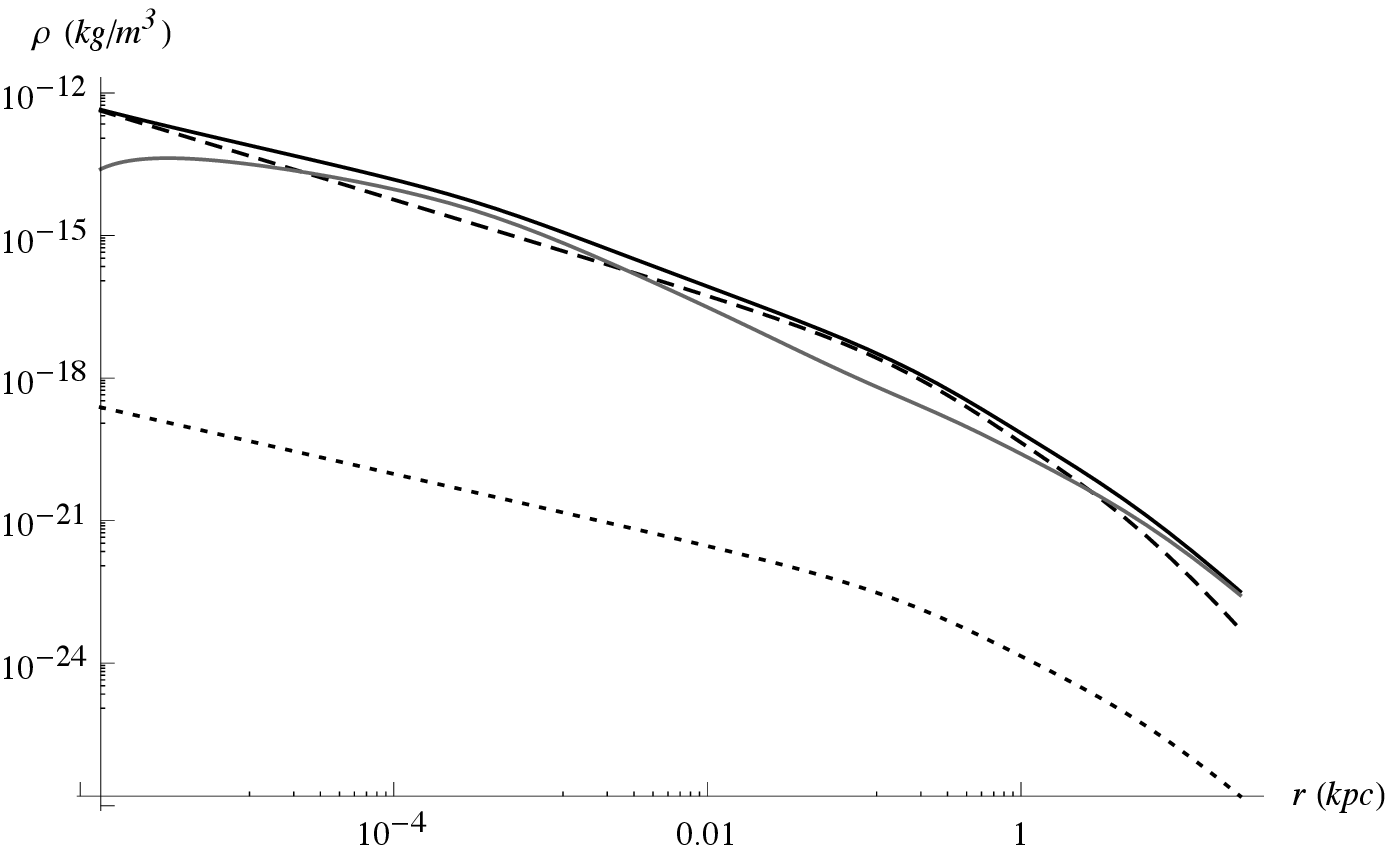} &
	\epsfxsize=7.5cm
	\epsffile{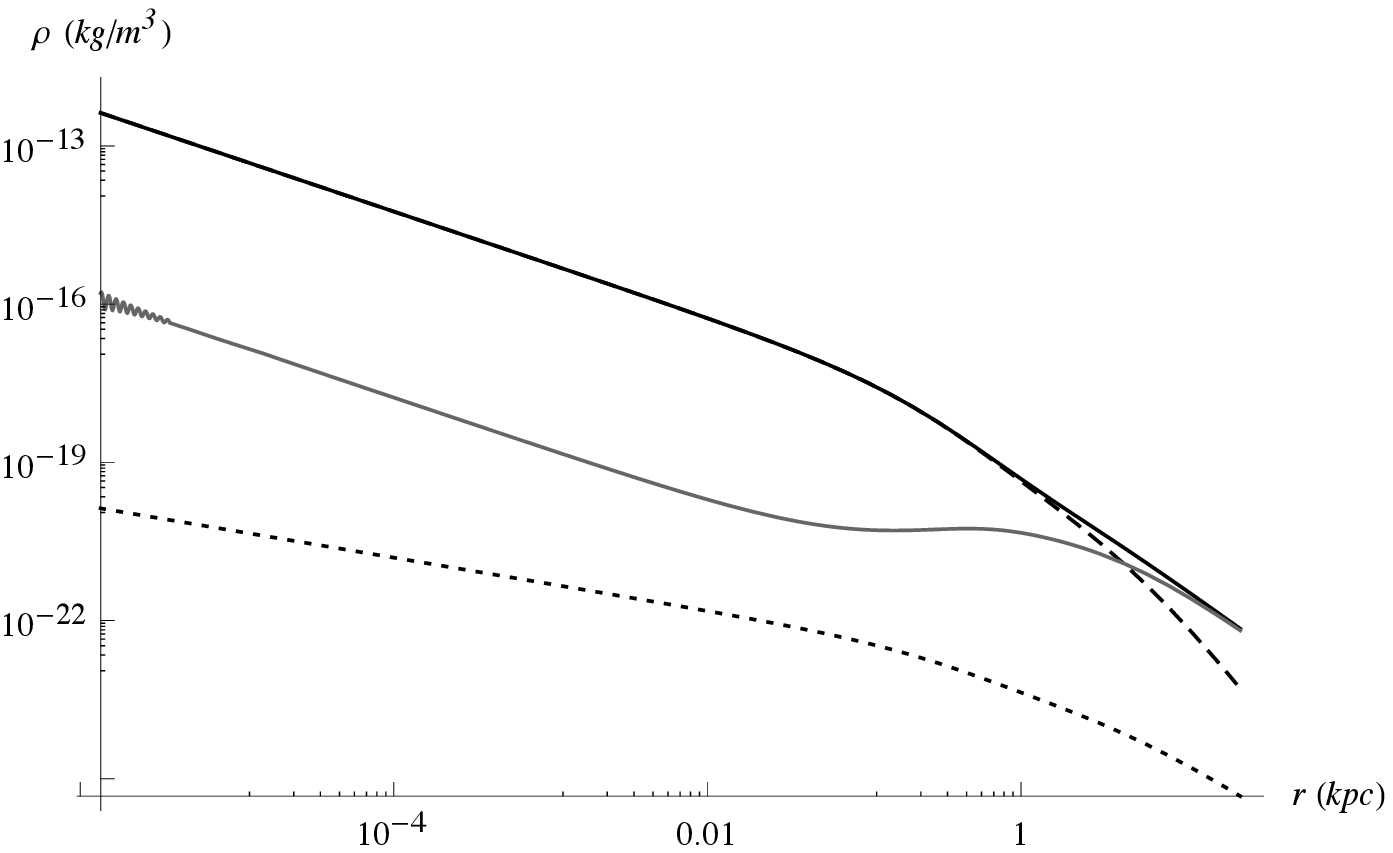} 

\end{array}$
\end{center}
\caption{Profiles for visible matter density (dashed), total density (full) and contributions arising from the gradient (full grey) and static (dotted) solutions with $n=-1/3$ (left) and $n=-1$ (right), for NGC 7145.\label{tablesingle}}

\label{figsingle}
\end{figure}

The corresponding rotation curves are depicted in Figs. \ref{fig3} and \ref{fig4}; as can be 
seen, the composition of the $n=-1$ and $n=-1/3$ scenarios (Figs. \ref{fig1} and \ref{fig2}) improves the quality of fit considerably. For illustration purposes, the visible matter and 
full density profiles are depicted for the NGC 7145 galaxy in Fig. \ref{figsingle}; the density 
distributions arising from the gradient and static solutions are also shown in these figures, with the tracking behaviour of the latter 
visible beyond the crossover between visible and dark matter dominance, for $r > r_c \sim 10~kpc$. This galaxy is chosen because 
it offers fits of similar quality with both the NFW and the isothermal sphere models.

%%%%%%%%%%%%%%%%%%%%%%%%%%%%%%%%%%%%%%%%%%%%%%%%%%%%%%
%%%%%%%%%%%%%%%%%%%%%%%%%%%%%%%%%%%%%%%%%%%%%%%%%%%%%%

%%%%%%%%%%%%%%%%%%%%%%%%%%%%%%%%%%%%%%%%%%%%%%%%%%%%%%
%%%%%%%%%%%%%%%%%%%%%%%%%%%%%%%%%%%%%%%%%%%%%%%%%%%%%%

%%%%%%%%%%%%%%%%%%%%%%%%%%%%%%%%%%%%%%%%%%%%%%%%%%%%%%
%%%%%%%%%%%%%%%%%%%%%%%%%%%%%%%%%%%%%%%%%%%%%%%%%%%%%%

\begin{figure}[h]
\begin{center}
$\begin{array}{cc}

	\epsfxsize=7.5cm
	\epsffile{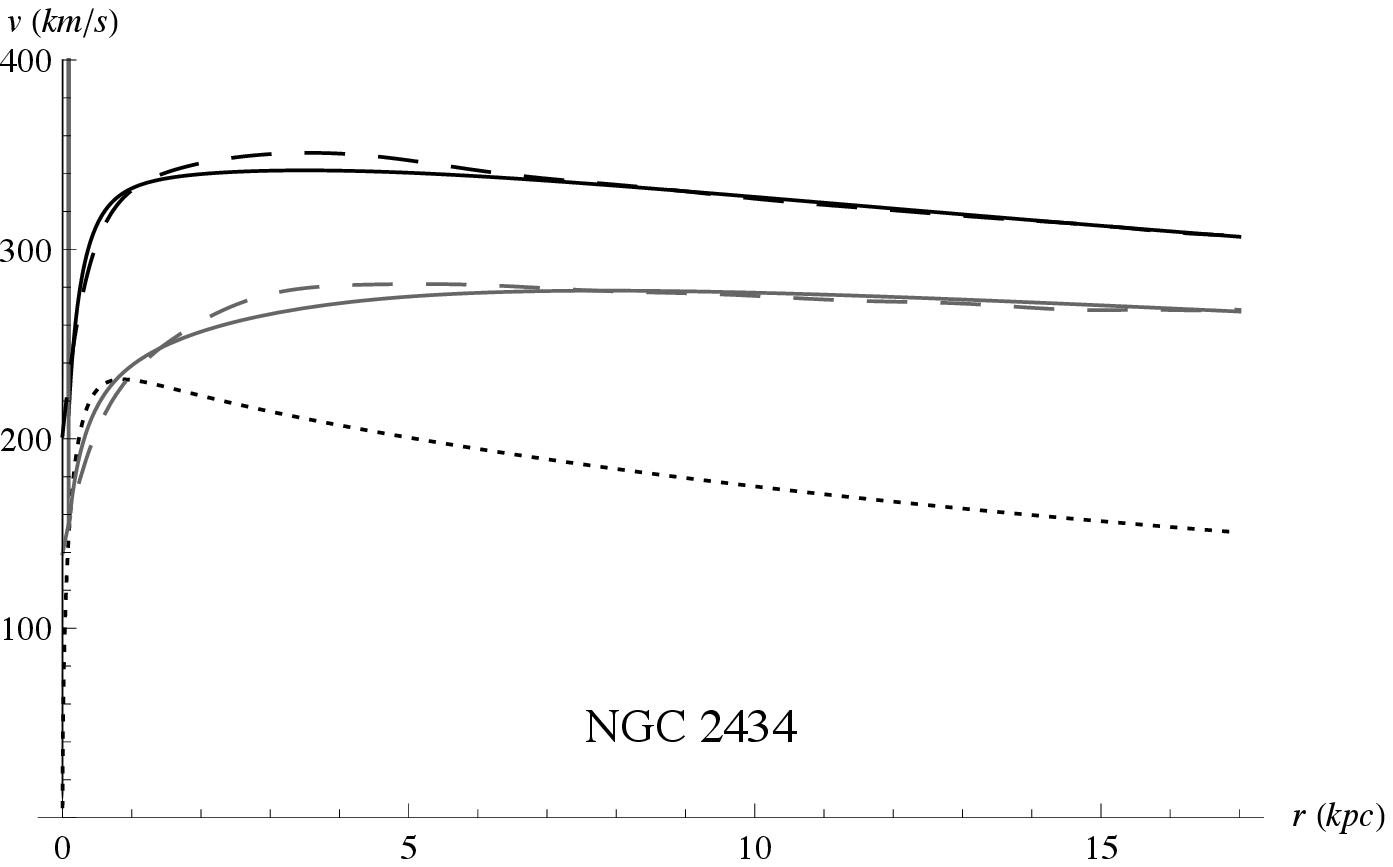} &
	\epsfxsize=7.5cm
	\epsffile{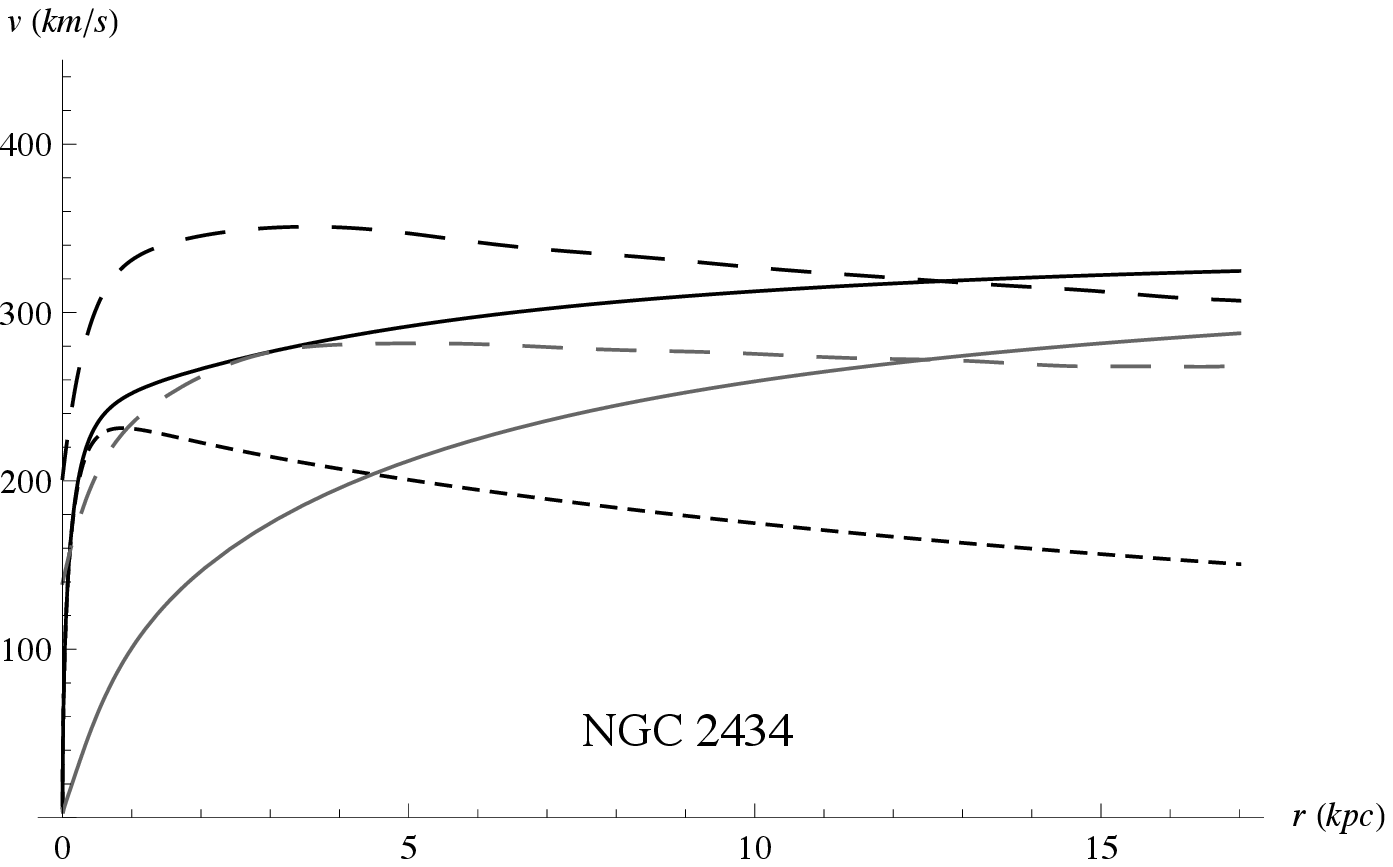} \\

	\epsfxsize=7.5cm
	\epsffile{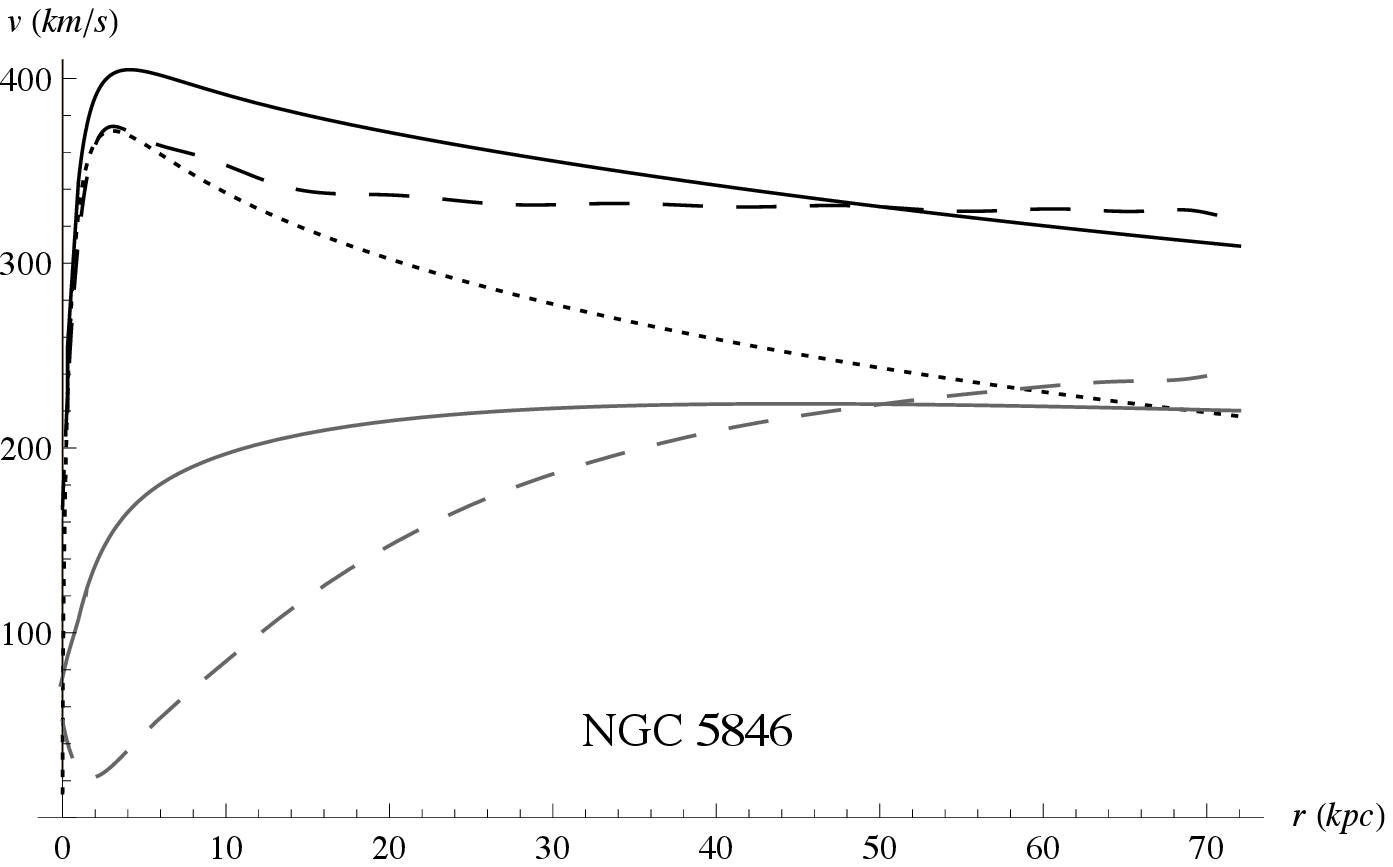} &
	\epsfxsize=7.5cm
	\epsffile{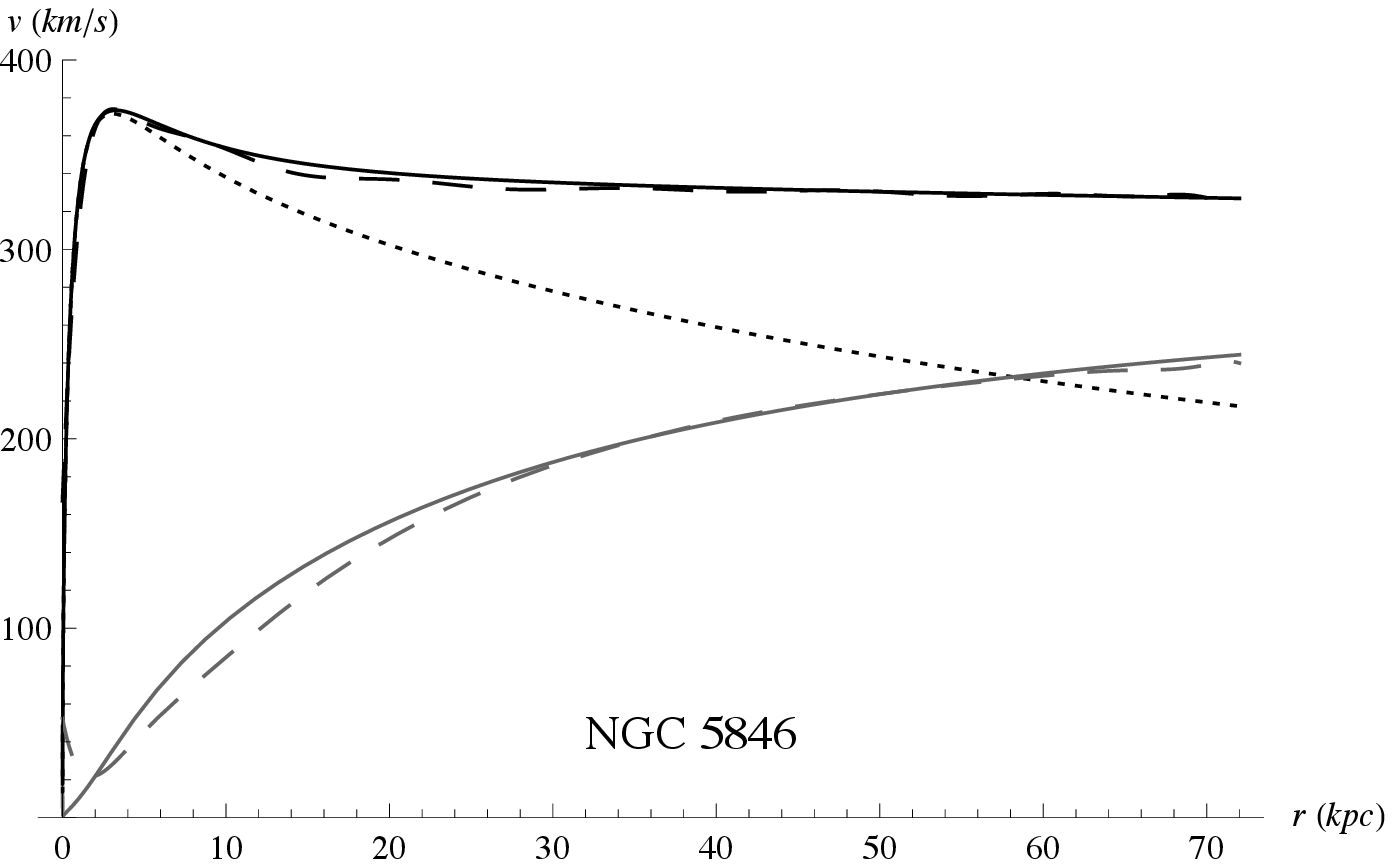} \\
	
	\epsfxsize=7.5cm
	\epsffile{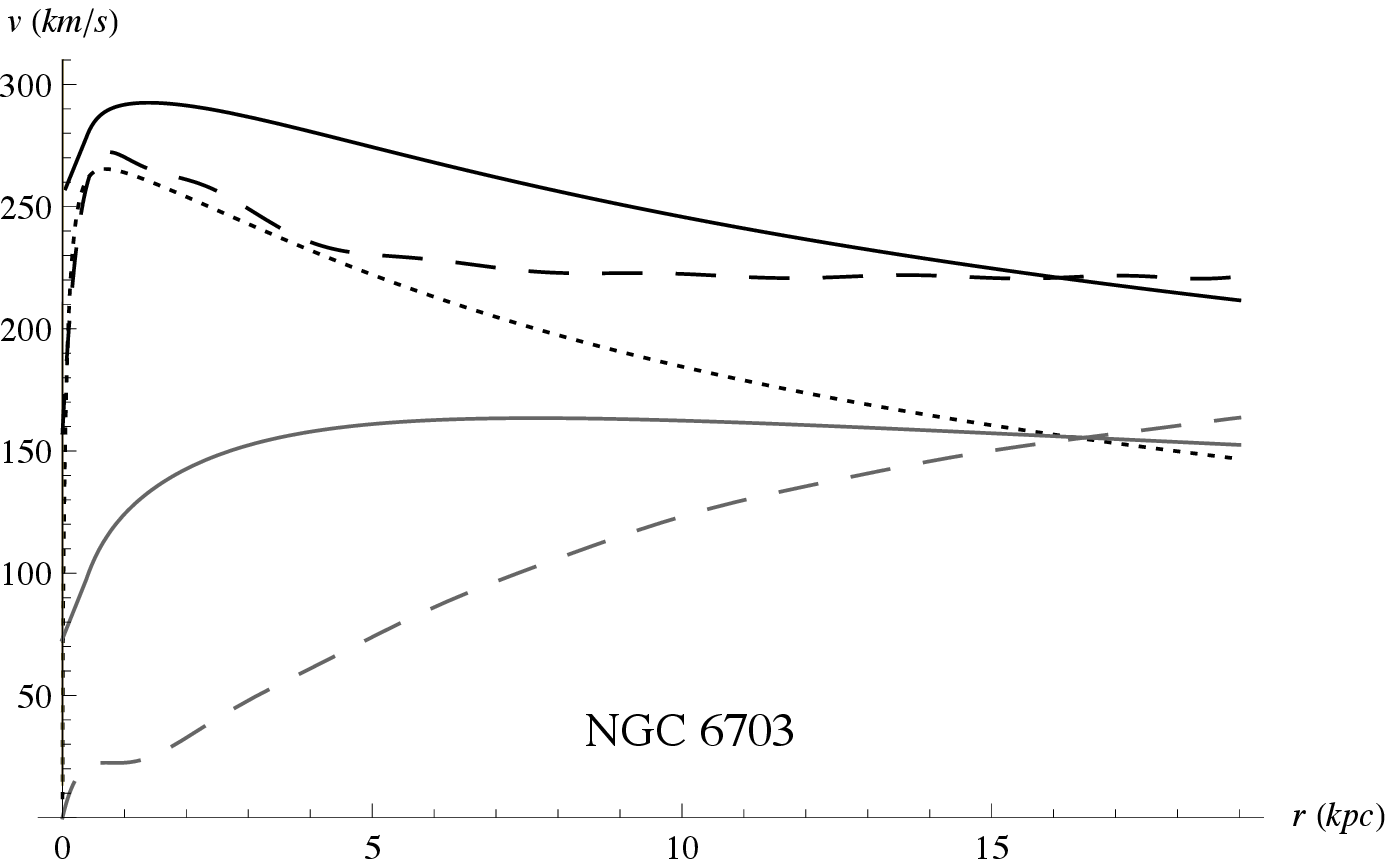} &
	\epsfxsize=7.5cm
	\epsffile{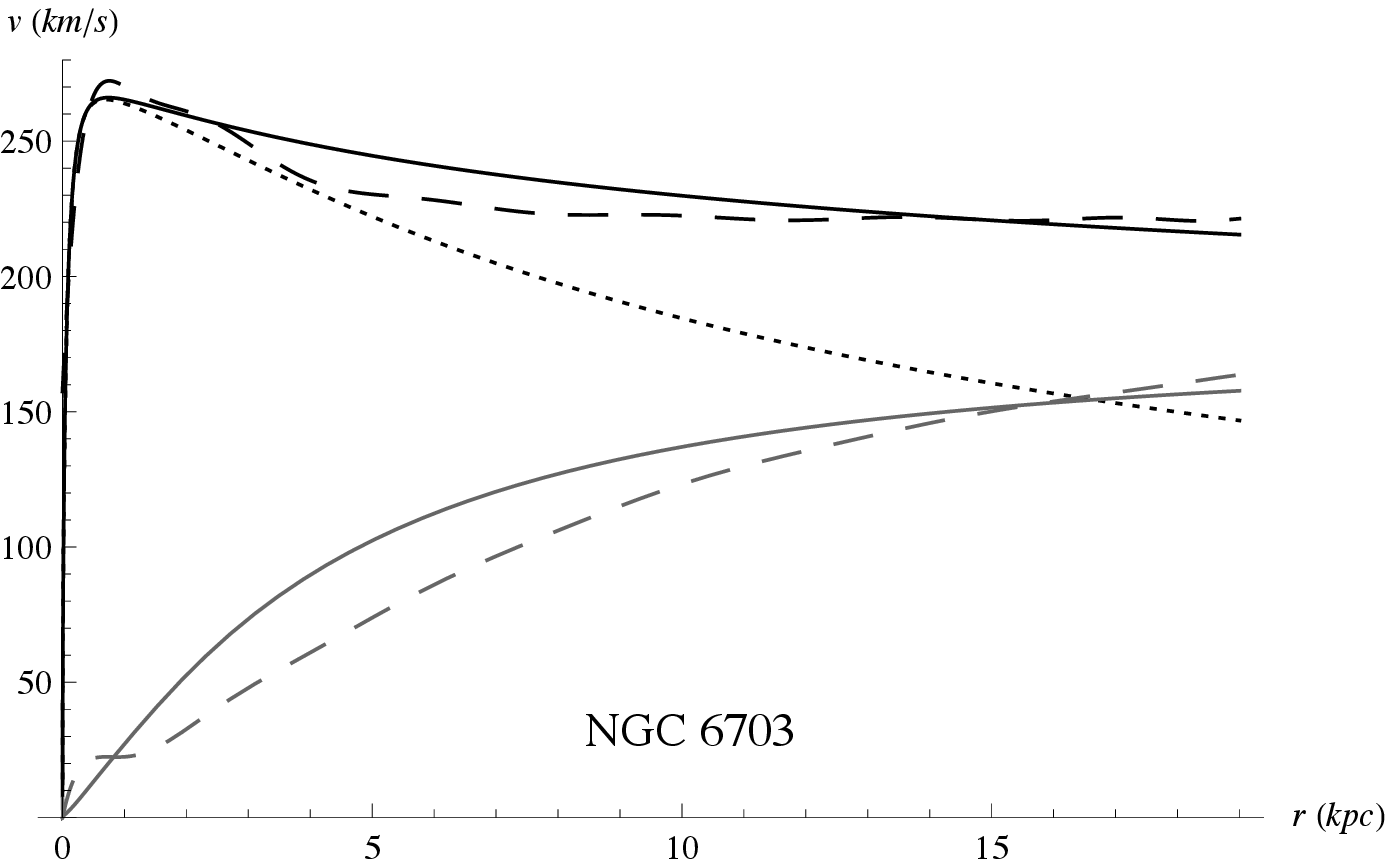} \\
		
	\epsfxsize=7.5cm
	\epsffile{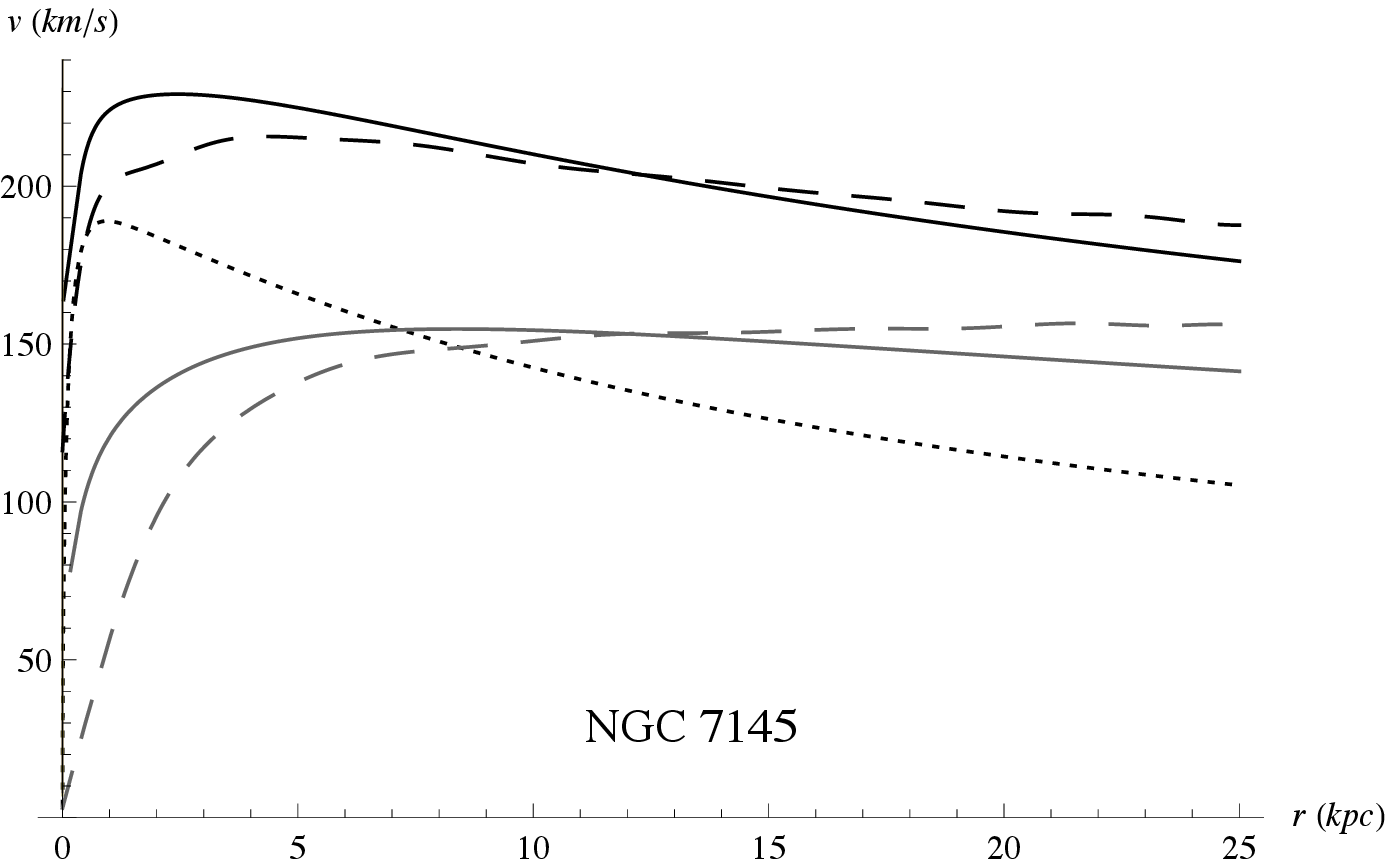} &
	\epsfxsize=7.5cm
	\epsffile{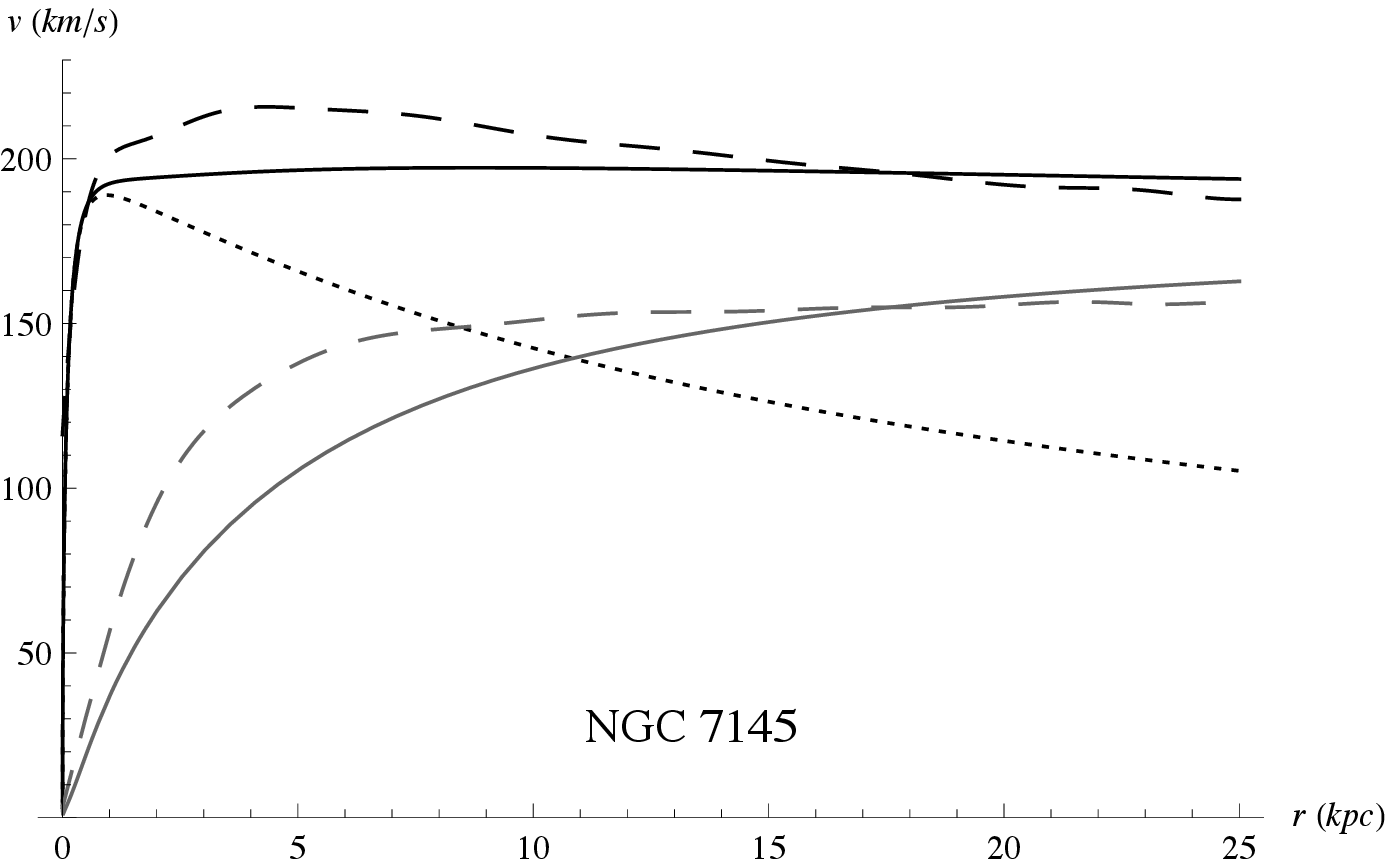} \\

\end{array}$
\end{center}
\caption{Observed rotation curve (dashed full), decomposed into visible (dotted) and dark matter (dashed grey) contributions  \cite{kronawitter}, superimposed with the mimicked dark matter profile (full grey) arising from the non-minimal coupling $f_2(R)=\sqrt[3]{R_3/R}$ (left) and $f_2(R)=R_1/R$ (right).}

\label{fig3}
\end{figure}

%%%%%%%%%%%%%%%%%%%%%%%%%%%%%%%%%%%%%%%%%%%%%%%%%%%%%%

\begin{figure}[h]
\begin{center}
$\begin{array}{cc}

	\epsfxsize=7.5cm
	\epsffile{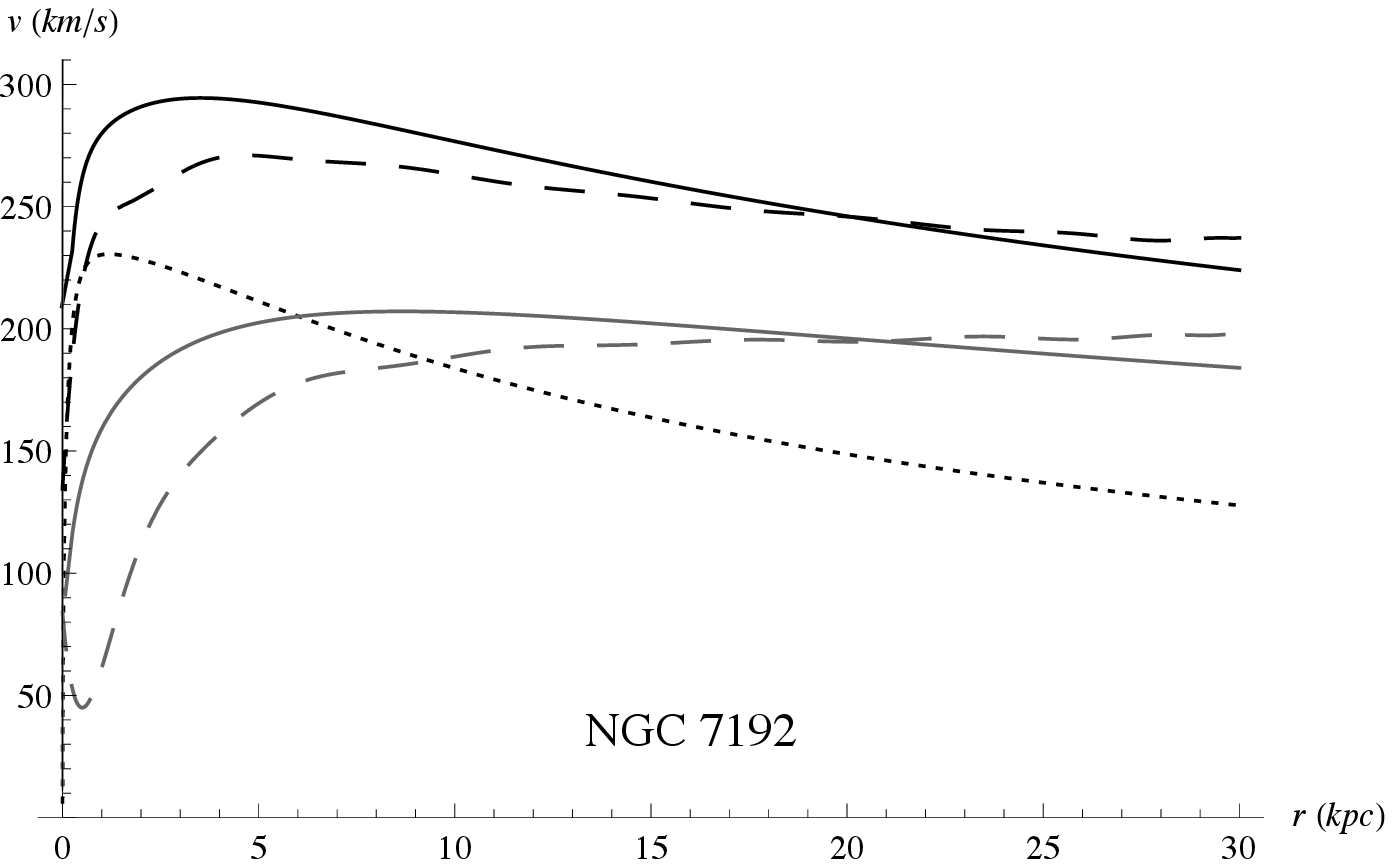} &
	\epsfxsize=7.5cm
	\epsffile{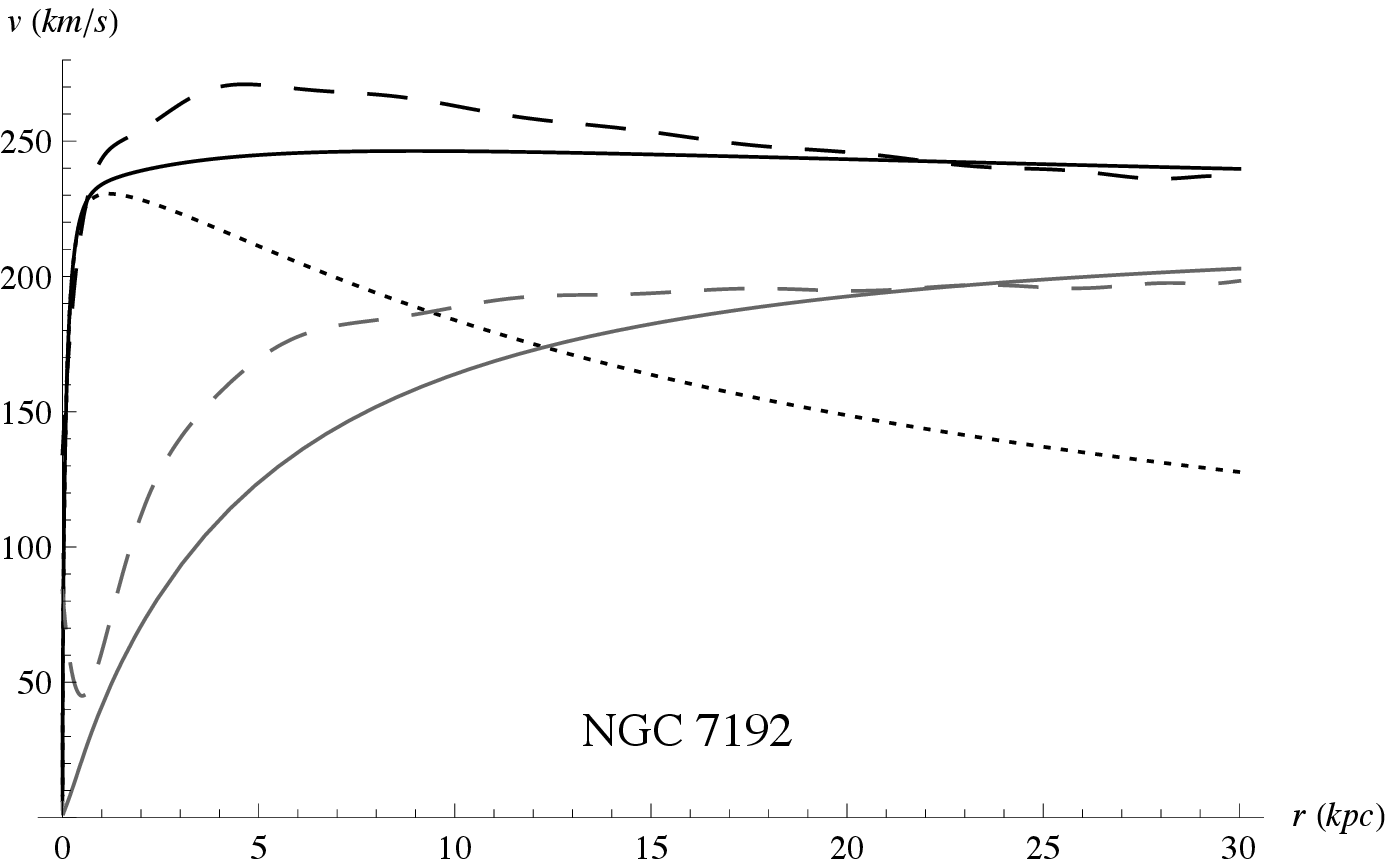} \\
		
	\epsfxsize=7.5cm
	\epsffile{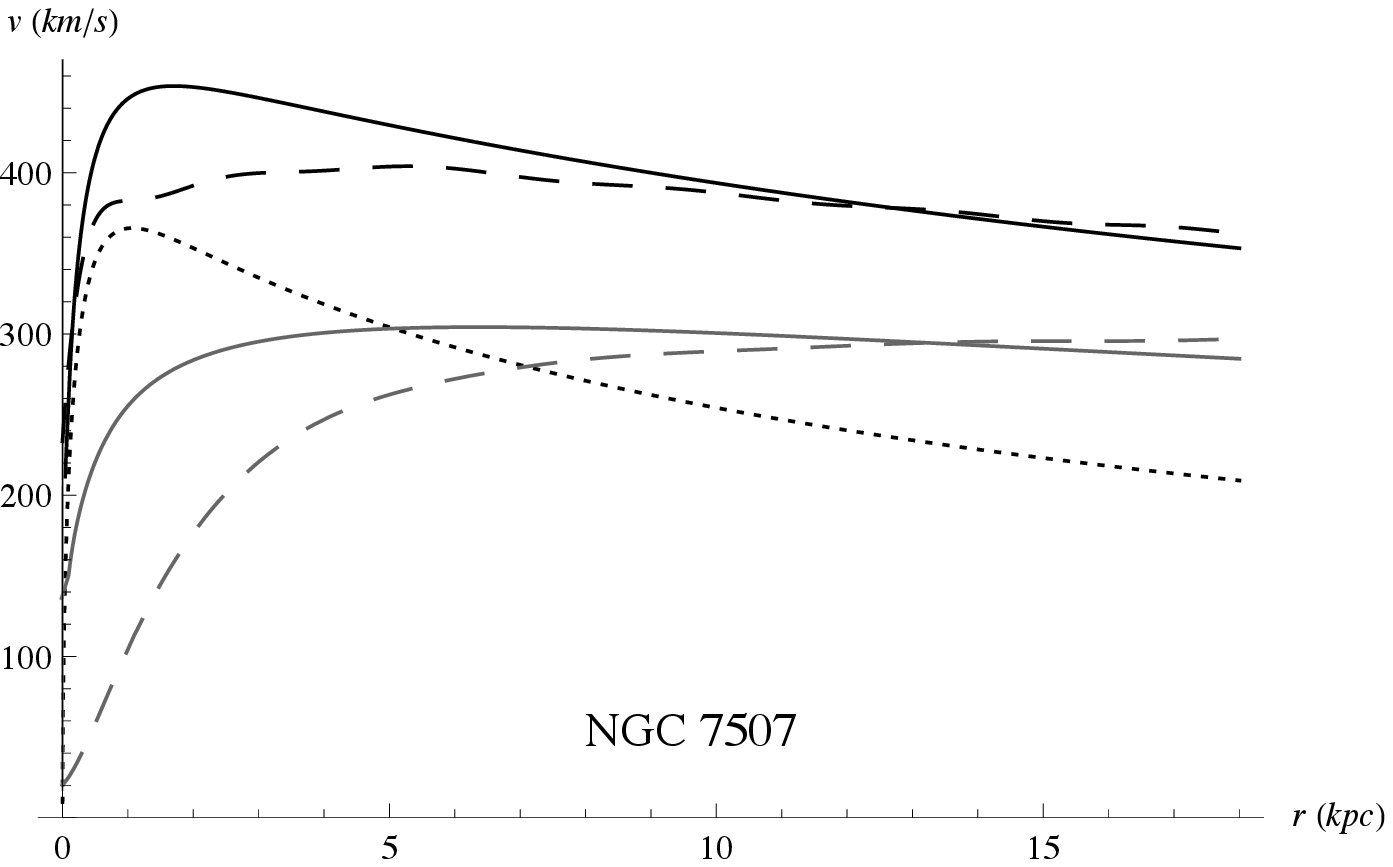} &
	\epsfxsize=7.5cm
	\epsffile{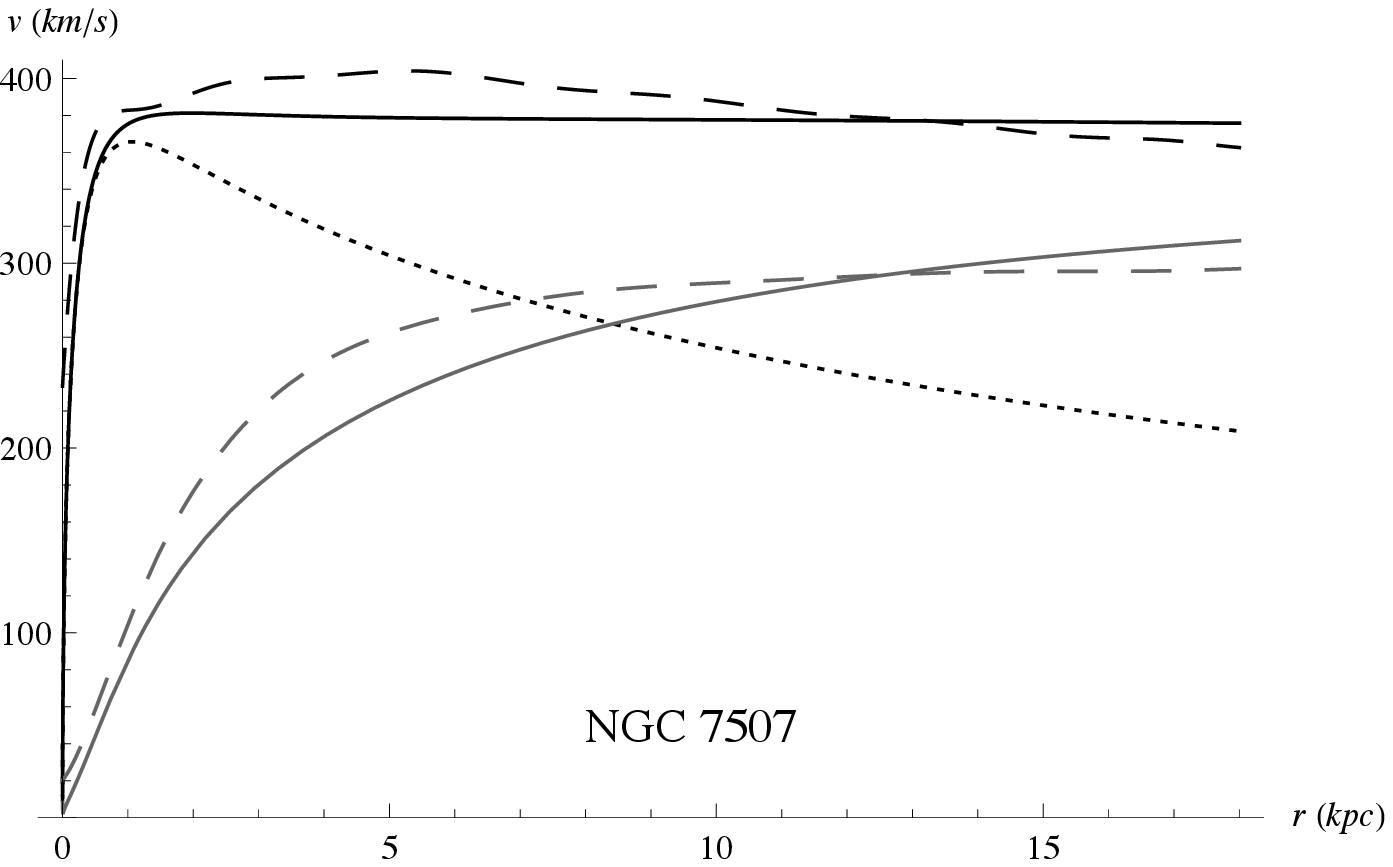} \\
		
	\epsfxsize=7.5cm
	\epsffile{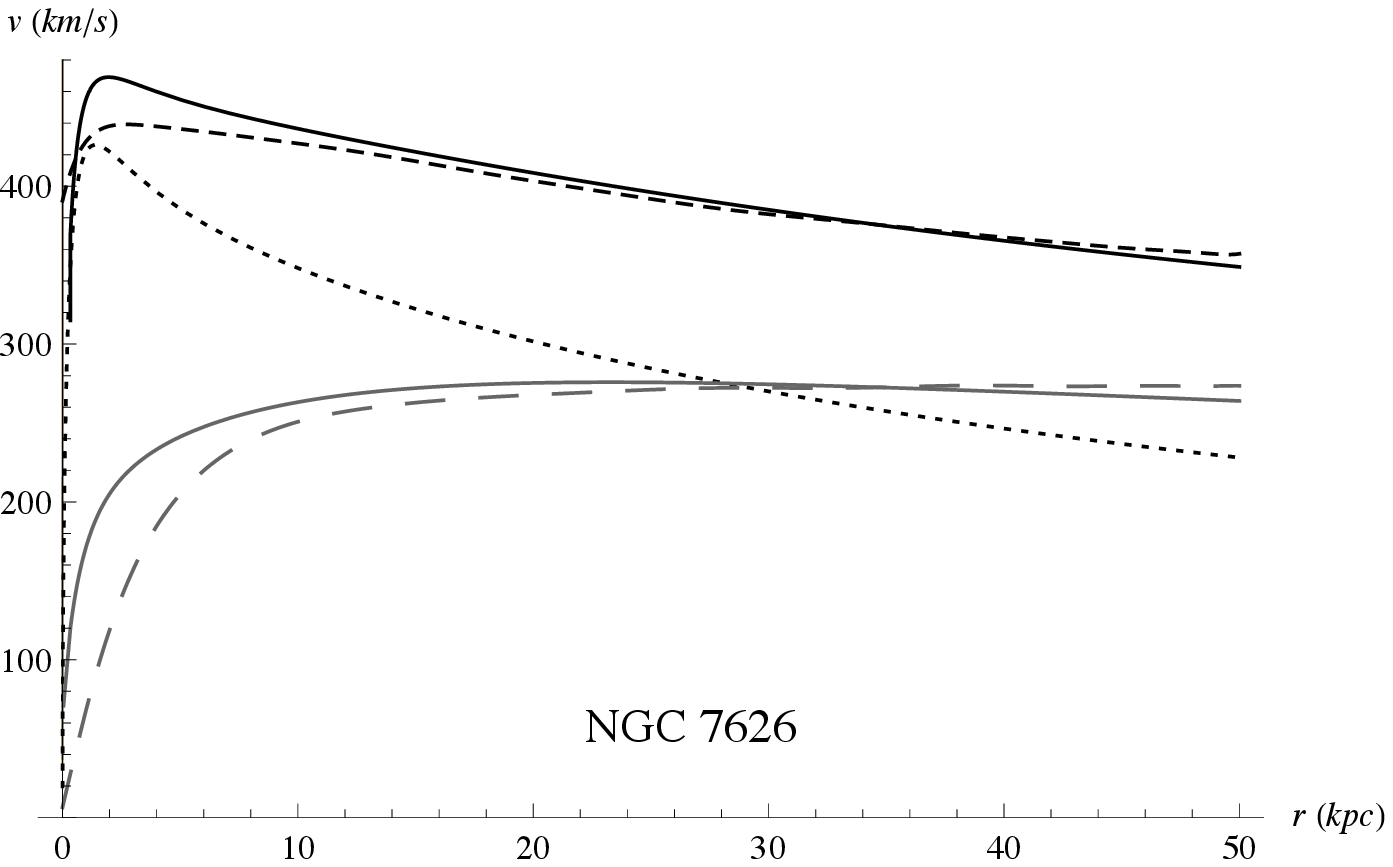} &
	\epsfxsize=7.5cm
	\epsffile{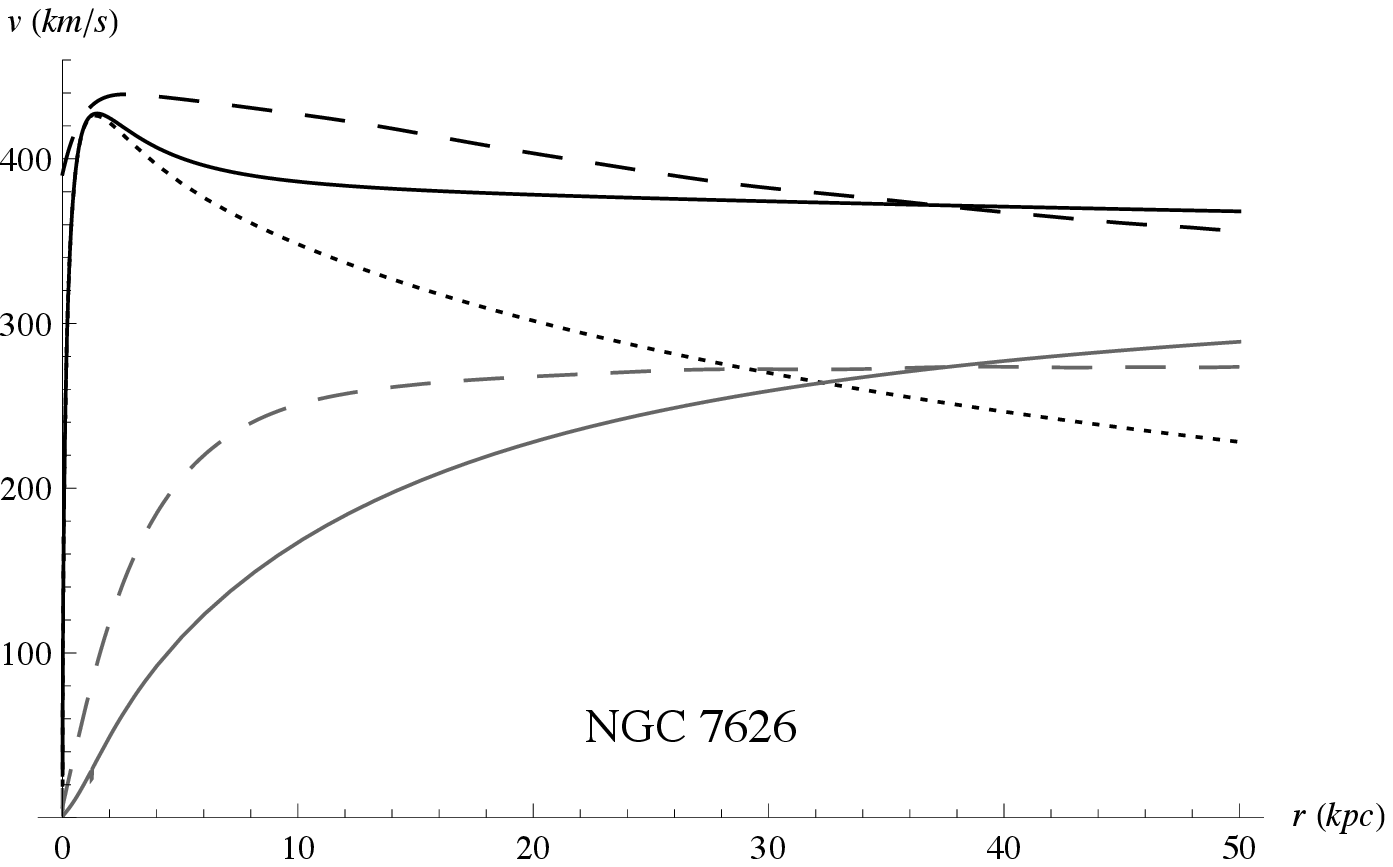} 

\end{array}$
\end{center}
\caption{Continuation of Fig. 4.}

\label{fig4}
\end{figure}

%%%%%%%%%%%%%%%%%%%%%%%%%%%%%%%%%%%%%%%%%%%%%%%%%%%%%%
%%%%%%%%%%%%%%%%%%%%%%%%%%%%%%%%%%%%%%%%%%%%%%%%%%%%%%

\subsection{Characteristic density and background matching and mimicked ``dark matter'' inter-dominance}

As can be seen, the rotation curves of the selected galaxies are fitted by values for the characteristic lengthscales of the order $r_1 
\sim 10~Gpc$ and $r_3 \sim 10^5-10^6~Gpc$. Keeping in mind that the analytic results obtained in the previous sections concern a 
single power-law non-minimal coupling only, not the composite form \eq{f2two}, one may nevertheless use \eq{matchingr} to obtain 
a quantitative figure for the background matching distances,

\beqa r_\infty &=& \left[{2\over 3} \times 3^n (1-2n) 
\right]^{1/4} \left( {c \over r_0 H } \right)^{(1-n)/2} \left( r_s r_0^2  a\right)^{1/4}  \nonumber \approx \\ 
&\simeq &\cases{
\left( {3.8~Gpc \over r_3} \right)^{2/3} \left( r_s r_3^2  a\right)^{1/4}   & , $n = -1/3$
\cr \left( {3.8~Gpc \over r_1} \right) \left( r_s r_1^2  a\right)^{1/4}  &,  $n=-1$
}, \eeqa

\noindent after inserting the Hubble constant $H_0 = 70.8 ~(km/s)/Mpc$. Similarly to the caveat above, one also notices that this 
expression was obtained with the assumption of a single Hernquist profile for the visible matter distribution, not the two component 
model used in the current fitting session. Nevertheless, one can compute the matching distances $r_\infty$ for the obtained values of 
$r_1$ and $r_3$ for each galaxy, taking again $a \sim a_1$; the result is displayed in the last columns of Table \ref{table}. 

Recalling the interpretation of $r_\infty$ as an indicator of the radius of the mimicked ``dark'' matter haloes, Table \ref{table} clearly 
shows that the obtained $n=-1/3$ NFW haloes are always smaller than those conforming to the $n=-1$ isothermal sphere profile 
(except for NGC 2434, which presents a best fit scenario with a single NFW ``dark matter'' component). Also, notice that $r_
{\infty~1}$ is always larger than the range of the rotation curves, indicating that the isothermal sphere ``dark matter'' component 
extends further than observed; on the contrary, the value of $r_{\infty~3}$ goes from about one half to less than twice the endpoint 
distance $L$ of the rotation curves ($r_{\infty~3}/L = 0.57 $ for NGC 7145 and $1.73$ for NGC 7507), indicating that the NFW halo is 
indeed less relevant than the isothermal sphere one in the composite scenario, where both contribute to the overall 
``dark matter'' component. Even the odd case of NGC 2434, which displays no dynamically generated isothermal sphere 
component, yields a matching distance $r_{\infty~3} = 33.1~kpc$ of only about twice the endpoint of the corresponding rotation 
curve, $L \simeq 17~kpc$.

As expected, the matching with the cosmological background density for the $n=-1/3$ NFW scenario occurs before that of the 
isothermal sphere, $r_{\infty~3} < r_{\infty~1}$, since the latter falls as $r^{-2}$, while the former drops more steeply as $r^{-3}$. 
Adding to this, the difference in scale between the characteristic densities $\rho_1 \gg \rho_3$ (since $\rho_i = 2\ka R_i = 2\ka / 
r_i^2$) should enhance this difference, not attenuate it. To ascertain this, one resorts to the dependence of the mimicked ``dark'' 
matter densities $\rho_{dm~1}$ and $\rho{dm~3}$ on the latter: recalling \eq{rhodm}, one has that

\beq \rho_{dm~1} = \rho_1^{1/2} \rho_c^{1/2} \left({a \over r}\right)^2 \qquad,\qquad \rho_{dm~3} = \rho_3^{1/4} \rho_c^{3/4} \left({a 
\over r}\right)^3,\eeq

\noindent where $\rho_c = M/(2\pi a^3 ) = $ sets the visible matter density profile. Since $\rho_1^{1/2} \propto 1/r_1$ and $\rho_3^
{1/4} \propto 1/\sqrt{r_3}$, one concludes that 

\beq {\rho_{dm~3} \over \rho_{dm~1}} = \left({r_1^2 \over a r_3 }\right)^{1/2} \left({r_s\over a}  \right)^{1/4} \left({a \over r}\right) \sim {1 
\over 10} {a \over r}  ,\eeq

\noindent using $r_s = 2GM$ and inserting typical values $r_s \sim 10^{-2}~pc$, $a \sim10~kpc$, $r_1 \sim 10~Gpc$ and $r_3 \sim 
10^6~Gpc$. Hence, the NFW component of the mimicked ``dark matter'' typically becomes dominated by the isothermal sphere halo 
for $r > a/10 \sim 1~kpc$: this can be interpreted as an indication that the considered galactic rotation curves are ``mostly flat'' (again, 
with the exception of NGC 2434).

\subsection{Possible cosmological relevance}

The above discussion serves to assess the validity of the proposed mechanism of dynamical ``dark matter'' arising from a non-minimal coupling: 
the matching distances naturally assume values within the astrophysical context. This does not go against the previous claim that the 
obtained values for $R_1$ and $R_3$ could give credence to the hypothesis of a relevant cosmological effect of the proposed 
model, thus enabling a possible unification of dark matter and dark energy.

For simplicity, one retains the lengthscales $r_1$ and $r_3$ as fundamental, and convert the cosmological background density $
\rho_{\infty}= 3 H^2 / 8 \pi G$ by defining the corresponding lengthscale $ (\rho_{\infty}/2\ka)^{-1/2} = r_H/\sqrt{3}$, with $r_H = c /H = 
4.2~Gpc$ the Hubble radius. With this notation, follows from the discussion of paragraph \ref{subsubsectionmatching} that the non-
minimal coupling \eq{f2two} reads, in a cosmological context, as

\beq f_2 = \sqrt[3]{R_3 \over R} + {R_1 \over R} \simeq \left( { r_H \over r_3} \right)^{2/3} + \left( { r_H \over r_1} \right)^2 \sim 10^{-4} + 
10^{-1} , \eeq

\noindent inserting the typical values $r_1 \sim 10~Gpc$ and $r_3 \sim 10^5-10^6~Gpc$.

Given the form of the modification \eq{action}, the above values hint that the $n=-1/3$ non-minimal coupling should not have a 
strong manifestation at a cosmological level; the strength of the $n=-1$ power-law coupling, however, approaches unity and could 
lead to potentially interesting cosmological effects,  i.e. a dynamically generated model for dark energy. Following the dominance of 
the gradient term in the {\it r.h.s.} of \eq{eqmotiontwo} in the astrophysical setting, it should be remarked that both couplings could 
possibly be enhanced by the corresponding time-derivative terms in the Friedmann equations. This possibility will be developed elsewhere.

\subsection{Universality of the model}

As depicted above, the obtained values for $r_1$ and $r_3$ show a variation between galaxies that is clearly not compatible with 
the predicted universality: $r_1$ averages $\bar{r}_1=21.5 ~Gpc$ with a standard deviation $\si_1=10.0~Gpc$, while $r_3$ 
presents $\bar{r}_3 = 1.69 \times 10^6~Gpc$ and $\si_3 = 1.72 \times 10^6 ~Gpc$. Although undesired, this is not unexpected, as 
there are several causes that might contribute to this deviation: although the selected type E0 galaxies are spherical at large, the 
small asymmetry due to the semi-axis difference $1 - b/a \lesssim 0.1$ could translate into the difference in obtained values for $r_i$. 
Furthermore, localized features and inhomogeneities of galaxies could contribute for the breaking of spherical symmetry and, if 
dense enough, serve as ``seeds'' for the dynamical generation of the mimicked ``dark matter'' component --- or, due to the 
dominance of the gradient term in the {\it r.h.s.} of \eq{eqmotiontwo}, if they vary significantly on a short lengthscale.

Even if spherical symmetry were fully enforced, the real visible matter density profile might differ slightly from that obtained from the 
fitting of the visible component of the rotation curves, either due to a misconstruction of the latter or from an inadequacy of the choice 
of the Hernquist profile as a fitting function. Notice that this mismatch does not need to be large, so that the quality of the rotation 
curves reported in Ref. \cite{kronawitter} or of the related two-component Hernquist fits here used should not be held as an immediate 
culprit. However, one can envisage that a small deviation between the real and the fitted density profiles could include a large 
deviation of its first and second derivatives in very localized regions: again, since the {\it r.h.s.} of \eq{eqmotiontwo} is dominated by 
the gradient term, this could lead to a large deviation of the obtained ``dark'' matter component.

Even if the simple modelling procedure followed in this study is sufficiently accurate, more fundamental issue could be responsible 
for the variation of the obtained values for $r_1$ and $r_3$: first of all, the non-minimal coupling \eq{f2two} could be too simplistic, 
and a more generalized model of the form

\beq f_2(R) =\sqrt[3]{R_3 \over R}\left[1+g_1(R)\right] + {R_1 \over R}\left[1+g_2(R)\right] , \eeq

\noindent could yield the desired universality for the related parameters $R_1$ and $R_3$. In this sense, the current proposal aims 
at the the general depiction of the dark matter mimicking mechanism, and the functions $g_i(R)$ could be read from a more 
thorough adjustment with galactic rotation curves. 

One could also assume that the assumption of a linear curvature term $f_1(R) = 2\ka R$ might lead to the observed deviation from universality: indeed, power-law terms $f_1(R) \propto R^n$, as reported in Ref. \cite{CapoLSB}, would also contribute to the dynamical generation of ``dark matter''. If the exponents are positive $n>0$, this effect would be strongly felt at regions with higher curvature (i.e. density), closer to the galactic core: the failure to include these could perhaps account for the missing ``dark matter'' in the inner region of the obtained rotation curves.
 
A second, perhaps more interesting issue could lie behind the obtained discrepancy: the non-minimal function $f_2(R)$ couples 
with the visible matter Lagrangian density, which is assumed to describe a perfect fluid form, ${\cal L}_m = -\rho$. This can be 
relaxed in two ways: firstly, the perfect fluid model could be proven insufficient, and one should opt for a more adequate form ${\cal 
L}_m = {\cal L}_m(\rho)$ (or one depending on more variables related to the thermodynamical description of the matter distribution).

Moreover, and perhaps more interestingly, one could invoke another related possibility, one with rather strong implications at a fundamental level: the non-minimal 
coupling does explicitly break the equivalence principle at large scales. This is not in conflict with laboratory determination of this principle, which are eminently local and follow the reasoning that a violation of this principle is expected, for instance, due to the interaction of dark matter and dark energy \cite{interaction2,interaction1}, or due to infrared gravity effects with some bearing on the cosmological constant problem \cite{guide}. Since this interaction (which might be modelled with non-trivial terms $f_1(R)$ or $f_2(R)$, as discussed before) should depend on the size of the galaxy itself, this could lead to the variation of the $r_1$ and $r_3$ parameters.

\section{Conclusions}

In this work one has addressed the issue of obtaining a solution to the dark matter puzzle, as embodied by the flattening of 
galaxy rotation curves. This is done by following the main phenomenological implications of models endowed with a non-minimal 
coupling of matter to curvature.

In order to do so, one has first addressed the possibility of obtaining an adequate extra force from the non-conservation of the 
energy-momentum tensor of matter, for a completely flat and a decaying rotation velocity curve. In the former case, 
it was concluded that this required a logarithmic coupling of the form $\la f_2(R) = -v^2/m \log (R/
R_*)$ (where $m$ is the outer slope of the visible matter density $\rho$), which might by fitted by a simpler power-law $\la f_2(R) \approx (R_*/R)^\al$, with the asymptotic velocity given by $v_\infty^2 = \al m$. However, this solution suffers 
from a lack of flexibility that is not observed in nature, given the observed variety of the velocity dispersion of rotation curves, while the obtained solution implies an almost universal $v_\infty$.

The second attempt was to solve this caveat, bearing in mind the phenomenological Tully-Fisher law, which roughly 
states a power-law relation between $v_\infty$ and the total visible mass $M$: this is obtained by assuming that geodesical motion 
is indeed preserved, $\nabla^\mu T_\mn = 0$, but that the metric itself is perturbed, thus providing the extra ``force'' 
required. In order to achieve that, one must solve the trace of the Einstein equations of motion, with the difference $R - 2 \ka \rho$ 
providing the dynamically generated ``dark matter''. We resorted to a power-law coupling with matter $f_2(R) = (R/R_0)^n$: since we 
expect the added dynamics to be manifest at large distances ({\it vis a vis} low density and curvature), this requires a negative index 
$n$. 

This approach yields a dark matter equivalent density directly related to a specific power of the visible matter density, thus 
accounting 
for the Tully-Fisher law in a natural way. Furthermore, the obtained dark matter component exhibits a negative pressure, a 
feature typical of dark energy models: this might hint at the potential of the considered model to unify both dark components 
of the Universe.

Two different scenarios were considered: the mimicking of the NFW and the isothermal sphere dark matter profiles ($n=-1/3$ 
and $n=-1$, respectively). Since a separate fit of the rotation curves with each power-law function $f_2(R) = (R_0/R)^n$ did not 
produce results deemed satisfactory, a composite model for this non-minimal coupling was considered, allowing for a greatly 
improved adjustment.

The value of the characteristic lengthscales $r_1$ and $r_3$ was obtained for each galaxy, instead of a simultaneous fit to all 
rotation curves: this yielded the order of magnitudes $r_1 \sim 10~Gpc$ and $r_3 \sim 10^5 ~Gpc$. This result allowed for the 
computation of the cosmological background matching distances for each galaxy (which depend on their characteristic lengthscale); 
the obtained astrophysical range served to validate the proposed mechanism and conclude that the $n=-1$ isothermal sphere 
``dark matter'' halo dominates the $n=-1/3$ NFW component.

Furthermore, the $n=-1$ scenario was also shown to possess possible cosmological implications, since $r_1 \sim r_H$, the latter 
being the Hubble radius. This lends support to the proposed physical description, as well as to the hypothesis of a successful 
unification of the dark matter and dark energy components of the Universe.

Given the lack of the desirable universality in the model parameters $R_1$ and $R_3$, several possible causes for the obtained 
variation were discussed: asides from the validity of the assumption of spherical symmetry, more fundamental issues were 
discussed --- including a more complex form for the non-minimal coupling between geometry and matter or the curvature term $f_1(R)$, or for the Lagrangian 
density of the latter. A possible violation of the equivalence principle that depends on the physical size of a galaxy could also play a role in this variation.

Although this work can be regarded as a first step, one concludes that the rich phenomenology stemming from a 
non-minimal coupling between matter and curvature enables an alternative description to dark matter. This should be contrasted 
with other candidate models, namely the MOdified Newtonian Dynamics (MOND), which is by itself purely phenomenological, or 
based upon an extensive paraphernalia of vector and scalar fields (the underlying Tensor-Vector-Scalar (TeVeS) model) 
\cite{MOND1,MOND2,MOND3,MOND4} (see Ref. \cite{MONDcritic} for a critical assessment).

Finally, we remark that the proposed model should not be held as a competing proposal to standard $f(R)$ theories (that is, with $f_1(R) = f(R)$ and $f_2(R) = 1$), but as complementary: it has been shown that a general action with non-trivial $f_1(R)$ and $f_2(R)$ dependences is equivalent to a scalar-tensor theory exhibiting the proposed non-minimal coupling (as studied in Ref. \cite{scalar} and discussed in Appendix B), so that both may be regarded as limiting cases. Indeed, in the metric formalism (where the Christoffel symbols are assumed to depend on the metric), $f(R)$ gravity is equivalent to a Brans-Dicke (BD) theory with a vanishing BD parameter $\om = 0$ and a suitable potential, together with a non-minimal gravitational coupling of the BD scalar field.

From a physical standpoint, the possibility of describing galaxy rotation curves via a non-minimal gravitational coupling (as obtained here and in Refs.\cite{CapoLSB,LoboLog}) may be interpreted straightforwardly: traditionally, the discrepancy between the observed velocity dispersion and the lower predicted value is due to the addition of some missing ``dark'' matter form. By resorting to a non-minimal gravitational coupling, either explicitly or via the discussed scalar-tensor equivalence, one is instead asserting that this disagreement is solved not by an additive mechanism, but a multiplicative one: the influence of the existing visible matter is suitably increased in the outer galactic region by the non-minimal coupling. The identification of this heightened effect as dark matter, albeit artificial, then serves the purpose of contextualizing the obtained results within the vast literature on the subject.

\appendix{}
\section{Power-law density profiles}

A plethora of different density profiles exist for both visible as well as dark matter components, arising both from observations as well 
as N-body simulations. The key feature of the  proposed approach is the ability 
to mimic dark matter in the outer region, where the curvature is low enough so that the effect of 
the non-trivial, inverse power-law gravitational coupling $f_2(R)$ becomes manifest. For this reason, it is 
not necessary to fully account the specificity of each of the competing models: it is sufficient that 
these exhibit a behaviour which is dominated by a power-law for distances above a certain threshold $a$.

Several models exhibit this feature: amongst these, one of the most discussed are the so-called 
generalized (spherical) cusped profile \cite{profiles1,profiles2,profiles3,profiles4}, given by

\beq \rho = {\rho_{cp} \over \left({r \over a} \right)^\ga \left(1 + {r \over a} \right)^{m- \ga} } , 
\label{cusped} 
\eeq

\noindent where $\rho_{cp}$ sets the density scale, $a$ the length and $\ga$, $m$ are the inner 
and outer 
slopes, respectively. 

Typical cusped density profiles include the $\ga = 1$, $m=3$ NFW profile for dark matter \cite{NFW}, and the $\ga = 1$, $m = 4$ 
Hernquist profile for visible matter \cite{Hernquist}. The latter yields a finite total mass, $ M = 2 \pi a^3 \rho_{cp}$; substituting into \eq
{cusped}, one also gets the mass enclosed within a sphere or radius $r$,

\beqa \rho (r) &=& {M \over 2\pi }{a \over r} {1 \over \left(r+a \right)^3} \rightarrow \\ \nonumber 
M(r) &=& M \left({r \over r+a } \right)^2,\eeqa

\noindent so that half of the total mass $M$ is contained within $r_{1/2} = (1+\sqrt{2})a $.

A more general model is the so-called $(\al,\be,\ga)$ model \cite{abc}, where the density is given by

\beq \rho(r) = \rho_v 2^{(\be-\ga)/\al} \left( {r \over a} \right)^{-\ga} \left[ 1 + \left( {r \over a } \right)^
\al 
\right]^{(\ga-\be)/\al}, \eeq

\noindent which, for $r \gg a$, assumes the simpler power-law $\rho(r) \propto r^{-\be}$.

The above models are all divergent as $r \rightarrow 0$, thus to alleviate this and accommodate the 
apparent 
flat density profile observed near the core, some other proposals have been put forward, 
including the 
Burkert profile \cite{Burkert}, given by

\beq \rho(r) = \rho_v {a^3 \over (r + a) (r^2 + a^2) } , \eeq

\noindent which, in the outer region $r \gg a$, behaves as $\rho(r) \propto r^{-3} $.

All the above models present a shift in behaviour at $r \approx a$, changing the slope of the 
inverse 
power-law from (assuming the notation of the Hernquist profile) $\ga$ to $m$. It should be 
remarked that 
the mechanism considered in this work for mimicking the outer density profile of dark matter is 
only valid 
for this region, since only by accident can the scaling law \eq{rhodm} yield the $r^{-m'}$ behaviour 
for the 
dark matter component from the $r^{-m}$ power law for visible matter, and simultaneously 
produce the 
$r^{-\ga'}$ law for the former, from the $r^{-\ga}$ profile for the latter (this is the case if one 
assumes the 
Hernquist profile for visible matter and the NFW profile for dark matter, as both exhibit $\ga = \ga' 
= 1$). 
This caveat from the current proposal, although relevant, does not hinder its main advantage: to 
reproduce 
the dark matter halo in the outer region, where it is most relevant.

Finally, one presents what is arguably the simplest of dark matter density profiles, the singular isothermal sphere, given by

\beq \rho(r) = {v_\infty^2 \over 2\pi G r^2},\eeq
 
\noindent where $v_\infty$ is the asymptotic velocity in the flattened region ( i.e. the velocity dispersion). Indeed, this 
allows for a straightforward computation of a flattened rotation curve, with $v(r) = v_\infty$ everywhere.

The isothermal sphere may be regarded as a generalized cusped profile \eq{cusped}, with an inner slope $\ga = 0$, outer 
slope $m = 2$, and vanishingly small $a \rightarrow 0$, while keeping $\rho_{cp}a^2 = v_\infty^2 / (2\pi G)$.

One might consider a different family of density profile models, which do not assume an inverse 
power-law 
behaviour at all, but instead rely on a modified exponential decay, usually depicted through 
one of two closely related models: the Einasto $R^{1/p}$ density profile \cite{Einasto}

\beq \rho(r) = \rho_v \exp \left\{ -d_p \left[ \left({r \over a} \right)^{1/p} - 1 \right] \right\},\eeq

\noindent where $d_p$ is an adequate function of $p$, such that $r=a$ encloses half of the total 
mass; 
and the deprojected S\'ersic law \cite{Sersic1,Sersic2,Sersic3}

\beq \rho(r) = \rho_v \left( {r \over a} \right)^{-p_m} \exp \left[ -d_m \left({r \over a} \right)^{1/m} 
\right],
\eeq

\noindent with $p_m$ and $d_m$ functions of $m$.

Most simulations assume a power-law density for visible matter (most often, the Hernquist profile), 
while 
the Einasto and deprojected S\'ersic law compete with the NFW density profile in describing 
dark 
matter halos. Simulations using a deprojected S\'ersic profile predict that the exponents 
for 
visible and dark matter are $m_* = 4.29$ and $m_{dm} = 2.96$, respectively  \cite{both}.

It is straightforward to see that exponential density profiles such as those depicted above would 
also be 
obtained for the dark matter component, if one assumed similar model for visible matter to 
begin with; 
however, the scaling law \eq{rhodm} will only change the overall value of $p_m$ and $d_m$, 
through

\beq p_{dm} = p_* {1 \over 1-n} ~~, \qquad ~d_{dm} = d_* {1 \over 1-n}, \eeq

\noindent while the exponent $1/m$ will remain unaltered by the scaling. Given that $p_m$ and 
$d_m$ are 
directly dependent on this quantity, one concludes that the proposed mechanism cannot 
satisfactorily 
reproduce the above exponential density profile for dark matter starting from a similar model for 
visible 
matter. This is a caveat of the current work --- although it should be stated that the specific choice 
$f_2(R) 
= (R/R_0)^n$ was tailored to deal with the power-law scaling, and a more appropriate choice 
might resolve 
this issue. This will not be discussed here, since the underlying idea would remain the same: to 
mimic 
dark matter through an appropriately driven dynamical effect arising from the non-trivial 
gravitational 
coupling.

\section{Equivalence with multi-scalar theory}

Following Ref. \cite{scalar}, one may rewrite the two couplings obtained in sections \ref{geo} and \ref{metric} as a multi-scalar 
theory \cite{damour} with action 

\beq S = \int  \sqrt{-g} d^4x  \bigg[ 2 \ka \left[ R - 2g^\mn \si_{ij} \varphi^i_{,\mu} \varphi^j_{,\nu} - 4 
U(\varphi^1,\varphi^2) \right] + f_2(\varphi^2){\cal L^*} \bigg], \eeq

\noindent where $\varphi^1$ and $\varphi^2$ are scalar fields, $\si_{ij}$ is the field-metric

\beq \si_{ij} = \left(\begin{array}{cc}1 & 1 \\ -1 & 0\end{array}\right) ,\eeq

\noindent the potential is given by 

\beq U(\varphi^1,\varphi^2) =  {1 \over 4} \exp \left( -{2 \sqrt{3}\over 3} \varphi^1 \right) \left[\varphi^2 - 
{f_1(\varphi^2 )\over 2\ka }  \exp \left( -{2 \sqrt{3}\over 3} \varphi^1 \right)  \right], \eeq

\noindent and  ${\cal L^* } = \exp [-(4\sqrt{3}/3)\varphi^1]$.

The two scalar fields are related with the scalar curvature and the non-trivial $f_1(R)$ and $f_2(R)$ functions:

\beq \varphi^1 = {\sqrt{3}\over2} \log \left[ {F_1(R) + F_2(R) {\cal L} \over 2\ka} \right] ~~,\qquad  \varphi^2 = R , \eeq

\noindent where $F_i \equiv d f_i / dR$. Inserting ${\cal L} = \rho$, $f_1(R) = 2\ka R $ and $f_2(R) = (R/R_0)^n$, one gets

\beq \varphi^1 = {\sqrt{3}\over2} \log \left[ 1 + n\left( {R \over R_0}\right)^{n-1} {\rho \over \rho_0} \right] = 
{\sqrt{3}\over2} \log \left[ 1 + n \varrho ^{n-1} \right]. \eeq

\noindent Hence, one concludes that the dimensionless function $\varrho$ not only simplifies treatment of \eq{trace2}, but 
embodies the added degree of freedom resulting from considering a non-minimal coupling between matter and curvature.

\bibliography{mimicking}{}
\bibliographystyle{JHEP}

\end{document}